\pdfoutput=1 
\documentclass[
twocolumn,
 amsmath,amssymb,
 aps,
prb,
]{revtex4-2}

\usepackage{mathrsfs} 
\usepackage{graphicx}
\usepackage{tikz}
\usepackage{color}
\usepackage{dcolumn}
\usepackage{multirow}
\usepackage{bm}
\usepackage{braket} 
\usepackage{siunitx} 
\usepackage{comment}

\usepackage[
colorlinks=true,
citecolor=gray,linkcolor=gray, 
]{hyperref}

\newcommand{\Tr}{\mathrm{Tr}\ }
\newcommand{\diag}{\mathrm{diag}\ }
\newcommand{\Abs}[1]{\left|#1\right|}

\begin{document}

\title{
Spin Relaxation, Diffusion and Edelstein Effect in Chiral Metal Surface
}
\author{Yuta Suzuki}
\email{suzuki@vortex.c.u-tokyo.ac.jp}
\affiliation{
 Department of Physics, The University of Tokyo, Bunkyo, Tokyo 113-0033, Japan}
\author{Yusuke~Kato}
 \email{yusuke@phys.c.u-tokyo.ac.jp}
\affiliation{Department of Basic Science, The University of Tokyo, Meguro-ku, Tokyo 153-8902, Japan}

\date{\today}
\begin{abstract}
We study electron spin transport at spin-splitting surface of chiral-crystalline-structured metals
and Edelstein effect at the interface,
by using the Boltzmann transport equation beyond the relaxation time approximation.
We first define spin relaxation time and spin diffusion length
for two-dimensional systems with anisotropic spin--orbit coupling through the spectrum of the integral kernel in the collision integral. 
We then explicitly take account of the interface between the chiral metal and 
a nonmagnetic metal with finite thickness. For this composite system, we derive analytical expressions for
efficiency of the charge current--spin current interconversion as well as other coefficients 
found in the Edelstein effect.
We also develop the Onsager's reciprocity in the Edelstein effect along with experiments 
so that it relates local input and output, which are respectively defined in the regions separated by the interface. 
We finally provide a transfer matrix corresponding to the Edelstein effect through the interface, 
with which we can easily represent the Onsager's reciprocity as well as
the charge--spin conversion efficiencies we have obtained.
We confirm the validity of the Boltzmann transport equation in the present system starting from the Keldysh formalism in the supplemental material.  
Our formulation also applies to the Rashba model and other spin-splitting systems.
\end{abstract}

\maketitle


\section{Introduction}

Over the last three decades, there has been considerable interest in spin generation, spin detection and spin transport in surfaces, interfaces, and noncentrosymmetric crystal structures. 
One of the well-studied phenomena in this field is charge--spin interconversion by 
the Edelstein effect (EE)~\cite{Edelstein1990,Aronov1989,Kato2004a} 
and its reciprocal effect, i.e., the inverse Edelstein effect (IEE)~\cite{Ganichev2002,Shen2014}.
They have been explored both theoretically~\cite{Silsbee2004,Gambardella2011,Zhang2016,Dey2018} and experimentally at 
the Rashba spin-splitting surface~\cite{Sanchez2013,Zhang2015a,Lesne2016} or 
topological insulator surfaces~\cite{Shiomi2014,Sanchez2016}, where spin and momentum are perpendicularly coupled.

Recently, current-induced magnetization and its inverse effect have been observed in chiral-crystalline-structured metals, 
bringing a new perspective to the field of spin transport.
Experiments on paramagnetic phase in a chiral metal $\text{CrNb}_3\text{S}_6$~\cite{Inui2020,Nabei2020} and nonmagnetic chiral metals $\text{TaSi}_2$ and
$\text{NbSi}_2$~\cite{Shiota2021,Shishido2021} with $D_6$ (622) point group display 
parallel coupling of charge current and spin polarization in the direction of the principal axis,
associated with an external electric field or spin current injection to the metals.
The relative sign of the current and polarization depends on chirality of the metal,
which makes sure that the observed effects are unique to the chiral crystalline structure
\footnote{
Note that the observed spin polarization unique to the chiral metals cannot be explained as 
a linear spin Hall effect, as described in Ref.~\cite{Roy2022}.
The spin Hall conductivity relates spin current and electric field, which are odd under spatial inversion. 
It follows that the spin Hall conductivity itself 
is independent of whether the spatial inversion is included in the point group or not;
in particular, no linear spin Hall effect is unique to chiral crystals.

More precisely, in the point group 622 without magnetic orders, 
spin Hall conductivity vanishes when the electric field and spin polarization direction are parallel~\cite{Seemann2015}.
The spin Hall effect is thus unrelated to the observed spin polarization that is parallel to the applied electric field.
}.

The parallel current-induced magnetization, allowed in chiral systems~\cite{Furukawa2021,Yoda2015},
may be understood based on a microscopic spin--orbit coupling (SOC), which includes
parallel coupling of spin and momentum around the $\Gamma$ point~\cite{Frigeri2005}. 
However, there are two distinctive features within spin polarization in the chiral metals:
high efficiency of the charge--spin conversion and long-range spin transport.
They make this effect intriguing but challenging from a theoretical standpoint.
For the former feature, the current-induced magnetization is reported to be so large~\cite{Nabei2020}
that the spin polarization has been detected 
simply by attaching nonmagnetic metals onto the surface of chiral metals~\cite{Inui2020,Shiota2021,Shishido2021}.
That process has been phenomenologically explained as a spin diffusion across the interface.
Such a large polarization is not shown in elemental tellurium~\cite{Furukawa2017,Furukawa2021},
and may be characteristic to the chiral metals. 
For the latter feature, the chiral metals are reported to have robust spin polarization, 
which persists over millimeters even in the absence of net charge current. 
That length scale is much longer than typical spin diffusion length in metals,
and the origin of such nonlocality is still under discussion~\cite{Tatara2022a,Roy2022}.

With these backgrounds, highly required is a theoretical scheme (a) having a firm basis and (b) capable of dealing with non-local spin transport in the presence of anisotropic SOC as well as (c) the charge--spin interconversion through an interface between a chiral metal and nonmagnetic achiral metal with finite thickness. We aim to present a prototypical model satisfying those conditions.

\begin{figure}[htbp]
\centering
\includegraphics[width=0.7\columnwidth]{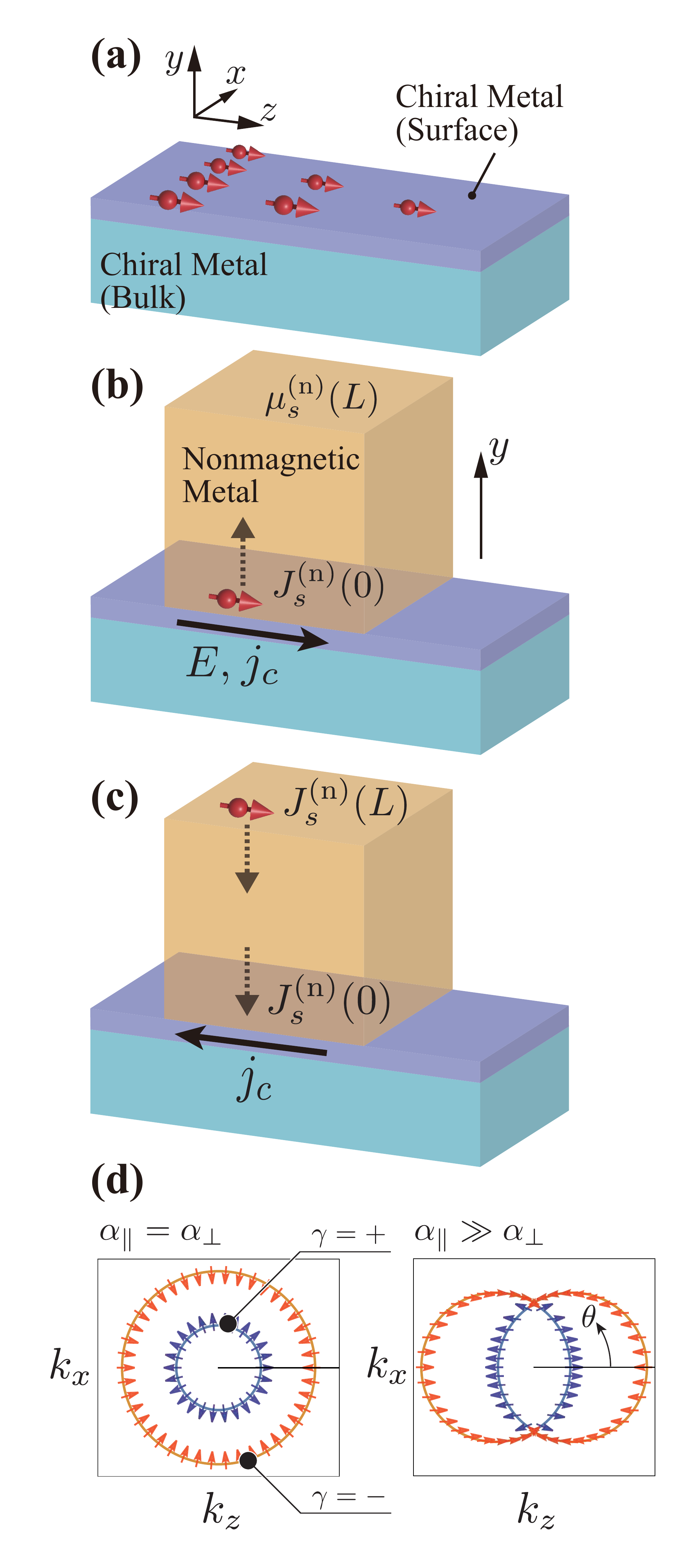}
\caption{(Color online)
Schematic pictures of temporal spin relaxation and spatial spin diffusion~[(a)] at the chiral metal surface,
direct Edelstein effect~[(b)], inverse Edelstein effect~[(c)] across the interface, 
and radial spin texture of the Fermi contours at the surface~[(d)].
Arrows denote spin polarization.
(d)~The spin--orbit coupling is set to be isotropic (left) and 
anisotropic (right, $\delta= \pi/64$) with the same strength of SOC $\alpha/v_{\rm F}\hbar = 0.6 $.}
\label{fig: schematic pic spin texture}
\end{figure}

In this paper, we study spin and charge transports in two-dimensional (2D) metals 
with an anisotropic SOC with weak disorder due to nonmagnetic impurities 
in Sect.~\ref{sec: Spin transport in chiral metal surface} (Fig.~\ref{fig: schematic pic spin texture} (a)) and spin transport between this 2D system, 
and a three-dimensional nonmagnetic metal with a finite thickness in Sect.~\ref{sec: Charge--spin interconversion at the interface} ((Fig.~\ref{fig: schematic pic spin texture} (b), (c)).
The 2D systems can be regarded as a chiral metal surface, a mimic of three-dimensional chiral metals, or an equivalent to the Rashba system or a Rashba--Dresselhaus system via 90-degree rotation in spin space.  
We model the interface as a tunnel junction with a nonmagnetic bulk metal that follows 
spin diffusion equation~\cite{Valet1993}~(Sect.~\ref{subsec: Formulation for the interface}). 
We make full use of the Boltzmann transport equation (BTE) beyond the relaxation time approximation, 
which provides accurate transport properties when the 2D electron system at the surface 
is clean enough or spin splitting caused by SOC is large enough~\cite{SupplementalMaterial}.
That is just the case when the Edelstein effect becomes evident.
Derivation of the BTE based on the Keldysh formalism is given 
in the Supplemental Material~\cite[Sect.~S4]{SupplementalMaterial}.  

Our contributions are three-fold:
(i)~We have defined spin diffusion length for each spin component and diffusion direction in the chiral metal surface, 
as well as spin relaxation time for each spin component~(Sect.~\ref{subsec: Spin relaxation} and \ref{subsec: Spin diffusion}).
Our definition does not rely on spin-dependent chemical potential,
which is conventionally employed but is ill-defined under strong SOC.
We have also clarified how these spin relaxation time and diffusion length depend on the anisotropy of the SOC, 
since some chiral crystals have strongly anisotropic SOC, such as elemental tellurium~\cite{Furukawa2017}.
(ii)~We have obtained analytical expressions for the conversion efficiencies at the chiral metal interface,
from charge current to spin current, and vice versa~(Sect.~\ref{subsec: Comparison between Edelstein effect and its inverse}).
Here in accordance with the spin-current-injection experiment, 
we take account of a finite thickness of the three-dimensional (3D) nonmagnetic metal 
and consider the spin current density at an edge of the 3D metal as a controllable parameter.
Such a realistic description has been done for the first time by this study, 
in contrast to previous theoretical studies on the EE and IEE~\cite{Shen2014,Zhang2016,Dey2018}.
The Rashba--Edelstein effect at interfaces also follows our analytical results, 
which practically supports a phenomenological model proposed by Ref.~\cite{Isshiki2020}.
(iii)~We have developed the Onsager's reciprocity between the EE and IEE, originally given by 
Ref.~\cite{Shen2014} at a surface, to the interface system~(Sect.~\ref{subsec: Reciprocal relationship}). 
Along with experiments, the reciprocity we found relates local input and output, which are defined in the regions separated by the interface.
The reciprocity, as well as the charge--spin conversion efficiencies
are finally summarized in terms of a transfer matrix method, which reflects
the nature of the composite systems.


\section{\label{sec: Spin transport in chiral metal surface}
Spin transport in chiral metal surface}
\subsection{\label{subsec: Formulation for the surface}
Formulation for the surface}
We start with a two-band effective model for electrons in the chiral metal surface
\begin{equation}
H^{\text{2D}}(\bm{k})= \frac{\hbar^2(k_z^2 + k_x^2)}{2m} 
+ H_{\text{SO}},
\label{eq: 2dchiral Hamilt}        
\end{equation}
with $z,x$-axes in the surface plane. The second term of SOC is described as
\begin{equation}
H_{\text{SO}} = 
\alpha_{\|} k_z\sigma_z + \alpha_{\perp}k_x\sigma_x
= \bm{g}(\bm{k})\cdot\bm{\sigma},
\label{eq: 2dchiral SOC}
\end{equation}
with two SOC parameters 
$(\alpha_\|, \alpha_\perp) =  (\alpha\cos\delta, \alpha\sin\delta)$, 
standing for its strength $\alpha$ and anisotropy $\delta$.
We here put the momentum 
$\bm{k}=(k_z, k_x)= (k\cos\theta, k\sin\theta)$ and 
spin denoted by Pauli matrices $\bm{\sigma} = (\sigma_z, \sigma_x)$.
That parallel coupling of spin and momentum~\eqref{eq: 2dchiral SOC} is obtained after we eliminate
the freedom of motion along the $y$-axis normal to the surface  
from the SOC $\alpha_{\|} k_z\sigma_z + \alpha_{\perp}(k_x\sigma_x + k_y\sigma_y)$
around $\Gamma$ point
in the bulk chiral metals with $D_6$ (622) point group
~\cite{Frigeri2005,Onuki2014}.
The surface model may also require
the Rashba SOC~\cite{Bychkov1984} and other SOC terms due to the structural inversion asymmetry, but
the whole SOC in that case is also reduced to the same expression as Eq.~\eqref{eq: 2dchiral SOC}
under appropriate rotations~\cite[Sect.~S1]{SupplementalMaterial}.
In particular, simple Rashba SOC corresponds to an isotropic case of that coupling~\eqref{eq: 2dchiral SOC}
$\alpha_\| = \alpha_\perp = \alpha/\sqrt{2}$;
the Rashba model is obtained after we rotate the spin by $90$ degrees while leaving the momentum space unchanged
\footnote{Our SOC model can also be converted to a 2D system with both Rashba and Dresselhaus SOCs
under appropriate rotations~\cite[Sect.~S1]{SupplementalMaterial}.}.

The vector $\bm{g}(\bm{k})$, written in polar coordinate as
\begin{equation}
\bm{g}(\bm{k})=|\bm{g}(\bm{k})| \cdot \hat{\bm{g}}(\bm{k})
\equiv \frac{\Delta (\bm{k})}{2}\bm{(}\cos\Theta (\bm{k}), \sin\Theta (\bm{k}) \bm{)},
\end{equation}
serves as an effective magnetic field 
in the momentum space.
Spin-degenerated states are then lifted into two bands
\begin{equation}
\epsilon(\bm{k},\pm)=\frac{\hbar^2 k^2}{2m}\pm \frac{\Delta (\bm{k})}{2},
\end{equation}
with band indices $\gamma=+,-$.
The spin-splitting energy is accordingly $\Delta (\bm{k})$. 
The spin wave function for the state $(\bm{k},\gamma)$ 
is expressed in the basis of eigenstates of spin $\sigma_z$ as
\begin{subequations}
\begin{align}
\ket{\bm{k},+}
&= \cos\frac{\Theta(\bm{k})}{2}\ket{\bm{k},\uparrow}
+ \sin\frac{\Theta(\bm{k})}{2}\ket{\bm{k},\downarrow}, \\
\ket{\bm{k},-}
&= \sin\frac{\Theta(\bm{k})}{2}\ket{\bm{k},\uparrow}
- \cos\frac{\Theta(\bm{k})}{2}\ket{\bm{k},\downarrow},
\end{align}
\end{subequations}
which has spin polarization in 
$\bm{S}(\bm{k},\gamma)\equiv \braket{\bm{k},\gamma|\bm{\sigma}|\bm{k},\gamma}= \gamma\cdot \hat{\bm{g}}(\bm{k})$.
As a result, hedgehog spin texture is formed on Fermi contours
at Fermi energy $\epsilon_\text{F} =\epsilon (\bm{k},\gamma) > 0$
(Fig.~\ref{fig: schematic pic spin texture}(d)).
The radii of the Fermi contours
can be typically measured by $k_\text{F}\equiv \sqrt{2m\epsilon_\text{F}}/\hbar$,
but is modulated for each band $\gamma = \pm$ and direction $\theta$.
In highly anisotropic SOC case when $\delta\to 0$ or $\pi/2$, 
the whole spin texture tends to face in the same direction, and that component of spin becomes nearly conserved.

The BTE for the 2D electron system is given by
\begin{equation}
 \frac{\partial f}{\partial t}+ \bm{v}\cdot \frac{\partial f}{\partial \bm{r}} + \frac{(-e)\bm{E}}{\hbar}\cdot \frac{\partial f}{\partial \bm{k}}
=\frac{df}{dt}\Big|_{\text{col}} ,
\label{eq: Naive BTE}
\end{equation}
with charge of the electron $(-e)$ and
group velocity $\bm{v} = \bm{v}(\bm{k},\gamma) = \hbar^{-1}\bm{\nabla}_{\bm{k}} \epsilon (\bm{k},\gamma)$.
Here non-equilibrium distribution function
$f = f(t, \bm{r}, \bm{k}, \gamma)$ is
the number of electrons in a band $\gamma$ 
occupying the volume of the phase space 
$d\bm{r}d\bm{k}$ at a time $t$.
In equilibrium, it is identical to 
the Fermi distribution function
$f = f_0\bm{(}\epsilon(\bm{k},\gamma)\bm{)}$.
Here we assume that the system is
in the low temperature $k_\mathrm{B}T\ll \epsilon_\mathrm{F}$.
The chemical potential $\mu$ then satisfies $\mu\simeq \epsilon_\text{F}$.

The right-hand side of Eq.~\eqref{eq: Naive BTE}
is a collision integral for nonmagnetic impurity scattering.
We assume that the impurity potential
is like $\delta$-function with strength $v_0$, randomly distributed 
with the density $n_{\rm imp}$ 
in the 2D system with the areal volume $V$.
The collision integral is then 
derived along the Fermi's golden rule~\cite[Sect.~S2]{SupplementalMaterial} as
\begin{multline}
\frac{df}{dt}\Big|_{\text{col}}
= \frac{2\pi v_0^2n_{\text{imp}}}{\hbar V}\sum_{\bm{k}', \gamma'}
\Abs{\Braket{\bm{k}', \gamma'|\bm{k}, \gamma}}^2\\
[f(\bm{k}',\gamma') - f(\bm{k},\gamma)]
\cdot \delta\bm{ (}\epsilon (\bm{k}', \gamma') - \epsilon (\bm{k}, \gamma)\bm{) }.
\label{eq: Boltzmann col int}
\end{multline}
The factor $\Abs{\Braket{\bm{k}', \gamma'|\bm{k}, \gamma}}^2$, represented as
\begin{equation}
\Abs{\Braket{\bm{k}' ,\gamma'|\bm{k}, \gamma}}^2
= \frac{1+{\bm{S}}(\bm{k},\gamma)\cdot {\bm{S}}(\bm{k}',\gamma')}{2},
\end{equation}
measures the relative angle of spin polarization between states before and after the spin-conserving scattering~\cite{Silsbee2001}.
Spin relaxation and diffusion in this system are thus associated with
the spin-dependent transition probability caused by the noncolinear spin texture in the momentum space.
The collision integral~\eqref{eq: Boltzmann col int} also indicates a typical impurity scattering rate,
i.e. an inverse of quasiparticle lifetime 
\begin{equation}
 \frac{1}{\tau_\text{p}}\equiv \frac{2\pi v_0^2n_{\text{imp}}\cdot N_0/2}{\hbar}.\label{eq: taup definition}
\end{equation}
Here $N_0 = m/(\pi\hbar^2)$ is an exact density of states 
of this 2D system~\cite[Sect.~S2]{SupplementalMaterial}.

The validity of the BTE shown above for spin-splitting bands
is supported by a derivation from the Keldysh Green's function method~\cite[Sect.~S4]{SupplementalMaterial},
which tells us that the BTE is valid in a clean limit
$\hbar/\tau_{\text{p}}\ll \Delta_{\text{F}}\ll\epsilon_\text{F}$.
Here $\Delta_\text{F} = \Delta (k_\text{F})$
is the spin-splitting energy gap around the Fermi energy $\epsilon_\text{F}$.
This condition validates the BTEs for each band, which are coupled through the collision integral.

In addition to the BTE, we must consider the Gauss' law, described as
\begin{equation}
\bm{\nabla}\cdot\bm{E}(t,\bm{r})=\frac{-e}{\varepsilon_0 d_y V}\sum_{\bm{k},\gamma}
\left[f(t,\bm{r},\bm{k},\gamma)-f_0\bm{(}\epsilon (\bm{k},\gamma)\bm{)}\right].
\label{eq: GaussLaw}
\end{equation}
Here we denote by $d_y$ a typical length normal to the surface.
The right-hand side stands for the charge density
induced by the shift of the Fermi contours.

In the rest of Sect.~\ref{sec: Spin transport in chiral metal surface},
we apply Eqs.~\eqref{eq: Naive BTE}--\eqref{eq: GaussLaw} to
the transport at the surface slightly out of equilibrium.
We consider the following two cases in the absence of external fields in order to extract spin relaxation time and spin diffusion length: 
spatially uniform relaxation and temporary stationary diffusion.
We also consider the linear response to a uniform stationary electric field
in the Supplemental Material~\cite[Sect.~S3]{SupplementalMaterial}.

\subsection{\label{subsec: Spin relaxation}
Spin relaxation time}
When we consider the relaxation of a spatially uniform non-equilibrium state, the left-hand side of the BTE~\eqref{eq: Naive BTE} is reduced to only a time-derivative term,
\begin{equation}
\frac{\partial f}{\partial t} = \frac{df}{dt}\Big|_{\text{col}}.
\label{eq: BTE-relaxation}  
\end{equation}
We are interested in the relaxation of low-energy states. We thus  assume that  electron distribution $f = f(t,\bm{k}, \gamma)$ is displaced around the Fermi energy $\epsilon_{\rm F}$ and seek for a solution to Eq.~\eqref{eq: BTE-relaxation} in the form 
\begin{equation}
 f(t,\bm{k},\gamma) = f_0\bm{(}\epsilon (\bm{k},\gamma)\bm{)} + e^{-t/\tau} 
\varphi_{\tau}(\bm{k},\gamma)\left(-\frac{\partial f_0 (\epsilon)}{\partial \epsilon}\right),
\label{eq: relaxation distribution}
\end{equation}
with $-\partial_{\epsilon} f_0 (\epsilon) \simeq \delta \bm{(}\epsilon (\bm{k},\gamma)-\epsilon_{\mathrm{F}}\bm{)}$.
The BTE after substitution of this assumption~\eqref{eq: relaxation distribution} results in an eigenvalue problem 
around the Fermi contours $\epsilon (\bm{k},\gamma)=\epsilon_\text{F}$, written as
\begin{equation}
\lambda_\tau\varphi_{\tau}(\bm{k},\gamma)=\sum_{\bm{k}',\gamma'}
M_{\text{col}}(\bm{k},\gamma,\bm{k}',\gamma')\varphi_{\tau}(\bm{k}',\gamma').
\label{eq: relaxation eig prob}
\end{equation}
Here we defined eigenvalue $\lambda_\tau= -\tau_{\text{p}}/\tau$ and a symmetric matrix
between the states $(\bm{k},\gamma)$ and $(\bm{k}',\gamma')$
\begin{multline}
M_{\text{col}}(\bm{k},\gamma,\bm{k}',\gamma')=\frac{1}{VN_0/2}
\sum_{\bm{k}'',\gamma''} |\langle\bm{k}'',\gamma''|\bm{k},\gamma\rangle|^2\\
(\delta_{\bm{k}', \bm{k}''}\delta_{\gamma', \gamma''}-\delta_{\bm{k},\bm{k}'}\delta_{\gamma,\gamma'})
\cdot \delta\bm{(}\epsilon(\bm{k}'',\gamma'')-\epsilon(\bm{k},\gamma)\bm{)},
\label{eq: Mmatrix-def}
\end{multline}
which we shall refer to as the \emph{relaxation matrix} (this can be regarded as the integral kernel because the collision integral is an integral transform of the distribution function). 
Let $\tau(j)$ be the relaxation time of $j$-th eigenvector $\varphi_{\tau(j)}$ of Eq.~\eqref{eq: relaxation eig prob}. The general solution (restricted to the low energy state) to Eq.~\eqref{eq: BTE-relaxation} is given in the form 
 \begin{align}
f(t,\bm{k},\gamma) &= f_0\bm{(}\epsilon (\bm{k},\gamma)\bm{)}\nonumber\\
&+ \sum_j c(j)e^{-t/\tau(j)} 
\varphi_{\tau(j)}(\bm{k},\gamma)\left(-\frac{\partial f_0 (\epsilon)}{\partial \epsilon}\right),
\label{eq: general solution to relaxation}
\end{align} 
where the coefficients $c(j)$ are determined by the initial distribution function in a relaxation process. The $\mu(=x,z)$ component of the spin density at time $t$ is given by
 \begin{align}
&\sum_{\bm{k},\gamma}
\frac{\langle \bm{k},\gamma|\sigma_\mu|\bm{k},\gamma\rangle}{V} f(t,\bm{k},\gamma) \nonumber\\
&=\langle s_\mu\rangle_{\rm eq}+\sum_j c(j) e^{-t/\tau(j)} s_{\mu}^{\rm relax}(j),
\label{eq: expectation value of sxsz in relaxation}
\end{align} 
with
\begin{equation}
s_{\mu}^{\rm relax}(j)=\frac{1}{V}\sum_{\bm{k},\gamma} \langle \bm{k},\gamma|\sigma_\mu|\bm{k},\gamma\rangle\varphi_{\tau(j)}(\bm{k},\gamma)\left(-\frac{\partial f_0 (\epsilon)}{\partial \epsilon}\right). 
\label{eq: Smuj}
\end{equation}
When $s_{\mu}^{\rm relax}(j)\ne 0$, we say that the $j$-th eigenmode {\it carries} $s_{\mu}$. 
We identify the longest relaxation time $\tau(j)$ among those of eigenmodes $j$ carrying $s_{\mu}$ with the spin relaxation time for $s_{\mu}$. 

\begin{figure}[htbp]
 \centering
\includegraphics[width=0.7\columnwidth]{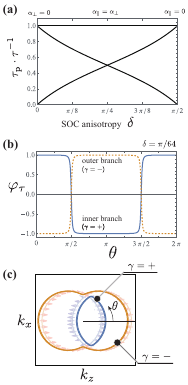}
\caption{(Color online) Eigenmode analysis of the Boltzmann transport equation for relaxation.
(a)~Inverse of the relaxation time $\tau$ plotted for all modes and for various anisotropy 
$\alpha_{\perp}/\alpha_{\|} = \tan\delta$.
All eigenmodes but two are degenerate in $\tau = \tau_{\text{p}}$.
(b)~Deviation from equilibrium in the distribution function around the Fermi contours
$\varphi_{\tau}(\theta,\gamma)$ that has the longest relaxation time when $\delta = \pi/64$.
The amplitude is put in arbitrary units. 
(c)~Schematic illustration of the deviation of the electron distribution~(b), drawn as a shift of the Fermi contours.
}
\label{fig: SpinTextureRelaxationMode}
\end{figure}

The eigenvalue spectrum $\tau_{\text{p}}/\tau$
is plotted with varying anisotropy of the SOC
$\delta$ in Fig.~\ref{fig: SpinTextureRelaxationMode}(a).
Here one trivial mode $\varphi_{\tau}= \text{constant}$ with $\tau_{\text{p}}/\tau = 0$
is omitted from the plots since it violates
the charge neutrality condition~\eqref{eq: GaussLaw} with $\bm{E}=0$~\footnote{
Such an extremely slowly decaying mode stems from the conservation of charge.
},
while the other eigenmodes automatically satisfy that condition~\cite[Sect.~S3]{SupplementalMaterial}.
Most eigenvalues are located at
$\tau= \tau_{\text{p}}$ but two eigenmodes have relaxation times longer than the others. We will focus the latter two modes.

We confirm that the mode with the relaxation time diverging as $\delta\to 0$ ($\delta\to \pi/2$) carries the $z$ ($x$)-component of spin. It can be naturally understood from the fact that 
spin component in the $z$ ($x$)-direction is conserved
in the highly anisotropic limit $\delta\to 0$ ($\pi/2$), i.e.
$\alpha_{\perp}\to 0$ ($\alpha_{\|}\to 0$).
%
%
Figure~\ref{fig: SpinTextureRelaxationMode}(b) shows the deviation $\varphi_{\tau}(\bm{k},\gamma) = \varphi_{\tau}(\theta, \gamma)$ in distribution function of the slowest mode from the equilibrium,
when spin $s_z$ is almost conserved ($\alpha_{\|}\gg \alpha_{\perp}$).
The inner and outer Fermi contours are shifted in opposite directions, which induces
nonzero spin density in the $z$-direction, as shown in Fig.~\ref{fig: SpinTextureRelaxationMode}~(c).
Indeed, $\varphi_{\tau}(\bm{k}, \gamma)$ is analytically expressed as 
$\varphi_{\tau}(\bm{k}, \gamma)\propto \gamma\cos\Theta (\theta) 
= \braket{\bm{k},\gamma|\sigma_z|\bm{k},\gamma}$~\cite[Sect.~S3]{SupplementalMaterial}. 
It follows that this slow mode $\varphi_{\tau}$ has nonzero $s_z^{\rm relax}(j)$ defined in Eq.~\eqref{eq: Smuj}, i.e., carries $z$-component of  spin density, regardless of the anisotropy $\delta$.
%
We obtain the analytical expressions for the two spin relaxation times shown in Fig.~\ref{fig: SpinTextureRelaxationMode}(a), 
 as a function of the anisotropy angle $\delta$ and we find that they are independent of 
the SOC strength $\alpha$~(The explicit expressions are available in~\cite[Sect.~S3]{SupplementalMaterial}). 
Such a characteristic spin relaxation time is attributed to the BTE scheme, which is valid in the clean limit or strong SOC case.
In a region $\Delta_{\rm F}\tau_{\text{p}}/\hbar \ll 1$~\cite{Szolnoki2017}, on the other hand,
the semiclassical picture of the spin-splitting bands breaks down.
The spin relaxation time then follows the Elliott-Yafet~\cite{Elliott1954,Yafet1963} 
and D'yakonov-Perel'~\cite{Dyakonov1972} mechanisms, instead of our description here.
\subsection{\label{subsec: Spin diffusion}
Spin diffusion length}
Spin diffusion length is extracted in the same way as the spin relaxation times, except for the treatment of the Gauss' law.
The spatially-inhomogeneous charge distribution accompanied by the diffusion induces an internal electric field.
The stationary diffusion thus follows both the BTE and Gauss' law
\begin{subequations}
\label{eqs: diffusion BTE Gauss}
\begin{gather}
\bm{v}\cdot \frac{\partial f}{\partial \bm{r}} + \frac{(-e)\bm{E}_{\text{in}}}{\hbar}\cdot \frac{\partial f}{\partial \bm{k}}
= \frac{df}{dt}\Big|_{\text{col}}, \\
\bm{\nabla}\cdot \bm{E}_{\text{in}} = \frac{-e}{\varepsilon_0 d_y V}
\sum_{\bm{k},\gamma}\left[
 f(\bm{r},\bm{k},\gamma) -f_0(\epsilon (\bm{k},\gamma))
\right],
\end{gather} 
\end{subequations}
where the internal electric field $\bm{E}_{\text{in}}$ works for charge screening effect.
We then assume that the distribution function $f(\bm{r}, \bm{k}, \gamma)$ 
and the electric field $\bm{E}_{\text{in}}$ 
decay in the $+z$-direction with a diffusion length $\ell > 0$, represented as 
\begin{subequations}
\label{eq: eigvec of diffusion}
 \begin{align}
 f(\bm{r},\bm{k},\gamma) &= f_0\bm{(}\epsilon (\bm{k},\gamma)\bm{)} + e^{-z/\ell} \varphi_{\ell}(\bm{k},\gamma)
\left(-\frac{\partial f_0 (\epsilon)}{\partial \epsilon }\right),\\
  E_{\text{in}, z}&  =  \frac{\mathcal{E}_{\ell}}{(-e) v_{\rm F} \tau_{\text{p}}} e^{-z/\ell},
 \end{align}
\end{subequations}
with $v_{\rm F} = \hbar k_{\rm F}/m$ a typical Fermi velocity.
Substitution of Eqs.~\eqref{eq: eigvec of diffusion} to Eqs.~\eqref{eqs: diffusion BTE Gauss}
yields a generalized eigenvalue problem with eigenvalues 
$-{v_{\rm F}\tau_{\text{p}}}/{\ell}$ and eigenvectors 
$\left(\{\varphi_{\ell} (\bm{k},\gamma)\}, \mathcal{E}_{\ell}\right)$ 
to be determined~\cite[Sect.~S3]{SupplementalMaterial}.

Similarly to the characterization of the relaxation eigenmode in the previous subsection, we say that 
the $j$-th eigenmode of spatial decaying {\it carries} $s_{\mu}$
when $s_{\mu}^{\rm diff}(j)\ne 0$ with 
\begin{equation}
s_{\mu}^{\rm diff}(j)=\frac1V\sum_{\bm{k},\gamma} \langle \bm{k},\gamma|\sigma_\mu|\bm{k},\gamma\rangle\varphi_{\ell(j)}(\bm{k},\gamma)\left(-\frac{\partial f_0 (\epsilon)}{\partial \epsilon}\right). 
\label{eq: Smuj-diffusion}
\end{equation}
We identify the longest decay length $\ell(j)$ among those of eigenmodes $j$ carrying $s_{\mu}$ with the spin diffusion length for $s_{\mu}$. 
 
In solving this problem, we have to fix a dimensionless parameter, i.e., a ratio of charge screening length to diffusion length
\begin{equation}
 \eta^{-1} \equiv (v_{\rm F} \tau_{\text{p}})^{2}\cdot 
\left(\frac{e^2N_0/2}{\varepsilon_0 d_y}\right)
\sim  (v_{\rm F}\tau_{\text{p}}q_{\text{TF}})^{2}
\sim 10^{6},
\end{equation}
with $v_{\rm F} \tau_{\text{p}}\sim \SI{e-8}{m}$ and Thomas--Fermi screening wavevector $q_{\text{TF}}\sim \SI{e11}{m^{-1}}$.
Numerical details of the dimensionless parameter $\eta$, however, give no striking difference in the results presented below.

\begin{figure}[htbp]
\centering
\includegraphics[width=0.7\columnwidth]{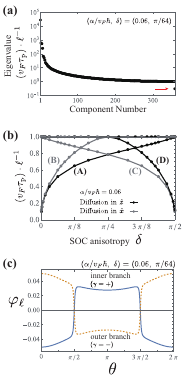}
\caption{(Color online) Eigenmode analysis of Boltzmann transport equation for diffusion.
(a)~Eigenvalues corresponding to inverse of the diffusion length $\ell > 0$ arranged in descending order.
The red arrow points to an isolated slowly decaying mode.
The total number of components reflects the number of angular meth points around a Fermi contour
(358) used in the numerical calculation.
(b)~Inverse of the diffusion length $\ell$ plotted for the isolated slowly decaying modes labeled (A)--(D) 
and for different anisotropy $\delta$. 
Both diffusion lengths in the $z$-direction and that in the $x$-direction are shown.
(c)~Deviation from equilibrium in distribution function around the Fermi contours
$\varphi_{\ell}(\theta,\gamma)$ that has the longest diffusion length when diffusing in the $z$-direction. 
The amplitude is put in arbitrary units. 
} 
\label{fig: SpinTextureDiffusionMode}
\end{figure}

The eigenvalue spectrum, which presents
the inverse of the diffusion length of each mode, is plotted 
in Fig.~\ref{fig: SpinTextureDiffusionMode}(a).
The spectrum is originally symmetric in $\pm \ell$
since diffusions in the $\pm z$-direction are equivalent.
We plot here only positive diffusion length $\ell >0$.
We also neglect a mode $v_{\rm F}\tau_{\text{p}}/\ell = 0$ 
since it describes homogeneous charge current density without external fields.

Most positive eigenvalues are continuously distributed and are concentrated around $\ell= v_{\rm F}\tau_{\text{p}}$.
There exist, however, isolated modes that have longer diffusion length than the others.

Figure~\ref{fig: SpinTextureDiffusionMode}(b) shows the inverse of such long diffusion lengths 
plotted for different anisotropy of the SOC.
The same figure also shows the result for the case when the $x$-direction is substituted for 
the $z$-direction in Eq.~\eqref{eq: eigvec of diffusion}.
We confirm that each of the slowly decaying modes (A)--(D) as spin diffusion  carries a different spin component into a different direction.
As for the diffusion in the $+z$-direction (branches (A) or (D)),
the diffusion length diverges at $\delta\to 0$ or $\pi/2$, which results from good conservation of spin $s_z$ or $s_x$, respectively.
It follows that branch (A) carries spin $s_z$, while (D) carries spin $s_x$ in the $+z$-direction.
That is supported by spin density calculation based on the shift of the Fermi contours 
(Fig.~\ref{fig: SpinTextureDiffusionMode}(c) depicts the branch (A)).
In the same manner, the branches (B), (C) are characterized as 
spin $s_z, s_x$ diffusion in the $+x$ direction, respectively.
Spin diffusion length for each spin component and each diffusion direction
is thus uniquely defined as the diffusion length of corresponding slowly decaying modes (A)--(D).

\section{\label{sec: Charge--spin interconversion at the interface}
Charge--spin interconversion at the interface}
\subsection{\label{subsec: Formulation for the interface}
Formulation for the interface}
We turn to the Edelstein effect (EE) and the inverse Edelstein effect (IEE)
at the 2D surface of the chiral metal attached with a 3D nonmagnetic metal.
The presence of an interface is treated as a boundary condition for the electron distribution in the 3D metal (See Figs.~\ref{fig: schematic pic spin texture} (b) and (c)), 
while it introduces an additional relaxation matrix and a driving term to the 2D system we have examined.
We now apply the electric field $\bm{E}$ in the $z$-direction on the surface (Fig.~\ref{fig: schematic pic spin texture} (b)),
or inject spins with polarization in the $z$-direction (Fig.~\ref{fig: schematic pic spin texture} (c)), both of which favor the $z$-component of spin polarization in the 3D metal. 
More generally, spin polarization of the 3D metal can point in an arbitrary direction, and we consider
such cases in the Supplemental Material~\cite[Sect.~S5, S7]{SupplementalMaterial}.
In the following, we assume that both the 2D and attached 3D metals
are electrically neutral and have a common chemical potential $\mu_0$, for simplicity.

We first explain the 3D nonmagnetic metal. 
As illustrated in Figs.~\ref{fig: schematic pic spin texture}~(b) and (c),
it is placed at $0\leq y\leq L$ with 
$y=0$ plane the interface at the chiral metal and $y=L$ plane an open end or interface with spin current source.
In the bulk, one-particle state is specified by wavevector $\bm{q}$ and $z$-component of spin $\sigma$,
denoted as $|\bm{q},\sigma)$.
The energy $\epsilon^{\rm (n)}(|\bm{q}|)$ of that state is degenerate with spin degrees of freedom and isotropic  
as a function of the modulus of $\bm{q}$. 
We denote the distribution function in the 3D nonmagnetic metal as $F(t,\bm{r},\bm{q},\sigma)$.

To describe the effect of the interface, we adopt a tunneling Hamiltonian
\begin{equation}
 \hat{H}_{\rm T}
=\sum_{\bm{k},\bm{q}}\sum_{\sigma = \uparrow, \downarrow}
\left[
T_{\bm{k}\bm{q}}\ket{\bm{k},\sigma} (\bm{q},\sigma| 
+ T^*_{\bm{k}\bm{q}}|\bm{q},\sigma)\bra{\bm{k},\sigma}\right],
\end{equation} 
which allows spin-independent transmission across the interface.
Here we consider that the interface between the two metals is rough enough 
to randomize momentum. That enables us to take $T_{\bm{k}\bm{q}} = T$.
The net transition rates into $\ket{\bm{k},\gamma}$ and $|\bm{q},\sigma)$ are, respectively, given by the Fermi's golden rule as
\begin{subequations}
 \label{eq: dfdt_dFdt_int}
 \begin{multline}
 \frac{d f(t,\bm{r},\bm{k},\gamma)}{d t}\Big|_{\text{int}}
 = \frac{2\pi|T|^2}{\hbar}\sum_{\bm{q},\sigma}
 |\braket{\bm{k},\gamma|\sigma}|^2\\
 [F(t,\bm{r},\bm{q},\sigma)-f(t,\bm{r},\bm{k},\gamma))] 
\delta\bm{(}\epsilon(\bm{k},\gamma)-\epsilon^{\rm (n)}(\bm{q})\bm{)}
 \label{eq: dfdt_int}
 \end{multline}
 and
 \begin{multline}
 \frac{d F(t,\bm{r},\bm{q},\sigma)}{d t}\Big|_{\text{int}}
 = \frac{2\pi|T|^2}{\hbar}\sum_{\bm{k},\gamma}
 |\braket{\bm{k},\gamma|\sigma}|^2\\
 [f(t,\bm{r},\bm{k},\gamma)-F(t,\bm{r},\bm{q},\sigma)]
 \delta\bm{(}\epsilon(\bm{k},\gamma)-\epsilon^{\rm (n)}(\bm{q})\bm{)},
 \label{eq: dFdt_int}
\end{multline}
\end{subequations}
with $\bm{r}$ in the $y=0$ plane.
These terms serve as extra collision terms in the BTE in the 2D and 3D metals.
They vanish in equilibrium, and we can replace $f$ and $F$ in Eqs.~\eqref{eq: dfdt_dFdt_int} by the deviation from the equilibrium
\begin{equation}
f(t,\bm{r},\bm{k},\gamma)-f_0\bm{(}\epsilon(\bm{k}, \gamma)\bm{)}
= -\frac{\partial f_0(\epsilon(\bm{k},\gamma))}{\partial \epsilon(\bm{k},\gamma)}\varphi(t,\bm{r},\bm{k},\gamma)
\label{eq: varphi}    
\end{equation}
and $F_1(t,\bm{r},\bm{q},\sigma)=F(t,\bm{r},\bm{q},\sigma)-f_0\bm{(}\epsilon^{\text{(n)}}(|\bm{q}|)\bm{)}$.
In the following, 
we consider that the whole system is stationary and spatially homogeneous in the $z$- and $x$-direction,
parallel to the interface (See Fig.~\ref{fig: schematic pic spin texture} (a), where $x,y,z$-directions are shown).
We also denote transmission rate across the interface~\cite{Dey2018} by
\begin{equation}
  \frac{1}{\tau_\text{t}}\equiv \frac{2\pi |T|^2 V^{\text{(n)}}N^{\text{(n)}}(\mu_0)/2}{\hbar}.\label{eq: taut definition}
\end{equation}
Here $N^{({\rm n})}(\epsilon)=2(V^{({\rm n})})^{-1}\sum_{\bm{q}}\delta\bm{(}\epsilon-\epsilon^{({\rm n})}(\bm{q})\bm{)}$ 
is the density of states per volume of the 3D system $V^{({\rm n})}$. 

Let us write down the electron distribution in the 3D metal. 
We assume that both the spin-conserving impurity scattering and much weaker spin-flip impurity scattering occur in the 3D metal.
The BTE in that nonmagnetic metal is examined by Valet and Fert~\cite{Valet1993},
and is briefly reviewed in \cite[Sect.~S6]{SupplementalMaterial}.
In the absence of external fields, the shift of the distribution $F_1$ they provide is expressed as
\begin{multline}
F_1(y, \bm{q},\sigma)
=-\frac{\partial f_0\bm{(}\epsilon^{\text{(n)}}(|\bm{q}|)\bm{)}}{\partial \epsilon}
\left\{\bar{\mu}- \mu_0 \right.\\
\left.
+ \frac{\sigma}{2}\left[
\mu^{\rm (n)}_{\rm s}(y)
+ \frac{2e^2\lambda^{\rm (n)}}{\sigma^{\rm (n)}}\frac{q_y}{|\bm{q}|}
J_{\rm s}^{\rm (n)}(y)\right]
\right\}
\label{eq: F-Valet-Fert-form}
\end{multline}
and higher multipole terms in $\bm{q}$ proportional to the Legendre polynomials $P_m(q_y/|\bm{q}|)$ with $m\ge 2$,
which are negligibly small. 
Here $\lambda^{\rm (n)}$ and $\sigma^{\rm (n)}$ are the electron mean free path
and the electrical conductivity summed over spins, respectively, while $\bar{\mu}$ is a constant to be determined later.
The two spatially-varying quantities
\begin{subequations}
\label{eq: mu_s and js}
\begin{align}
\mu^{\rm (n)}_{\rm s}(y)&=2(A e^{-y/\ell_{\rm sf}}+B e^{y/\ell_{\rm sf}}),\\
J_{\rm s}^{\rm (n)}(y)&=\frac{\sigma^{\rm (n)}}{e^2\ell_{\rm sf}}\left(A e^{-y/\ell_{\rm sf}}
-B e^{y/\ell_{\rm sf}}\right)
\label{eq: Js-n-y-edelstein}
\end{align}
\end{subequations}
are spin accumulation polarized in the $z$-direction and spin current density flowing parallel to the $y$-direction, respectively.
They follow spin diffusion equation with spin diffusion length $\ell_{\rm sf}$~\cite{Valet1993},
but two coefficients $A$, $B$ in them are still undetermined.
For later use, we here introduce a dimensionless parameter $\tau_{\text{p}}/\tau_{\text{3D}}$
that measures spin diffusion in the 3D metal with a typical rate 
\begin{equation}
\frac{1}{\tau_{\text{3D}}}
\equiv \frac{\sigma^{\rm (n)}}{e^2 \ell_{\rm sf}N_0}.
\label{eq: tau3D definition}
\end{equation}
When the spin-flip scattering relaxation time $\tau_{\text{sf}}$ is much longer than 
the spin-conserving scattering relaxation time $\tau_{\text{s}}$, 
this time scale $\tau_{\text{3D}}$ is given as
$\tau_{\text{3D}} \simeq 
N_0\cdot [N^{\text{(n)}}(\mu_0)v_{\rm F}^{\text{(n)}}]^{-1}\cdot \sqrt{3\tau_{\text{sf}}/(2\tau_{\text{s}})}$
with the Fermi velocity in the 3D metal $v_{\rm F}^{\text{(n)}}$~\cite[Sect.~S7]{SupplementalMaterial}.

To determine the electron distribution~\eqref{eq: F-Valet-Fert-form} with \eqref{eq: mu_s and js} described by Valet and Fert,
we consider boundary conditions to the 3D metal based on 
the extra collision term~\eqref{eq: dFdt_int} at the interface.
There are three conditions;
(i)~the absence of charge current through the interface,
(ii)~the continuity of spin current at the interface,
(iii)~the boundary condition on the other side of the 3D metal at $y=L$. 
The three parameters $\bar{\mu}$, $A$ and $B$ will be then expressed 
as the functionals of the distribution function in the 2D metal.

The first condition (i) is described as
\begin{equation}
\frac{1}{V}\sum_{\bm{q}, \sigma}\frac{d F (y=0, \bm{q}, \sigma)}{dt}\Big|_{\text{int}} = 0,\label{eq: first condition}
\end{equation}
with the left-hand side being the number of electrons flowing through the unit area of the interface 
from the 2D metal to the 3D metal per unit time.
This condition yields the balance of electrochemical potentials on both sides of the interface
\begin{equation}
\bar{\mu}-\mu_0
=\frac{1}{VN_0}\sum_{\bm{k},\gamma} \varphi(\bm{k},\gamma) \delta(\epsilon(\bm{k},\gamma)-\mu_0),
\label{eq: electrochemical potential balance}
\end{equation}
with the right-hand side net charge density of the 2D metal divided by $-e$ times the density of states.
As we have first assumed that the whole system is electrically neutral with the same chemical potential $\mu_0$, 
both sides of Eq.~\eqref{eq: electrochemical potential balance} are zero;
it follows that $\bar{\mu}=\mu_0$ and the charge neutrality condition for the 2D system.

We turn to the remaining conditions.
The condition~(ii) on 
the transmission of the $z$-component spin is described in the same way as Eq.~\eqref{eq: first condition}, 
\begin{equation}
\frac{1}{V}\sum_{\bm{q}, \sigma}\sigma \frac{d  F(y=0,\bm{q},\sigma)}{d t}\Big|_{\rm int}= J^{\text{(n)}}_\text{s}(y=+0),
\label{eq: Spin-density-from2to3}
\end{equation}
with the right-hand side given by Eq.~\eqref{eq: Js-n-y-edelstein}.
As for the condition~(iii), we assume either that the 3D metal has an open end at $y=L$
or that the 3D metal is attached with 
the source of spin current---such as a ferromagnetic metal or a metal with strong spin Hall effect
at this location $y=L$. In both cases, the boundary condition is expressed as  
\begin{equation}
J^{\text{(n)}}_\text{s}(y=L) = 0~\text{or}~J^{\text{ext}}_\text{s},\label{eq: bc-JsL}
\end{equation}
where $J^{\text{ext}}_\text{s}\neq 0$ implies the external spin current injected from the source.
The relations~\eqref{eq: Spin-density-from2to3} and \eqref{eq: bc-JsL} determine the parameters 
$A$ and $B$~\cite[Sect.~S7]{SupplementalMaterial}. 

The Valet--Fert solution $F(y,\bm{q},\sigma)$
is thus determined by the distribution function $\varphi(\bm{k},\gamma)$ in the 2D metal
and the spin current density at the boundary $J_{\rm s}^{\rm (n)}(L)$.
It follows that the electron in the 2D metal under an electric field is described by
\begin{equation}
(-e)E v_z(\bm{k},\gamma)\frac{\partial f_0(\bm{k},\gamma)}{\partial \epsilon(\bm{k},\gamma)}
=\frac{d f}{d t}\Big|_{\rm col}+\frac{d f}{d t}\Big|_{\rm int},
\label{eq: Inhomo-linear-added}
\end{equation}
with the second term on the right-hand side the extra Boltzmann collision term~\eqref{eq: dfdt_int}, 
which is written with $\varphi$ and $J_{\rm s}^{\rm (n)}(L)$.
This extra term also includes three dimensionless parameters $L/\ell_\text{sf}$, 
$\tau_\text{p}/\tau_{\text{t}}$ and $\tau_\text{p}/\tau_{\text{3D}}$,
which provide information on the 3D metal and the interface.
The effective BTE~\eqref{eq: Inhomo-linear-added} for 2D electron distribution is simplified as
\begin{equation}
b_{\text{EE}}(\bm{k},\gamma)+b_{\text{IEE}}(\bm{k},\gamma) =\sum_{\bm{k}',\gamma'}
M_{\rm tot}(\bm{k},\gamma,\bm{k}',\gamma')\varphi (\bm{k}',\gamma')
\label{eq: linear-inhomo-attached-Edelstein}
\end{equation}
for $\ket{\bm{k},\gamma}$ such that $\epsilon(\bm{k},\gamma)=\mu_0$~\cite[Sect.~S7]{SupplementalMaterial}. 
The two terms on the left-hand side are driving forces: One term is the electric field applied on the surface
\begin{subequations}
\begin{equation}
b_{\text{EE}}(\bm{k},\gamma)\equiv eE\tau_\text{p} v_z(\bm{k},\gamma),\label{eq: b} \end{equation}
which is a source of the EE---charge-to-spin conversion.
The other term is the spin current injected from the 3D metal
\begin{equation}
b_{\text{IEE}}(\bm{k},\gamma)\equiv \frac{2K\tau_{\rm p} J_{\rm s}^{\rm ext}}{N_0}\cdot S_z(\bm{k},\gamma),\label{eq: bPrime}
\end{equation}
which leads to the IEE---spin-to-charge conversion. 
\end{subequations}
Here we introduce a notation
\begin{equation}
2K \equiv [\cosh L/\ell_{\rm sf} + (\tau_\text{t}/\tau_\text{3D})\sinh L/\ell_{\rm sf}]^{-1}.
\label{eq: K-def}
\end{equation}
The IEE source term~\eqref{eq: bPrime} indicates that 
the injected spin current serves as a time-dependent magnetic field coupling to the spin 
$S_z(\bm{k}, \gamma) \equiv \braket{\bm{k},\gamma|\sigma_z|\bm{k},\gamma}$, 
inducing non-equilibrium state in the surface ~\cite{Silsbee2004,Shen2014}.

On the right-hand side of Eq.~\eqref{eq: linear-inhomo-attached-Edelstein},
the relaxation matrix is given by two contributions
$M_{\rm tot}= M_{\text{col}} + M_{\rm int}$
where the matrix $M_{\text{col}}$, provided in Eq.~\eqref{eq: Mmatrix-def}, stems from the impurity scattering
within the surface, while  
\begin{multline}
M_{\rm int}(\bm{k},\gamma,\bm{k}',\gamma')\equiv 
\frac{\tau_\text{p}}{\tau_\text{t}}
\Big\{ -\delta_{\bm{k},\bm{k}'}\delta_{\gamma,\gamma'}\\
+\frac{\delta(\epsilon(\bm{k},\gamma)-\epsilon(\bm{k}',\gamma')
)}{N_0V} [1+K'\bm{S}(\bm{k},\gamma)\cdot \bm{S}(\bm{k}',\gamma') ]\Big\},
\label{eq: Mprime-int}
\end{multline}
with $K' \equiv 2K \cosh L/\ell_{\rm sf}$ represents an effective scattering process mediated by the interface.

We here assume that no charge current is induced in the 3D metal without considering the penetration of 
an electric field applied on the 2D system into the 3D metal.
Such leakage of the electric field in the EE case is small 
only when the 3D metal has low conductivity and/or the thickness $L$ of the 3D metal
is sufficiently small.
It is thus generally needed to incorporate the charge current density in the 3D metal.
We, however, restrict ourselves to neglecting that effect 
as a starting point of the formulation of the EE.

\subsection{\label{subsec: Comparison between Edelstein effect and its inverse}
Comparison between the Edelstein effect and its inverse}
Let us compare the non-equilibrium distributions of the EE and the IEE.
In both effects, there exist charge current density and spin density induced in the 2D metal and spin accumulation and spin current density in the 3D metal.
The deviations of the electron distribution around the Fermi contours $\varphi (\bm{k},\gamma) = \varphi (\theta, \gamma)$
for the two effects are, however, different,
as can be seen from Figs.~\ref{fig: CS_SC_conversions}(a) and (b).
Their analytical expressions are available in the Supplemental Material~\cite[Sect.~S8]{SupplementalMaterial}.
In the EE, $\varphi = \varphi_{\text{EE}} (\theta,\gamma)$ driven by the external electric field $E\neq 0$ 
without spin current injection $J_{\rm s}^{\rm (n)}(L)=0$ 
is illustrated in Fig.~\ref{fig: CS_SC_conversions}(a).
\begin{figure}[htbp]
\centering
\includegraphics[width=0.65\columnwidth]{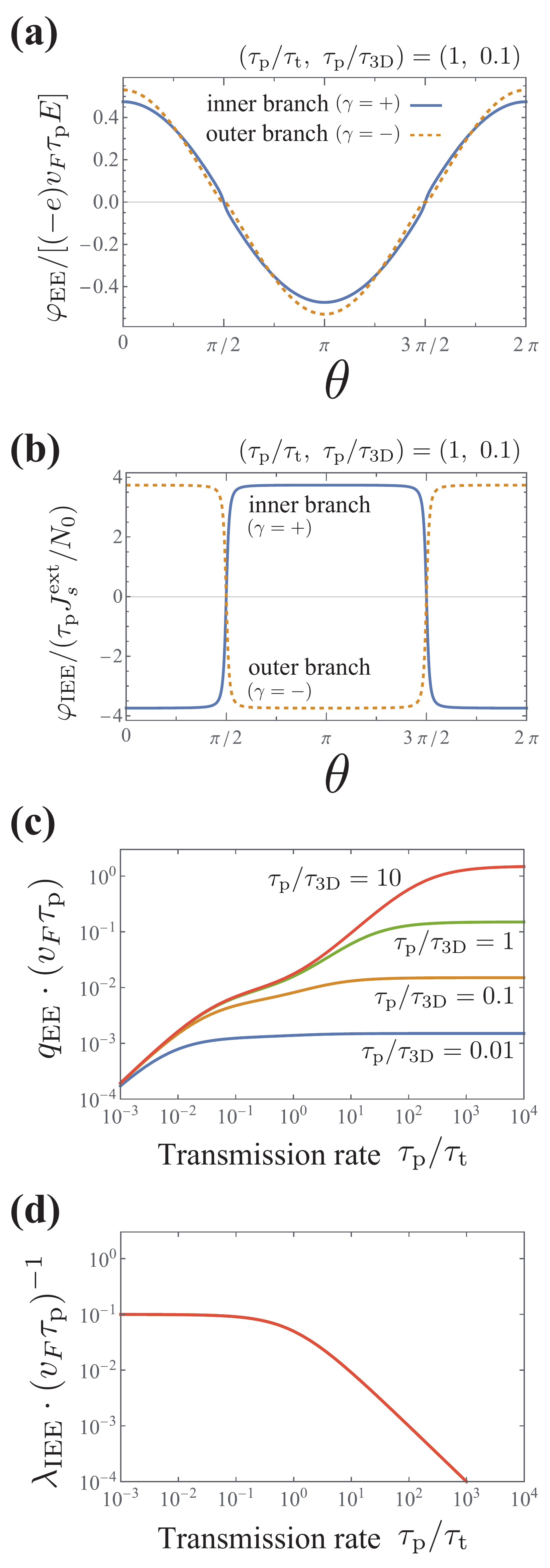}
\caption{(Color online)
The deviation from the equilibrium distribution function around the two Fermi contours for direct Edelstein effect [(a)] and inverse Edelstein effect [(b)], and associated 
charge-spin conversion efficiencies at
the interface $q_{\text{EE}}$ [(c)] and $\lambda_{\text{IEE}}$ [(d)].
The parameters are chosen as $(\alpha/v_{\rm F}\hbar, \delta, L/\ell_{\text{sf}})= (0.1, \pi/64, 1)$.
[(c), (d)]~The colored lines show different spin diffusion rates in the three-dimensional nonmagnetic metal
(defined in Eq.~\eqref{eq: tau3D definition}).
Those lines collapse onto a single curve in (d).}
\label{fig: CS_SC_conversions}
\end{figure}
The Fermi contours are basically shifted in the direction of the electric field with
a relaxation time $(1/\tau_\text{p} + 1/\tau_\text{t})^{-1}$, modified from the ordinary 
momentum relaxation time $\tau_\text{p}$ due to the interface transmission.
More precisely, the deviation from the equilibrium distribution function for the two bands differ of order $\alpha/v_{\rm F}\hbar$, which induces net spin density.
In the IEE, on the other hand,
the distribution function $\varphi = \varphi_\text{IEE}(\bm{k},\gamma)$ driven by
the external spin current $J_{\rm s}^{\rm (n)}(L) = J_{\rm s}^{\rm ext}$ without electric field $E=0$
is illustrated in Fig.~\ref{fig: CS_SC_conversions}(b).
The deviation changes its sign with respect to the band $\gamma = \pm$.
Indeed, it is found analytically that the deviation is proportional to the spin polarization $S_z(\bm{k},\gamma)$,
which is the same as the spin $s_z$ relaxation mode (Fig.~\ref{fig: SpinTextureRelaxationMode}(b)).
The different distribution functions given in Figs.~\ref{fig: CS_SC_conversions}(a) and (b)
show that the EE and IEE are in different non-equilibrium states.

Linear responses found in the EE and IEE are different, consequently.
We now consider charge--spin conversion efficiency---ratio between electric current density in the 2D metal 
and spin current density in the 3D metal at the interface~\cite{Sanchez2013,Lesne2016,Sanchez2016,Zhang2016,Dey2018,Isshiki2020}.
In the EE and IEE, that efficiency is defined as
\begin{subequations}
\label{eq: def qEE and lambdaIEE} 
 \begin{align}
q_{\text{EE}}&\equiv \left.\frac{
\left|J^{\text{(n)}}_\text{s}(y=0)\right|}{
\left|j_{\rm c}/(-e)\right|}\right|_{\text{EE}},\\
\lambda_{\text{IEE}} &\equiv \left.\frac{
\left|j_{\rm c}/(-e)\right|}{
\left|J^{\text{(n)}}_\text{s}(y=0)\right|}\right|_{\text{IEE}},
 \end{align}
\end{subequations}
with $j_{\rm c}=j_z^{\rm c}$ a charge current density in the surface of the chiral metal, flowing in the $z$-direction. 
The two efficiencies look similar but
behave differently with respect to both the interface transmission rate $1/\tau_\text{t}$
and the detail of the 3D metal, specified by the time scale of spin diffusion $\tau_\text{3D}$
and the thickness $L/\ell_\text{sf}$.
Figure~\ref{fig: CS_SC_conversions}(c) shows that 
the charge current-to-spin current conversion efficiency $q_{\text{EE}}$ decreases with $\tau_\text{3D}$,
while Fig.~\ref{fig: CS_SC_conversions}(d) shows that 
the spin current-to-charge current conversion efficiency $\lambda_{\text{IEE}}$ is independent of $\tau_\text{3D}$.
It is also found that $q_{\text{EE}}$ increases with the thickness of the 3D mental $L/\ell_\text{sf}$,
while $\lambda_{\text{IEE}}$ is independent of $L/\ell_\text{sf}$~\cite[Sect.~S9]{SupplementalMaterial}.

The efficiency $\lambda_{\text{IEE}}$ is determined only by the interface---the non-equilibrium state in the 2D metal 
for the IEE case
is affected only by the spin current injected from the interface $J^{\text{(n)}}_\text{s}(y=0)$.
This explains why the details of the 3D metal does not affect the conversion efficiency $\lambda_{\text{IEE}}$.
Indeed, $\lambda_{\text{IEE}}$ is proportional to
the modified relaxation time in the 2D metal by the interface transmission
$(1/\tau_\text{p} + 1/\tau_\text{t})^{-1}$~\cite{Dey2018}.
The increase of tunneling rate $1/\tau_\text{t}$ thus suppresses $\lambda_{\text{IEE}}$,
as shown in Fig.~\ref{fig: CS_SC_conversions}(d).

The behavior of $q_{\text{EE}}$, on the other hand, depends on the details of the 3D metal; the time scale $\tau_\text{3D}\propto (\tau_\text{sf}/\tau_\text{s})^{1/2}$ (see Eq.~\eqref{eq: tau3D definition} for definition of $\tau_\text{3D}$) increases and 
the thickness measured in units of the spin diffusion length $L/\ell_\text{sf}$ decreases
when the spin-flip scattering is negligibly small, i.e. $\tau_\text{sf},~ \ell_\text{sf}\to \infty$.
Stationary spin current density in the 3D metal, which is nearly constant in that case, is then suppressed 
since we impose the boundary condition $J^{\text{(n)}}_\text{s}(L)= 0$ for the EE case.
The efficiency into spin current $q_{\text{EE}}$ thus decreases with $\tau_\text{3D}$ but increases with $L/\ell_\text{sf}$.
Indeed, we find based on an analytical calculation
that $q_{\text{EE}}$ is roughly proportional to a rate $(\tau_\text{t} + \tau_\text{3D}\coth L/\ell_\text{sf})^{-1}$,
which indicates spin transmission rate across the interface into the 3D metal with finite thickness.

We now compare our results with previous studies, as summarized in Table~\ref{tab: qEE and lambdaIEE previous vs our study}.
\begin{table*}
\caption{\label{tab: previous works comparison}
Comparison between previous studies and the present study
with respect to the charge current--spin current conversion efficiency at the interface
for the EE and IEE.
``TI'' stands for topological insulator.
The definition of the efficiencies $q_\text{EE}$ and $\lambda_\text{IEE}$ are given in Eqs.~\eqref{eq: def qEE and lambdaIEE}.}
\label{tab: qEE and lambdaIEE previous vs our study}
\begin{ruledtabular}
\begin{tabular}{ccccc}
& \multirow{2}{*}{S.~Zhang and A.~Fert~\cite{Zhang2016}} 
& \multirow{2}{*}{R.~Dey \textit{et al.}~\cite{Dey2018}} 
& \multirow{2}{*}{H.~Isshiki \textit{et al.}~\cite{Isshiki2020}}
& The present study \\ 
& & & & ($\delta=\pi/4$, $L/\ell_{\text{sf}}\to\infty$) \\ \hline
 subject & TI surface     & TI surface & Rashba model     & Chiral metal surface \\
$\displaystyle q_{\text{EE}}$     
& $\displaystyle \frac{1}{v_{\rm F}(\tau_\text{t} + \tau_{\text{sf}})}$
\footnote{Eqs.~(13) and (19) in Ref.~\cite{Zhang2016}.}
& -  &   $\displaystyle \frac{\alpha_{\text{R}}}{v_{\rm F}\hbar}\cdot\frac{1}{v_{\rm F}\tau_{\text{t}}}$  
\footnote{Eqs.~(3), (6) and (7) in Ref.~\cite{Isshiki2020}.}
& $\displaystyle \frac{\alpha/\sqrt{2}}{v_{\rm F}\hbar}\cdot \frac{2 + \mathcal{O}\bm{(}(\alpha/v_{\rm F}\hbar)^2\bm{)}}{v_{\rm F}
\left[\tau_{\text{t}} + \tau_{\text{3D}} + (1/\tau_{\text{p}} +1/\tau_{\text{t}} )^{-1} \right]}$  \\
$\displaystyle \lambda_{\text{IEE}}$     
& $\displaystyle v_{\rm F}\tau_{\text{p}}$ \footnotemark[1]
& $\displaystyle \frac{v_{\rm F}}{1/\tau_\text{p} + 2/\tau_\text{t}}$ \footnote{Eq.~(18) in Ref.~\cite{Dey2018}.}          
&$\displaystyle \frac{\alpha_{\text{R}}}{v_{\rm F}\hbar}\cdot\frac{v_{\rm F}}{1/\tau_\text{p} + 1/\tau_\text{t}}$
\footnotemark[2]
&$\displaystyle \frac{\alpha/\sqrt{2}}{v_{\rm F}\hbar}\cdot\frac{v_{\rm F}}{1/\tau_\text{p} + 1/\tau_\text{t}}$                     \\
\end{tabular}
\end{ruledtabular}
\end{table*}
Here the Rashba SOC $\alpha_\text{R}(k_z\sigma_x -k_x\sigma_z)$ 
can be regarded as the isotropic case of the SOC here~\eqref{eq: 2dchiral SOC}
with the Rashba parameter $\alpha_\text{R}= \alpha/\sqrt{2}$,
as we stated at the beginning of Sect.~\ref{subsec: Formulation for the surface}.
The different behaviors between $q_\text{EE}$ and $\lambda_\text{IEE}$ 
were discussed by Zhang and Fert~\cite{Zhang2016}, where they considered
charge--spin interconversion at the topological insulator surfaces.
Dey et al.~\cite{Dey2018} then found that the efficiency $\lambda_\text{IEE}$ is suppressed by
the interface transmission rate $1/\tau_\text{t}$, which is in good agreement with our results,
except for the difference in the factor $(\alpha/\sqrt{2})/v_{\rm F}\hbar$ due to the difference in the targeted systems;
indeed, $\tau_\text{t}/2$ in their paper is equivalent to $\tau_\text{t}$ in our study.
Isshiki et al.~\cite{Isshiki2020} also provided phenomenological calculation of  
$q_\text{EE}$ and $\lambda_\text{IEE}$.
Our analytical calculation practically supports their expression for $\lambda_\text{IEE}$.
A trade-off relation between the conversion efficiencies for the EE and IEE, proposed by Isshiki et al.~\cite{Isshiki2020}
is also found in general, which is expressed as
\begin{equation}
 q_\text{EE} \cdot \lambda_\text{IEE} < \frac{2\alpha_{\|}\alpha_{\perp}}{(v_{\rm F}\hbar)^2}.
\end{equation}

The efficiency $q_\text{EE}$ itself is, on the other hand,
obtained on the basis of the Boltzmann equation for the first time by the present study.
Indeed, $q_\text{EE}$ obtained by Zhang and Fert~\cite{Zhang2016}
is similar to our result in that $\tau_\text{3D}\propto (\tau_{\text{sf}}/\tau_\text{s})^{1/2}$
consists of spin-flip scattering relaxation time.
We, moreover, clarify the dependence of $q_\text{EE}$
on the momentum relaxation time $\tau_\text{p}$, the thickness $L/\ell_\text{sf}$
(in Table~\ref{tab: qEE and lambdaIEE previous vs our study}, we put it infinite), 
and the SOC anisotropy of the surface $\delta$ (in Table~\ref{tab: qEE and lambdaIEE previous vs our study}, 
$\delta = \pi/4$).
It also should be noted that the previous studies considered the spin accumulation 
at the interface as an external parameter in the formulation of both the EE and IEE.
According to the experiments on the IEE~\cite{Sanchez2013,Lesne2016,Sanchez2016}, 
and on the chiral metals~\cite{Inui2020,Shiota2021,Shishido2021}, 
however, controllable parameter---what we can exert directly---to the 3D nonmagnetic metal
is often spin current, injected or fixed to be zero at the open surface ($y =L$ plane in this study).
We here adopt a formulation close to these experimental situations.

We then consider other coefficients and linear responses in the EE and IEE.
We analytically obtained ratios between typical quantities---electric field applied in the $z$-direction $E$, 
2D spin density $s_z$, 2D charge current density flowing in the $z$-direction $j_c$,
3D spin accumulation $\mu^{\text{(n)}}_\text{s}(y)$ at $y=0,~L$ planes and
3D spin current density $J^{\text{(n)}}_\text{s}(y)$ at $y=0,~L$ planes.
These ratios are listed in Table~\ref{tab: EE coefficients} for the EE case
and Table~\ref{tab: IEE coefficients} for the IEE case,
where we used the following auxiliary variables:
\begin{turnpage}
\setlength{\tabcolsep}{10pt} 
\begin{table}[ht]
\centering
  \caption{
Ratios between physical quantities in the case of direct Edelstein effect (EE),
including the linear response to the electric field.
The entry in the row $X$ and column $Y$ gives the ratio $Y/X$.
The time scales $\tau_\text{p}$ and $\tau_\text{3D}$ are defined in Eq.~\eqref{eq: taup definition}
and Eq.~\eqref{eq: tau3D definition}, respectively, while
$\tau_\text{a}$, $\tau_\text{b}$, $\tilde{\alpha}$ and $r$ are defined in
Eqs.~\eqref{eq: auxiliary variables}.
}
\label{tab: EE coefficients}
\begin{tabular}{cc|ccccc}
\hline\hline
\multicolumn{2}{c|}{\multirow{2}{*}{$\displaystyle \frac{Y}{X}\Big|_{\text{EE}}$}} & \multicolumn{5}{c}{$Y$}  
\rule[-5mm]{0mm}{10mm}    
\\ \cline{3-7} 
\multicolumn{2}{l|}{}                     & $\displaystyle 2 s_z/N_0$ & $\displaystyle j_c/[(-e)v_\text{F}N_0/2]$ & $\displaystyle \mu^{(n)}_\text{s}(0)$ & $\displaystyle \mu^{(n)}_\text{s}(L)$ & $\displaystyle \tau_{\text{p}}J^{(n)}_\text{s}(0)/N_0$ 
\rule[-5mm]{0mm}{10mm}
\\ \hline
\multicolumn{1}{c|}{\multirow{8}{*}{$X$}} &
  \multirow{2}{*}{$\displaystyle (-e)v_\text{F}\tau_{\text{p}}E$} &
  $\displaystyle -2\tilde{\alpha}\sin\delta\cdot\frac{1/\tau_\text{p}}{1/\tau_\text{a} + 1/\tau_\text{b}}$ &
  $\displaystyle \frac{(\tau_\text{a} \tan\delta)\cdot r}{\tau_\text{p}}$ &
  \multirow{2}{*}{$\displaystyle -2\tilde{\alpha}\sin\delta\cdot\frac{(\tau_\text{3D}/\tau_\text{p})\coth L/\ell_\text{sf}}
{1 + \tau_\text{b} /\tau_\text{a}}$} &
  $\displaystyle \frac{-2\tilde{\alpha}\sin\delta (\tau_\text{3D}/\tau_\text{p})}
{(1+\tau_\text{b}/\tau_\text{a})\sinh L/\ell_\text{sf}}$ &
  \multirow{2}{*}{$\displaystyle \frac{-\tilde{\alpha}\sin\delta}{1 + \tau_\text{b}/\tau_\text{a}}$} 
\rule[0mm]{0mm}{10mm}
\\
\multicolumn{1}{l|}{}        &            & 
\multicolumn{1}{l}{$\displaystyle = \frac{2}{N_0}\cdot\frac{\langle S_z, v_z\rangle}{v_\text{F}}$}
&
\multicolumn{1}{l}{$\displaystyle = \frac{2}{N_0}\cdot\frac{\langle v_z, v_z\rangle}{v_\text{F}^2}$} &     & 
\multicolumn{1}{l}{$\displaystyle = -2\cdot\frac{\lambda_\text{recip}}{v_\text{F}\tau_\text{p}}$} &     
\rule[-10mm]{0mm}{10mm}
\\
\multicolumn{1}{l|}{}        & $\displaystyle 2 s_z/N_0$        & $1$ & $\displaystyle -\frac{(1+\tau_\text{a}/\tau_\text{b})r}{2\tilde{\alpha} \cos\delta}$ & $\displaystyle \frac{\tau_\text{3D}}{\tau_\text{b}}\coth L/\ell_\text{sf} $ & $\displaystyle \frac{\tau_\text{3D}/\tau_\text{b}}
{\sinh L/\ell_\text{sf}}$ & $\displaystyle \frac{\tau_\text{p}}{2\tau_\text{b}}$ 
\rule[-5mm]{0mm}{10mm}
\\
\multicolumn{1}{l|}{} &
  \multirow{2}{*}{$\displaystyle j_c/[(-e)v_\text{F}N_0/2]$} &
  \multirow{2}{*}{} &
  \multirow{2}{*}{$1$} &
  \multirow{2}{*}{$\displaystyle \frac{-2\tilde{\alpha}\cos\delta}{r}\cdot \frac{\tau_\text{3D}\coth L/\ell_\text{sf}}{\tau_\text{a} +\tau_\text{b}}$} &
  \multirow{2}{*}{$\displaystyle 
\frac{-2\tilde{\alpha}\cos\delta}{r\sinh L/\ell_\text{sf}}\cdot \frac{\tau_\text{3D}}{\tau_\text{a} +\tau_\text{b}}$} &
$\displaystyle \frac{-\tilde{\alpha}\cos\delta}{r}\cdot \frac{\tau_\text{p} }{\tau_\text{a} +\tau_\text{b}}$
\rule[0mm]{0mm}{10mm}
\\
\multicolumn{1}{l|}{}        &            &     &     &     &     & 
\multicolumn{1}{l}{$\displaystyle = -q_\text{EE}\cdot \frac{v_\text{F}\tau_\text{p}}{2}$}
\rule[-6mm]{0mm}{10mm}
\\
\multicolumn{1}{l|}{}        & $\displaystyle \mu^{(n)}_\text{s}(0)$        &     &     & $1$ & $\displaystyle \frac{1}{\cosh L/\ell_\text{sf} }$ & $\displaystyle \frac{\tau_\text{p}}{2\tau_\text{3D}\coth L/\ell_\text{sf} }$ 
\rule[-5mm]{0mm}{10mm}
\\
\multicolumn{1}{l|}{}        & $\displaystyle \mu^{(n)}_\text{s}(L)$        &     &     &     & $1$ & $\displaystyle \frac{\tau_\text{p}\sinh L/\ell_\text{sf} }{2\tau_\text{3D}}$ 
\rule[-5mm]{0mm}{10mm}
\\
\multicolumn{1}{l|}{}        & $\displaystyle \tau_{\text{p}}J^{(n)}_\text{s}(0)/N_0$        &     &     &     &     & $1$ 
\rule[-5mm]{0mm}{10mm}
\\ \hline\hline
\end{tabular}
\end{table}
\setlength{\tabcolsep}{6pt} 
\end{turnpage}
\begin{turnpage}
\setlength{\tabcolsep}{10pt} 
\begin{table}[ht]
 \centering
  \caption{
Ratios between physical quantities in the case of inverse Edelstein effect (IEE).
The entry in the row $X$ and column $Y$ gives the ratio $Y/X$.
The time scales $\tau_\text{p}$, $\tau_\text{t}$ and $\tau_\text{3D}$ are defined in Eq.~\eqref{eq: taup definition},
Eq.~\eqref{eq: taut definition} and Eq.~\eqref{eq: tau3D definition}, respectively, while
$\tau_\text{a}$, $\tau_\text{b}$, $\tau_\text{c}$, $\tilde{\alpha}$ and $r$ are defined in
Eqs.~\eqref{eq: auxiliary variables}.}
\label{tab: IEE coefficients}
\begin{tabular}{cc|ccccc}
\hline\hline \multicolumn{2}{c|}{\multirow{2}{*}{$\displaystyle \frac{Y}{X}\Big|_{\text{IEE}}$}}       
& \multicolumn{5}{c}{$Y$}     
\rule[-5mm]{0mm}{10mm}
\\ \cline{3-7} 
\multicolumn{2}{l|}{}                           & $\displaystyle 2s_z/N_0$ & $\displaystyle j_c/[(-e)v_\text{F}N_0/2]$ & $\displaystyle \mu^{(n)}_\text{s}(0)$ & $\displaystyle \mu^{(n)}_\text{s}(L)$ & $\displaystyle \tau_{\text{p}}J^{(n)}_\text{s}(0)/N_0$ 
\rule[-5mm]{0mm}{10mm}
\\ \hline
\multicolumn{1}{c|}{\multirow{9}{*}{$X$}} & $\displaystyle 2s_z/N_0$ & $1$ &     &     &     &     
\rule[-5mm]{0mm}{10mm}
\\
\multicolumn{1}{l|}{}                     & $\displaystyle j_c/[(-e)v_\text{F}N_0/2]$ & $\displaystyle \frac{-1}{\tilde{\alpha}\sin\delta}$ & $1$ &     &     &     
\rule[-5mm]{0mm}{10mm}
\\
\multicolumn{1}{l|}{} & \multirow{2}{*}{$\displaystyle \mu^{(n)}_\text{s}(0)$} & \multirow{2}{*}{$\displaystyle \frac{1}{1+\tau_\text{t} /\tau_\text{a}}$} & $\displaystyle \frac{-\tilde{\alpha}\sin\delta}{1+\tau_\text{t} /\tau_\text{a}}$ & \multirow{2}{*}{$1$} & \multirow{2}{*}{}    & \multirow{2}{*}{}    
\rule[0mm]{0mm}{10mm}
\\
\multicolumn{1}{l|}{}                     &     &     & 
\multicolumn{1}{c}{$\displaystyle = \frac{1}{Q_{12}v_\text{F}N_0}$} &     &     &     
\rule[-8mm]{0mm}{10mm}
\\
\multicolumn{1}{l|}{}                     & $\displaystyle \mu^{(n)}_\text{s}(L)$ & $\displaystyle \frac{\tau_\text{a}}{\tau_\text{c}\sinh L/\ell_\text{sf}}$ & $\displaystyle \frac{-\tilde{\alpha}\sin\delta\cdot\tau_\text{a}}{\tau_\text{c}\sinh L/\ell_\text{sf} }$ & $\displaystyle \frac{\tau_\text{t} +\tau_\text{a}}{\tau_\text{c}\sinh L/\ell_\text{sf} }$ & $1$ &     
\rule[-5mm]{0mm}{10mm}
\\
\multicolumn{1}{l|}{} & \multirow{2}{*}{$\displaystyle \tau_{\text{p}}J^{(n)}_\text{s}(0)/N_0$} & \multirow{2}{*}{$\displaystyle \frac{-2\tau_\text{a}}{\tau_\text{p}}$} & $\displaystyle 2\tilde{\alpha}\sin\delta\cdot\frac{\tau_\text{a}}{\tau_\text{p}}$ & \multirow{2}{*}{$\displaystyle (-2)\cdot\frac{\tau_\text{t} +\tau_\text{a}}{\tau_\text{p}}$} & \multirow{2}{*}{$\displaystyle 
\frac{-2\tau_\text{c}}{\tau_\text{p}}\cdot \sinh L/\ell_\text{sf}$} & \multirow{2}{*}{$1$} 
\rule[0mm]{0mm}{10mm}
\\
\multicolumn{1}{l|}{}                     &     &     & 
\multicolumn{1}{c}{$\displaystyle = 2\cdot \frac{\lambda_\text{IEE}}{v_\text{F}\tau_\text{p}}$} &     &     &     
\rule[-5mm]{0mm}{10mm}
\\
\multicolumn{1}{l|}{} & \multirow{2}{*}{$\displaystyle \tau_{\text{p}}J^{(n)}_\text{s}(L)/N_0$} & $\displaystyle \frac{-2 \tau_\text{3D}/\tau_\text{p}}{(1+\tau_\text{b}/\tau_\text{a})\sinh L/\ell_\text{sf}}$                  & $\displaystyle \frac{2\tilde{\alpha}\sin\delta\cdot (\tau_\text{3D}/\tau_\text{p})}
{(1+\tau_\text{b}/\tau_\text{a})\sinh L/\ell_\text{sf}}$ & \multirow{2}{*}{$\displaystyle \frac{-2\tau_\text{3D}/\tau_\text{p}}
{\sinh L/\ell_\text{sf}}\cdot \frac{\tau_\text{t} +\tau_\text{a}}{\tau_\text{a} +\tau_\text{b}}$} & \multirow{2}{*}{$\displaystyle 
\frac{-2\tau_\text{3D}}{\tau_\text{p}}\cdot\frac{\tau_\text{c}}{\tau_\text{a}+\tau_\text{b}}$} & \multirow{2}{*}{$\displaystyle \frac{\tau_\text{3D}}{(\tau_\text{a} +\tau_\text{b})\sinh L/\ell_\text{sf}}$} 
\rule[0mm]{0mm}{10mm}
\\
\multicolumn{1}{l|}{}                     &     & 
\multicolumn{1}{l}{$\displaystyle = \frac{2K''}{\tau_\text{p}}\cdot\langle S_z, S_z\rangle$} & 
\multicolumn{1}{l}{$\displaystyle = \frac{2\lambda_\text{recip}}{v_\text{F}\tau_\text{p}} = \frac{2K''}{\tau_\text{p}}\cdot\frac{\langle v_z, S_z\rangle}{v_\text{F}}$}
 &     &     &     
\rule[-5mm]{0mm}{10mm}
\\ \hline\hline
\end{tabular}
\end{table}
\setlength{\tabcolsep}{6pt} 
\end{turnpage}
\begin{subequations}
\label{eq: auxiliary variables}
\begin{gather}
 \frac{1}{\tau_\text{a}}\equiv \left(\frac{1}{\tau_{\text{p}}}+\frac{1}{\tau_\text{t}}\right)\tan\delta,\quad
 \tau_\text{b}\equiv \tau_\text{t} + \tau_\text{3D}\coth \frac{L}{\ell_\text{sf}},\\
 \tau_\text{c}\equiv \tau_\text{3D}+(\tau_\text{t} + \tau_\text{a})\coth \frac{L}{\ell_\text{sf}},\qquad
 \tilde{\alpha}\equiv \frac{\alpha}{v_{\rm F}\hbar},\\
 r\equiv 1 + \tilde{\alpha}^2\left(1- \frac{\sin 2\delta}{1 + \tau_\text{b}/\tau_\text{a}}\right). 
\end{gather}
\end{subequations}
For example, a current-induced spin polarization coefficient $\beta$ found in a relation $s_z = \beta j_c$ 
is obtained in Table~\ref{tab: EE coefficients} and Table~\ref{tab: IEE coefficients} as 
$\beta_{\text{EE}} = (ev_{\rm F})^{-1}
\cdot 2\tilde{\alpha}\cos\delta /[(1+\tau_\text{a}/\tau_\text{b})r]$
for the EE, and $\beta_{\text{IEE}}= (ev_{\rm F}\cdot\tilde{\alpha}\sin\delta)^{-1}$
for the IEE. They are different from each other,
reflecting the different non-equilibrium states between the EE and IEE.

\subsection{\label{subsec: Reciprocal relationship}
Reciprocal relationship}
We here derive the reciprocal relationship between the EE and IEE within the presented schemes.
Let $\mathscr{V}$ be the linear space of $\varphi(\bm{k},\gamma)$ 
and $G(\bm{k},\gamma,\bm{k}',\gamma')$ be the inverse matrix of 
$M_{\rm tot}(\bm{k},\gamma,\bm{k}',\gamma')$ in the quotient space $\mathscr{V}/{\rm Ker}(M_{\rm tot})$.
Then $\varphi(\bm{k},\gamma)$ is expressed as
\begin{equation}
\varphi(\bm{k},\gamma)=\sum_{\bm{k}',\gamma'}G(\bm{k},\gamma,\bm{k}',\gamma')
[b_{\text{EE}}(\bm{k}',\gamma')+b_{\text{IEE}}(\bm{k}',\gamma')].
\end{equation}
The charge current density and spin density are accordingly expressed as
\begin{equation}
\begin{bmatrix}
j^{\text{c}}_z/(-e)\\ s_z
\end{bmatrix}
= 
\begin{bmatrix}
\langle v_z, v_z  \rangle & \langle v_z, S_z\rangle \\
\langle S_z, v_z \rangle & \langle S_z, S_z \rangle 
\end{bmatrix}
\begin{bmatrix}
(-e) E_z \tau_{\text{p}}\\ 
K''\cdot J^{\text{ext}}_\text{s}
\end{bmatrix},\label{eq: current spin induced by E and Js}
\end{equation}
with $K'' = -2K\tau_{\text{p}}/N_0$ (The symbol $K$ has been defined in Eq.~\eqref{eq: K-def}). 
Here we defined a bilinear form
\begin{multline}
\langle X, Y\rangle \equiv
\frac{-1}{V}\sum_{\bm{k},\gamma, \bm{k}',\gamma'}
X(\bm{k},\gamma)
G(\bm{k},\gamma, \bm{k}',\gamma')
Y(\bm{k}',\gamma')\\
\cdot \delta (\epsilon(\bm{k},\gamma)-\mu_0).
\label{eq: inner product of BTE}
\end{multline}
As we consider the elastic scattering process,
the matrices $M_{\text{tot}}$ and its inverse $G$ have nonzero elements only between eigenstates with the same energy. 
We can thus replace $\epsilon(\bm{k},\gamma)$ in Eq.~\eqref{eq: inner product of BTE} by $\epsilon(\bm{k}',\gamma')$. 
In addition, $M_{\text{tot}}$ is a symmetric matrix and so is its inverse $G$.
The symmetry
\begin{equation}
\langle Y, X\rangle = \langle X, Y\rangle
\end{equation}
holds accordingly, which leads to equality between cross-coefficients: 
$\langle S_z, v_z \rangle = \langle v_z, S_z\rangle$.
This equality is also expressed as a ratio of quantities
in both sides of Eq.~\eqref{eq: current spin induced by E and Js}.
We can eliminate a factor $K''$ from this expression
by using a relation for the spin accumulation at the open end $\mu^{\text{(n)}}_\text{s}(L) = 4K s_z/N_0$
that holds for the EE.
A reciprocal relationship between the EE and IEE is then obtained as,  
\begin{equation}
\frac{-1}{2}\cdot 
\left.\frac{\mu^{\text{(n)}}_\text{s}(L)}{(-e)E}\right|_{\text{EE}}
= \left.\frac{j_\text{c}/(-e)
}{J^{\text{ext}}_\text{s}}\right|_{\text{IEE}}
~(~\equiv \lambda_{\text{recip}}~).
\label{eq: recip from BTE}
\end{equation}
Here the coefficients in both sides represent non-local responses separated by the interface;
$E$ and $J^{\text{ext}}_\text{s}$ are input, and
2D electric current $j_\text{c}$ can be detected as a voltage at the boundary of the 2D metal;
the spin accumulation at the edge $\mu^{\text{(n)}}_\text{s}(L)$ can also be detected by the Kerr effect.

The reciprocal relationship obtained above is a generalization of the reciprocal relationship at the surface
derived by K.~Shen et al.~\cite{Shen2014} to the interface system.
In their study, the spin current injection in the IEE was treated effectively as a time-dependent magnetic field
applying on the surface, though there remained ambiguity to read such external field to the spin current.
Our direct calculation on the spin current injection from the 3D metal
and resulting reciprocal relationship~\eqref{eq: recip from BTE} overcome that difficulty.

\begin{figure}[htbp]
\centering
\includegraphics[width=0.75\columnwidth]{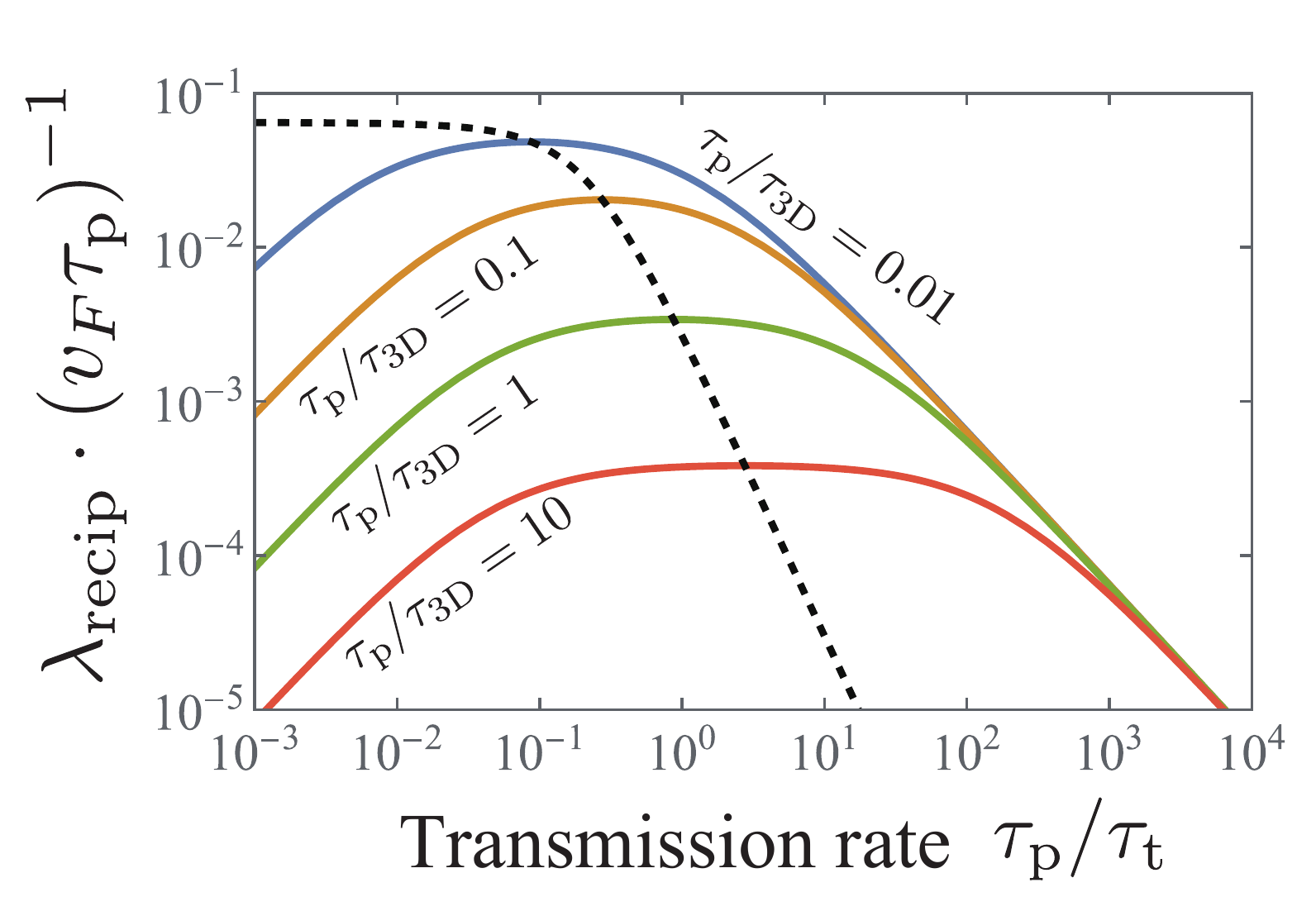}
\caption{(Color online)
Linear response specified by the reciprocal relationship of the Edelstein effect $\lambda_{\text{recip}}$
for varying the parameters of the interface and 3D metal.
The colored lines show different spin diffusion rates in the three-dimensional nonmagnetic metal.
The dashed black line shows the peak location of the response
with respect to the transmission rate $\tau_\text{p}/\tau_\text{t}$.
The parameters are chosen as $(\alpha/v_{\rm F}\hbar, \delta, L/\ell_{\text{sf}})= (0.1, \pi/64, 1)$.
}
\label{fig: reciprocity}
\end{figure}

The cross-coefficient given by the reciprocal relationship $\lambda_{\text{recip}}$ is shown in Fig.~\ref{fig: reciprocity}.
It is read that the linear response exhibits nonmonotonic behavior
with respect to the transmission rate across the interface $\tau_\text{p}/\tau_\text{t}$, 
while the smaller spin diffusion rate in the 3D metal
$\tau_\text{p}/\tau_\text{3D}$ gives a larger response.
In a low transmission rate $\tau_\text{p}/\tau_\text{t}\ll 1$, 
the response across the interface is governed by 
the spin transmission rate into or out of the 3D metal with finite thickness
$(\tau_\text{t} + \tau_\text{3D}\coth L/\ell_\text{sf})^{-1}$, as the same as $q_\text{EE}$.
Then we find $\lambda_\text{recip}\sim q_\text{EE}\cdot [\mu^{\text{(n)}}_\text{s}(L)/J^{\text{(n)}}_\text{s}(0)]_{\text{EE}}
\sim \tau_\text{3D}/(\tau_\text{t}\sinh L/\ell_\text{sf} + \tau_\text{3D}\cosh L/\ell_\text{sf})$, which explains
the behavior of $\lambda_\text{recip}$ in $\tau_\text{p}/\tau_\text{t}\ll 1$.
In a high transmission rate $\tau_\text{p}/\tau_\text{t}\gg 1$, on the other hand,
the bottleneck of the response is the Edelstein effect or its inverse at the 2D metal, 
which is roughly represented as the modified relaxation time in the 2D metal by the interface transmission
$(1/\tau_\text{p} + 1/\tau_\text{t})^{-1}$.
We thus find $\lambda_\text{recip}\sim (1/\tau_\text{p} + 1/\tau_\text{t})^{-1}$ in 
$\tau_\text{p}/\tau_\text{t}\gg 1$.
The response $\lambda_\text{recip}$ is maximal in the intermediate region, accordingly.

As the cross-coefficient given by the reciprocal relationship can be measured directly,
$\lambda_\text{recip}$ is an experimentally important ratio 
as well as $q_\text{EE}$ and $\lambda_\text{IEE}$.

We close this section by showing another representation of the reciprocal relationship 
that captures the essence of the composite system we have considered.
That is obtained by arranging the original linear transformation between 
the external forces $(E,~J^\text{(n)}_\text{s}(L))$ and responses $(j_c,~ \mu^\text{(n)}(L))$ 
into a new linear transformation between the pairs at both sides of the system
$(\mu^\text{(n)}_\text{s}(L),~J^\text{(n)}_\text{s}(L))$ and $(E,~j_c)$:
\begin{widetext}
\begin{equation}
\begin{bmatrix}
 \mu^\text{(n)}_\text{s}(L)/2 \\ -J^\text{(n)}_\text{s} (L)
\end{bmatrix} 
= 
\begin{bmatrix}
\cosh L/\ell_\text{sf} & \frac{e^2\ell_\text{sf}}{\sigma^\text{(n)}}\sinh L/\ell_\text{sf}\\ 
\frac{\sigma^\text{(n)}}{e^2\ell_\text{sf}}\sinh L/\ell_\text{sf} & \cosh L/\ell_\text{sf}
\end{bmatrix}
\begin{bmatrix}
 \mu^\text{(n)}_\text{s}(0)/2 \\ -J^\text{(n)}_\text{s} (0)
\end{bmatrix}
~,\qquad
\begin{bmatrix}
  \mu^\text{(n)}_\text{s}(0)/2 \\  -J^\text{(n)}_\text{s} (0)
\end{bmatrix} 
= 
\begin{bmatrix}
 Q_{11} & Q_{12}\\ Q_{21} & Q_{22}
\end{bmatrix}
\begin{bmatrix}
 (-e)E \\ j_c/(-e)
\end{bmatrix}.\label{eq: Qmatrix}
 \end{equation}
\end{widetext}
Here the former represents the spin diffusion equation in the 3D metal,
while the latter indicates a local charge--spin conversion at the interface.
The matrix elements $Q_{ij}$ are 
calculated based on the coefficients in Table~\ref{tab: EE coefficients} and Table~\ref{tab: IEE coefficients} as
\begin{subequations}
 \begin{align}
 Q_{11} & = (1 + Q_{12}Q_{21})/Q_{22},\\
 Q_{12}& = -(1+\tau_\text{t}/\tau_\text{a})/(v_{\rm F}N_0\tilde{\alpha}\sin\delta), \\
 Q_{21}&= (1 + \tilde{\alpha}^2)v_{\rm F}N_0/(2\tilde{\alpha}\cos\delta), \\
 Q_{22}&= -(v_{\rm F}\tau_\text{a}\tilde{\alpha}\sin\delta)^{-1} = -\lambda_\text{IEE}^{-1}.
 \end{align}
\end{subequations}
The Onsager's reciprocity~\eqref{eq: recip from BTE}
is equivalent to that the determinant of each matrix above is equal to 1,
in particular $\det Q =1$.

Moreover, the transfer matrix method expressed in Eqs.~\eqref{eq: Qmatrix}
serves as a powerful tool for computing transport coefficients via the Edelstein effect in the composite systems.
First, by using Eqs.~\eqref{eq: Qmatrix} and the matrix elements $Q_{ij}$,
we can derive not only the spin current-to-charge current conversion efficiency $\lambda_{\rm IEE} = -Q_{22}^{-1}$
but also the charge current-to-spin current conversion efficiency and cross-coefficient 
\begin{subequations}
 \begin{align} 
 q_{\rm EE} &= \left| Q_{11} + Q_{21}\cdot 
\frac{e^2\ell_\text{sf}}{\sigma^\text{(n)}} \coth \frac{L}{\ell_\text{sf}}\right|^{-1}\\
 \lambda_{\rm recip}
 &= - \left[Q_{12}\cdot\frac{\sigma^\text{(n)}}{e^2\ell_\text{sf}} \sinh \frac{L}{\ell_\text{sf}}
 + Q_{22}\cosh\frac{L}{\ell_\text{sf}} \right]^{-1}.
 \end{align}
\end{subequations}
The matrix elements $Q_{ij}$ thus play a fundamental role in the conversion ratio of the Edelstein effect.
Second, Eqs.~\eqref{eq: Qmatrix} allows us to consistently describe the EE and IEE 
not depending on the choice of controllable parameters.
For example, we can regard $(E, \mu^{\text{(n)}}_\text{s}(0))$, not $(E, J^{\text{(n)}}_\text{s}(L))$,
as a set of independent variables in Eqs.~\eqref{eq: Qmatrix}, i.e. input for the EE and IEE,
which is consistent with the previous studies~\cite{Zhang2016,Dey2018}.
Third, the transfer matrix method is useful for systematic calculations of transport coefficients
when another system, such as a spin Hall material, is attached on the $y=L$ plane 
(shown in Figs.~\ref{fig: schematic pic spin texture}~(b), (c)). 
In that new composite system, another transfer matrix would be multiplied to the vector at $y=0$ plane
$(\mu^\text{(n)}_\text{s}(L)/2, -J^\text{(n)}_\text{s}(L))$, 
which relates other parameters at another end of the attached system.

\section{Discussion}
We have presented a theoretical scheme capable of dealing with spin relaxation, nonlocal spin transport of a metal with strong SOC and charge--spin interconversion at an interface between a metal with strong SOC and a nonmagnetic metal. An important 
 direction of a future study is an application/ a generalization of the present scheme to the 3D chiral metal with the SOC expressed as 
$\alpha_{\|} k_z\sigma_z + \alpha_{\perp} (k_x\sigma_x + k_y\sigma_y)$. It will unravel the underlying mechanisms in  transport properties found in \cite{Inui2020,Shiota2021,Shishido2021}.  

Besides, our results are important as they stand in the sense that they can be translated to those on the 2D systems with the Rashba SOC or Rashba--Dresselhaus SOC. Among them, particularly, the analytical solution without using the relaxation approximation to the Boltzmann equation for the composite systems of 2D metal with SOC and 3D nonmagnetic metal will help us to understand and control spin transport through the interface between those systems and metals.

In generalizing the scheme in Sect.~\ref{sec: Charge--spin interconversion at the interface} to the bulk chiral metals, along with the experimental setup,
we need to calculate the spatial distribution of the charge current density and spin density 
in the chiral metal from the interior to the interface with another nonmagnetic metal, 
which may describe charge--spin interconversion more consistent with the experiments.

Along the experimental situations, we have to consider also spin-flip scattering process, which we neglected at the surface in the present study. It 
can contribute to these spin relaxation time and spin diffusion length of the conduction electrons in general. 
That scattering process is due to spin--orbit interaction from impurity potentials, lattice vibrations, and 
the hyperfine interaction~\cite{Overhauser1953,Yafet1963,Zutic2004}.

%
%
%
%
\section{Conclusions}
We have described spin transport in a spin-splitting model of the chiral metal surface and interface,
making full use of the Boltzmann transport equation beyond the relaxation time approximation.
The condition if we can safely use the Boltzmann transport equation for that two-band system 
is also discussed based on the Keldysh formalism in the Supplemental Material,
which endorses the validity of the following results.
We have first extracted slow modes responsible for spin relaxation and spin diffusion in the surface,
respecting conservation laws.
That enables us to define spin relaxation time and spin diffusion length
without using the conventional idea of spin-dependent chemical potentials.
Our definition applies to the systems with strong spin--orbit coupling in the clean limit 
when the Edelstein effect becomes evident, and 
it will serve as a foundation for discussing the non-local spin transport in the bulk chiral metal.
We have then clearly addressed the charge--spin interconversion efficiency at the interface,
which has been treated phenomenologically in previous studies.
In particular, we have derived the analytical expression for 
the charge current-to-spin current conversion efficiency $q_\text{EE}$ for the first time, 
which is found to depend on the details of the 3D nonmagnetic metal attached on the chiral metal surface. 
We have finally developed the Onsager's reciprocal relationship for Edelstein effect~\eqref{eq: recip from BTE} 
that relates local input and local output spatially separated by the interface.
Comparing the Edelstein effect and its inverse effect,
their distribution functions help us to understand the non-equilibrium states.
In addition, expressions for various transport coefficients that we have obtained analytically
would provide a powerful tool to 
evaluate the accuracy of measurements of the Edelstein effect, or to calculate what cannot be measured directly 
in the Edelstein effect.


\begin{acknowledgments}
We are grateful to Y.~Togawa, H.~Shishido, J.~Ohe, Y.~Fuseya and T.~Kato for constructive discussions on the subject.
We wish to thank J.~Kishine, H.~M.~Yamamoto, H.~Kusunose, E.~Saitoh and S.~Sumita for their helpful comments. 
Y.S. is supported by World-leading Innovative Graduate Study Program for Materials
Research, Industry, and Technology (MERIT-WINGS) of the University of Tokyo. 
Y.S. is also supported by JSPS KAKENHI Grant Number 22J12348. 
Y.K. is supported by JPSJ KAKENHI Grant Number 20K03855 and 21H01032.
This research was supported by Special Project by Institute for Molecular Science (IMS program 21-402).
\end{acknowledgments}



%

\onecolumngrid
\clearpage 

\renewcommand{\theequation}{S\arabic{equation}}
\renewcommand{\thefigure}{S\arabic{figure}}
\renewcommand{\thetable}{S\arabic{table}}
\renewcommand{\thesection}{S\arabic{section}}
\setcounter{equation}{0}
\setcounter{figure}{0}
\setcounter{table}{0}
\setcounter{section}{0}
\setcounter{page}{1}
\makeatletter

\begin{center}
\textbf{\large
Supplemental Material for \\
``Spin Relaxation, Diffusion and Edelstein Effect in Chiral Metal Surface''
}
\end{center}
\vspace{5pt}

\twocolumngrid

\section{Relation between our model and other SOC models}
The spin--orbit coupling (SOC) we assumed in the main text with two SOC parameters
$H_{\text{SO}} = \alpha_{\|} k_z\sigma_z + \alpha_{\perp}k_x\sigma_x$
can represent various types of SOC that is linear in $\bm{k}$.
We give some examples in this section.

We first take up the Rashba SOC, expressed as $\alpha_\text{R}(k_z\sigma_x -k_x\sigma_z)$.
Let us introduce new spin operators (with tilde) in a rotated frame by 90 degrees as 
$(\tilde{\sigma}_z, \tilde{\sigma}_x)= (-\sigma_x, \sigma_z)$.
The isotropic case of $H_\text{SO}$ with $\alpha_\| = \alpha_\perp = \alpha/\sqrt{2}$
is then expressed as the Rashba SOC
\begin{equation}
H_\text{SO}= \frac{\alpha}{\sqrt{2}}(k_z\sigma_z+k_x\sigma_x)
= \frac{\alpha}{\sqrt{2}}(k_z\tilde{\sigma}_x-k_x\tilde{\sigma}_z)~.
\end{equation}
Note that the rotation here is only applied 
to the spin space, and leaves the real space and momentum space unchanged.

We then discuss another example, the two-dimensional system with both 
the Rashba and Dresselhaus SOCs
$\alpha_{\text{R}} (k_z\sigma_x - k_x\sigma_z) + \alpha_{\text{D}} (k_z\sigma_z -k_x\sigma_x)$.
This model is identical to our model; if we apply
$45$-degrees rotations in opposite direction on the real (momentum) space and spin space such that
\begin{align}
    (\tilde{k}_z\quad \tilde{k}_x)&=
    ({k}_z\quad {k}_x)\frac{1}{\sqrt{2}}
    \begin{pmatrix}
     1 & -1\\ 1 & 1
    \end{pmatrix}
    \\
    (\tilde{\sigma}_z\quad \tilde{\sigma}_x)&=
    ({\sigma}_z\quad {\sigma}_x)\frac{1}{\sqrt{2}}
    \begin{pmatrix}
     1 & 1\\ -1 & 1
    \end{pmatrix}
~,
\end{align}
our model $H_\text{SO}$ in the original frame is then expressed as
the Rashba--Dresselhaus SOC with $\alpha_{\text{R}} = (\alpha_{\|} + \alpha_{\perp})/2, ~
\alpha_{\text{D}} = (\alpha_{\|} - \alpha_{\perp})/2$ in the rotated frame
\begingroup
\allowdisplaybreaks
\begin{align}
&H_\text{SO}=
\alpha_{\|} k_z\sigma_z + \alpha_{\perp}k_x\sigma_x\\
&= \frac{\alpha_{\|} + \alpha_{\perp}}{2}
(\tilde{k}_z\tilde{\sigma}_x - \tilde{k}_x\tilde{\sigma}_z) 
+\frac{\alpha_{\|} - \alpha_{\perp}}{2}
(\tilde{k}_z\tilde{\sigma}_z -\tilde{k}_x\tilde{\sigma}_x). 
\end{align}
\endgroup

We finally consider every SOC term allowed in 
the surface ($zx$ plane) of the chiral metals with point group $D_6$ (622) with $z$-axis being 
the principal axis of the bulk.
As the original $D_6$ symmetry is reduced to $C_{2y}$-rotation within the surface,
the SOC terms are written with four independent parameters~\cite{Frigeri2005}
\begin{equation}
 H^{\text{general}}_{\text{SO}} = 
\begin{pmatrix}
 k_z & k_x
\end{pmatrix}
\begin{pmatrix}
 \alpha_{zz} & \alpha_{zx}\\
 \alpha_{xz} & \alpha_{xx}
\end{pmatrix}
\begin{pmatrix}
 \sigma_z\\\sigma_x
\end{pmatrix}
= \bm{k}^{\top} \underline{\alpha} \bm{\sigma}
\label{eq: general SOC terms}
\end{equation}
The Rashba SOC is also included in the expression above.
The whole SOC terms $H^{\text{general}}_{\text{SO}}$ can also be reduced to our SOC model $H_\text{SO}$
under proper rotations and reflection in the real space and spin space.
The real $2\times 2$ matrix $\underline{\alpha}$ denoting SOC parameters in Eq.~\eqref{eq: general SOC terms}
is factorized by the singular value decomposition as
$\underline{\alpha} = U^{\top}\Sigma V$, where $U$ and $V$ are real orthogonal matrices
and $\Sigma$ is a diagonal matrix with non-negative real numbers on the diagonal.
It follows that the SOC terms
\begin{equation}
 H_{\text{SO}}= 
\left(U\bm{k}\right)^{\top} 
\begin{pmatrix}
 \Sigma_{zz} &\\ & \Sigma_{xx}
\end{pmatrix}
 \left(V\bm{\sigma}\right)~.
\end{equation}
described by new variables $\bm{k}'\equiv U\bm{k}$ and $\bm{\sigma}'\equiv V\bm{\sigma}$
are considered to be a parallel coupling of spin and momentum with SOC parameters $\Sigma_{zz}$ and $\Sigma_{xx}$.
Note that the rotations and reflections by $U$ and $V$ are different in general,
and they separately act on the real space and spin space, respectively.

\section{Derivation of basic quantities in the chiral metal surface model}
In this section, we detail analytical calculations on some basic 
quantities of the two-dimensional (2D) spin-splitting system, omitted at Sect.~II in the main text.
We consider the collision integral, radii of the Fermi contours, density of states, summation over states on Fermi contours,
and group velocity for each band.
These analytical expressions are also used in the following sections of this Supplemental Material.

We first consider the Boltzmann collision integral due to the impurity scattering $\displaystyle \frac{df}{dt}\Big|_\text{col}$.
The potential due to randomly distributed nonmagnetic impurities with the density 
$n_{\rm imp}$ in the two-dimensional system with the areal volume $V$ is set to be 
\begin{equation}
V_{\rm imp}(\bm{r})=v_0 \sum_{j=1}^{N_{\rm imp}}\delta(\bm{r}-\bm{r}_j)~.
\label{eq: imp-scatt potential}
\end{equation}
Here $N_{\rm imp}(=n_{\rm imp}V)$ is the number of impurity centers and $v_0$ is the strength of each impurity. The $j$-th impurity is located at $\bm{r}_j$.
After we take the average over the impurity location, the expression for the collision term
is obtained as follows, by using the Fermi's golden rule:
\begingroup
\allowdisplaybreaks
\begin{widetext}
\begin{align}
\frac{df(\bm{k},\gamma)}{dt}\Big|_{\text{col}}&=
\frac{2\pi}{\hbar}\sum_{\bm{k}', \gamma'}\Abs{\Braket{\psi_{\bm{k}',\gamma'} |V_{\text{imp}}|\psi_{\bm{k},\gamma}}}^2 
\cdot \left[f(\bm{k}', \gamma')(1-f(\bm{k},\gamma)) - f(\bm{k}, \gamma)(1-f(\bm{k}',\gamma'))\right] 
\delta \bm{(}\epsilon (\bm{k}', \gamma') - \epsilon (\bm{k}, \gamma)\bm{)}\\
&= \frac{2\pi v_0^2n_{\text{imp}}}{\hbar V}\sum_{\bm{k}', \gamma'}
\Abs{\Braket{\bm{k}', \gamma'|\bm{k}, \gamma}}^2
[ f(\bm{k}',\gamma') - f(\bm{k},\gamma) ]
\cdot \delta\bm{ (}\epsilon (\bm{k}', \gamma') - \epsilon (\bm{k}, \gamma)\bm{) }
\label{eq: col term from G.R.}\\
&=\frac{1}{\tau_{\text{p}}} \sum_{\bm{k}', \gamma'} M_{\text{col}}(\bm{k},\gamma,\bm{k}',\gamma') f(\bm{k}', \gamma')~.
\label{eq: collision integral as relaxation mat}
\end{align}  
\end{widetext}
\endgroup
Here we put the Bloch state as
$ \ket{\psi_{\bm{k},\gamma}} = V^{-1/2} e^{i\bm{k}\cdot \hat{\bm{r}}}\ket{\bm{k}, \gamma}$
with $\hat{\bm{r}}$ position operator.

We then turn to follow the derivation of the density of states
$N_0 = N_0(\epsilon_\text{F}) = m/(\pi\hbar^2)$ for $\epsilon_\text{F} > 0$, 
which is included in the definition of the typical impurity scattering rate $\tau_\text{p}^{-1}$.
To simplify the sum of the states, we first derive two radii of the Fermi contours 
from an equation $\epsilon(k,\theta,\gamma) = \epsilon_\text{F}$, i.e. 
\begin{multline}
\frac{\hbar^2 k^2}{2m} + \gamma \alpha k \sqrt{\cos^2\delta\cos^2\theta+\sin^2\delta\sin^2\theta}\\
= \epsilon_\text{F} ~
\left(
= \frac{\hbar^2 k_\text{F}^2}{2m} = \frac{mv_\text{F}^2}{2}
\right) 
\end{multline}
where $k_\text{F}$ and $v_\text{F}$ are the Fermi momentum and Fermi velocity in the absence of the SOC ($\alpha = 0$), respectively.
The quadratic equation with respect to $k = k(\theta, \gamma, \epsilon_\text{F})$ provides the radii of the Fermi contours
for two bands $\gamma = \pm $, as shown in Fig.~\ref{fig: dispersion simple ver},
\definecolor{mycolor1}{HTML}{5D7EAB}
\definecolor{mycolor2}{HTML}{D6933C}
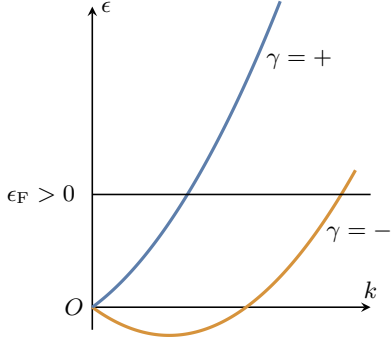
\begin{figure}
\centering\begin{tikzpicture}
\draw[->,>=stealth,semithick](0,0)--(3.7,0)node[above]{$k$};
\draw[->,>=stealth,semithick](0,-0.3)--(0,4)node[right]{$\epsilon$};
\draw(0,0)node[left]{$O$};
\draw[very thick,mycolor2,samples=100,domain=0:3.5]plot(\x,{(2*(2.2*\x/3)^2-3*(2.2*\x/3))/3});
\draw[very thick,mycolor1,samples=100,domain=0:2.5]plot(\x,{(2*(2.2*\x/3)^2+3*(2.2*\x/3))/3});
\draw[semithick](3.7,1.5)--(0,1.5)node[left]{$\epsilon_\text{F} > 0\ $};
\draw(2.2,3.3)node[right]{$\gamma = +$};
\draw(3.0,1.0)node[right]{$\gamma = -$};
\end{tikzpicture}
 \caption{
Dispersion of two bands $\epsilon = \epsilon(k,\theta,\gamma =\pm 1)$
splitting by the SOC depicted for a fixed angle $\theta$. 
The origin $\epsilon =k= 0$ corresponds to the band-crossing point.}
\label{fig: dispersion simple ver}
\end{figure}
\begin{equation}
 \tilde{k}(\theta,\gamma)= \sqrt{1+\tilde{\alpha}^2A^2(\theta)} - \gamma \tilde{\alpha }A(\theta)
\label{eq: radius Fermi contour}
\end{equation}
with dimensionless wavevector $\tilde{k}$ and spin--orbit coupling constant $\tilde{\alpha}$ as well as 
a direction-dependent factor~$A(\theta)$
\begin{subequations}
\label{eq: tilde k tilde alpha}
 \begin{gather}
 \tilde{k} \equiv \frac{k(\theta, \gamma, \epsilon_\text{F})}{k_\text{F}}, \quad
 \tilde{\alpha}\equiv \frac{\alpha}{v_\text{F}\hbar},\\
 A(\theta)\equiv \sqrt{\cos^2\delta\cos^2\theta+\sin^2\delta\sin^2\theta}.
 \end{gather}
\end{subequations}

We now calculate the sum of the states around the Fermi contours
\begin{equation}
 N_0(\epsilon_\text{F}) = \frac{1}{V}\sum_{\bm{k},\gamma}\delta \bm{(}\epsilon (\bm{k},\gamma) -\epsilon_\text{F}\bm{)}~.
\label{eq: DOS}
\end{equation}
The delta function is expanded as
\begin{multline}
\delta \bm{(} \epsilon (\bm{k},\gamma) - \epsilon_\text{F} \bm{)}\\ = 
\Abs{\left.\frac{\partial \epsilon (k,\theta,\gamma)}{\partial k}\right|_{k=k_\text{F}\cdot \tilde{k}(\theta, \gamma)}}^{-1}\cdot 
\delta \bm{(} k - k_\text{F}\cdot \tilde{k}(\theta, \gamma) \bm{)} 
\end{multline}
by using the solution of $\epsilon_\text{F} = \epsilon(k,\theta,\gamma) > 0$ obtained just above.
The prefactor corresponding to the Fermi velocity is
\begin{align}
&  \Abs{\left.\frac{\partial \epsilon (k,\theta,\gamma)}{\partial k}\right|_{k=k_\text{F}\cdot \tilde{k}(\theta, \gamma)}}
= \Abs{\frac{\hbar^2k_\text{F}\cdot \tilde{k}(\theta, \gamma)}{m}+ \gamma \alpha  A(\theta)}\\
&= \frac{\hbar^2 k_\text{F}}{m}\sqrt{
1 + \tilde{\alpha}^2 A^2(\theta)}. 
\end{align}
Replacing the sum~\eqref{eq: DOS} with an integral leads to the expression for the 2D density of states 
\begingroup
\allowdisplaybreaks
\begin{align}
&N_0(\epsilon_\text{F})= \sum_{\gamma = \pm} \int \frac{k dk d\theta}{(2\pi)^2} 
\frac{\delta \bm{(} k - k_\text{F}\cdot \tilde{k}(\theta, \gamma) \bm{)}}{
(\hbar^2 k_\text{F}/m)\sqrt{1 + \tilde{\alpha}^2 A^2(\theta)}}\\
&= \frac{m}{2\pi\hbar^2}
 \sum_{\gamma = \pm} \int^{2\pi}_0\frac{d\theta}{2\pi}
\left[
1-\frac{\gamma \tilde{\alpha}A(\theta)}{\sqrt{1+\tilde{\alpha}^2A^2(\theta) }}
\right]\\
&= \frac{m}{\pi\hbar^2}~. 
\end{align}
\endgroup

In the same way, the average of an arbitrary function $O(\bm{k},\gamma)= O(k,\theta, \gamma)$
over the Fermi contours in the 2D system (2DFC), defined as
\begin{equation}
 \left\langle O\right\rangle_{\text{2DFC}} \equiv \frac{1}{N_0 V}
\sum_{\bm{k},\gamma} O(\bm{k},\gamma)\cdot \delta \bm{(} \epsilon (\bm{k},\gamma) - \epsilon_\text{F} \bm{)}~,
\label{eq: def average over 2DFC}
\end{equation}
is arranged to the following expression:
\begin{equation}
 \left\langle O\right\rangle_{\text{2DFC}}
=  \frac{1}{2} \sum_{\gamma = \pm} \int^{2\pi}_0\frac{d\theta}{2\pi}
\left[
1-\frac{\gamma \tilde{\alpha}A(\theta)}{\sqrt{1+\tilde{\alpha}^2A^2(\theta) }}
\right]O(\theta, \gamma) 
\label{eq: average over 2DFC expression}
\end{equation}
where $O(\theta, \gamma) \equiv O\bm{(}k=k_\text{F}\cdot \tilde{k}(\theta, \gamma), \theta, \gamma \bm{)}$ 
is the expression at the Fermi energy.
The Boltzmann collision integral~\eqref{eq: col term from G.R.} is also written as
\begin{widetext}
 \begin{equation}
  \frac{df(\bm{k},\gamma)}{dt}\Big|_{\text{col}}
= \frac{1}{\tau_{\text{p}}} \sum_{\gamma' = \pm} \int^{2\pi}_0\frac{d\theta'}{2\pi}
\left[ 1-\frac{\gamma' \tilde{\alpha}A(\theta')}{\sqrt{1+\tilde{\alpha}^2A^2(\theta') }} \right]
 \cdot\frac{1+\gamma\gamma'\cos[\Theta (\theta) - \Theta (\theta')]}{2}
[ f(\theta',\gamma') - f(\theta, \gamma) ]\label{eq: explicit collison integral} 
 \end{equation}
\end{widetext}
with $\epsilon (\bm{k},\gamma)= \epsilon_\text{F}$.
We will make use of both expressions~\eqref{eq: average over 2DFC expression} and \eqref{eq: explicit collison integral}
in the following.

We finally calculate the $z$-component of the group velocity:
\begingroup
\allowdisplaybreaks
\begin{align}
& \hbar v_{z}(\bm{k},\gamma) 
=\frac{\partial \epsilon (\bm{k},\gamma )}{\partial k_z}
= \frac{\partial k}{\partial k_z} \frac{\partial \epsilon (k,\theta , \gamma )}{\partial k} +
\frac{\partial \theta}{\partial k_z} \frac{\partial \epsilon (k,\theta , \gamma )}{\partial \theta}\\
 &= \cos \theta  \frac{\partial \epsilon (k,\theta , \gamma )}{\partial k} - \frac{\sin\theta}{k}
\frac{\partial \epsilon (k,\theta , \gamma )}{\partial \theta}\\
&= \frac{\hbar^2k}{m}\cos\theta
+ \frac{\gamma \alpha}{A(\theta)}
\left[
\cos\theta A^2(\theta) -
\frac{\sin\theta}{2} \frac{ \partial A^2(\theta)}{\partial \theta}
\right]~.
\end{align}
\endgroup
Here the second term with $A^2(\theta)$ is simplified as
\begingroup
\allowdisplaybreaks
\begin{align}
&\cos\theta A^2(\theta) -
\frac{\sin\theta}{2} \frac{ \partial A^2(\theta)}{\partial \theta}\\
&=\cos\theta (\cos^2\delta\cos^2\theta + \sin^2\delta\sin^2\theta)\nonumber \\
&\quad - \frac{\sin\theta}{2}\frac{\partial}{\partial \theta}\left(
\cos^2\delta\cos^2\theta + \sin^2\delta\sin^2\theta
\right)\\
& = \cos^2\delta \cos\theta~, 
\end{align}
\endgroup
which yields
\begin{equation}
 v_{z}(\bm{k},\gamma) = 
 \frac{\hbar k}{m}\cos\theta  + \frac{\gamma \alpha}{\hbar}\cdot
 \frac{\cos^2\delta \cos\theta }{A(\theta)}\\ 
\end{equation}
At the Fermi energy, 
the group velocity normalized by the Fermi velocity in the absence of the SOC $v_\text{F} = \hbar k_\text{F}/m$ is 
\begin{equation}
 v_{z}(\theta,\gamma)/v_\text{F} = 
\tilde{k}(\theta,\gamma)\cos\theta +\gamma \tilde{\alpha}\cos\delta\cos\Theta (\theta)\label{eq: v_z around Fermi contours}
\end{equation}
with the dimensionless radii of the Fermi contours $\tilde{k}(\theta, \gamma)$
and SOC parameter $\tilde{\alpha}$, shown in Eqs.~\eqref{eq: radius Fermi contour} and
\eqref{eq: tilde k tilde alpha}.

\section{Detail of the Boltzmann equation analysis at Chiral Metal Surface}
In this section, we describe details of the analysis based on 
the Boltzmann transport equation (BTE) that are omitted at Sect.~II in the main text.
In Sect.~\ref{subsec: Edelstein effect in uniform steady state},
we first consider linear responses to an external electric field applied on the surface of the chiral metal.
In Sect.~\ref{subsec: Relaxation in time with spatially-homogeneous state},
we then review the relaxation case and derive analytical expression for both the spin relaxation time
and corresponding deviation of the distribution function.
In Sect.~\ref{subsec: Diffusion in space with temporally stationary state},
we finally give the details of the diffusion case, such as the choice of parameters in 
the numerical calculation of the eigenmode analysis.

\subsection{\label{subsec: Edelstein effect in uniform steady state}
Edelstein effect in uniform steady state}

In the presence of static uniform electric field $E$ parallel to the $z$-direction, the BTE is expressed as
\begin{equation}
 \frac{(-e)E}{\hbar}\frac{\partial f}{\partial k_z} = \frac{df}{dt}\Big|_{\text{col}}~.
\end{equation}
Here the left-hand side is 
\begin{align}
 \frac{(-e)E}{\hbar}\frac{\partial f}{\partial k_z}
&= \frac{(-e)E}{\hbar}v_{z}(\bm{k},\gamma)
\frac{\partial f_0\bm{(}\epsilon(\bm{k},\gamma)\bm{)}}{\partial \epsilon(\bm{k},\gamma)}\\
&\simeq -\frac{(-e)E}{\hbar}v_{z}(\bm{k},\gamma)
\delta\bm{(}\epsilon(\bm{k},\gamma) -\epsilon_\text{F}\bm{)} 
\end{align}
in the low temperature $k_B T\ll \epsilon_\text{F}$, which shows
the deviation from equilibrium in the distribution function only occurs around the Fermi energy $\epsilon_\text{F}$.
We thus put the distribution function as
\begin{equation}
f(\bm{k},\gamma)=f_0\bm{(}\epsilon(\bm{k},\gamma)\bm{)}
+\varphi (\bm{k},\gamma)
\left(-
\frac{\partial f_0\bm{(}\epsilon(\bm{k},\gamma)\bm{)}}{\partial \epsilon(\bm{k},\gamma)}\right),
\end{equation}
where the deviation $\varphi (\bm{k},\gamma)$ is of first order in the electric field $E$.

The BTE is then reduced to an inhomogeneous linear equation 
\begin{subequations}
 \label{eq: linear-algebra-inhomo}
\begin{equation}
b_\text{EE}(\bm{k},\gamma)=\sum_{\bm{k}',\gamma'}M_\text{col}(\bm{k},\gamma,\bm{k}',\gamma')\varphi(\bm{k}',\gamma')~,
\end{equation}
for states $(\bm{k},\gamma)$ at the Fermi energy $\epsilon (\bm{k},\gamma)= \epsilon_\text{F}$.
Here inhomogeneous term $b_\text{EE}(\bm{k},\gamma)=eE\tau_{\text{p}} v_{z}(\bm{k},\gamma)$ arising from the electric field
drives the shift of Fermi contours.
More explicit expression for this equation is obtained by means of Eq.~\eqref{eq: explicit collison integral} as
\begin{widetext}
\begin{multline}
  -\left[\sqrt{1+\tilde{\alpha}^2A^2(\theta)} - \gamma \tilde{\alpha }A(\theta)\right]\cos\theta 
-\gamma \tilde{\alpha}\cos\delta\cos\Theta (\theta) \\
= \sum_{\gamma' = \pm} \int^{2\pi}_0\frac{d\theta'}{2\pi}
\left[ 1-\frac{\gamma' \tilde{\alpha}A(\theta')}{\sqrt{1+\tilde{\alpha}^2A^2(\theta') }} \right]
\frac{1+\gamma\gamma'\cos[\Theta (\theta) - \Theta (\theta')]}{2}
\left[
\frac{\varphi(\theta',\gamma')}{(-e)Ev_\text{F}\tau_{\text{p}}}
- \frac{\varphi(\theta, \gamma)}{(-e)Ev_\text{F}\tau_{\text{p}}}
\right]~.\label{eq: BTE of EE in 2D analytic} 
\end{multline}
\end{widetext}
\end{subequations}

That linear equation~\eqref{eq: linear-algebra-inhomo} still has a redundancy in its solutions;
a new solution $\varphi^{\text{new}} (\bm{k},\gamma)$ can be created by adding an arbitrary constant to
an existing one $\varphi^{\text{new}} (\bm{k},\gamma)= \varphi^{\text{old}} (\bm{k},\gamma) + \text{constant}$.
We can remove that redundancy by considering the charge neutrality condition.
That condition is expressed as
\begin{equation}
 \left\langle{\varphi}(\bm{k},\gamma) \right\rangle_{\text{2DFC}} = 0
\label{eq: charge neutrality cond}
\end{equation}
with the average over the 2D Fermi contours $\langle \cdots \rangle_\text{2DFC}$ given by Eq.~\eqref{eq: def average over 2DFC}.

\begin{figure}
 \centering
\includegraphics[width=0.8\columnwidth]{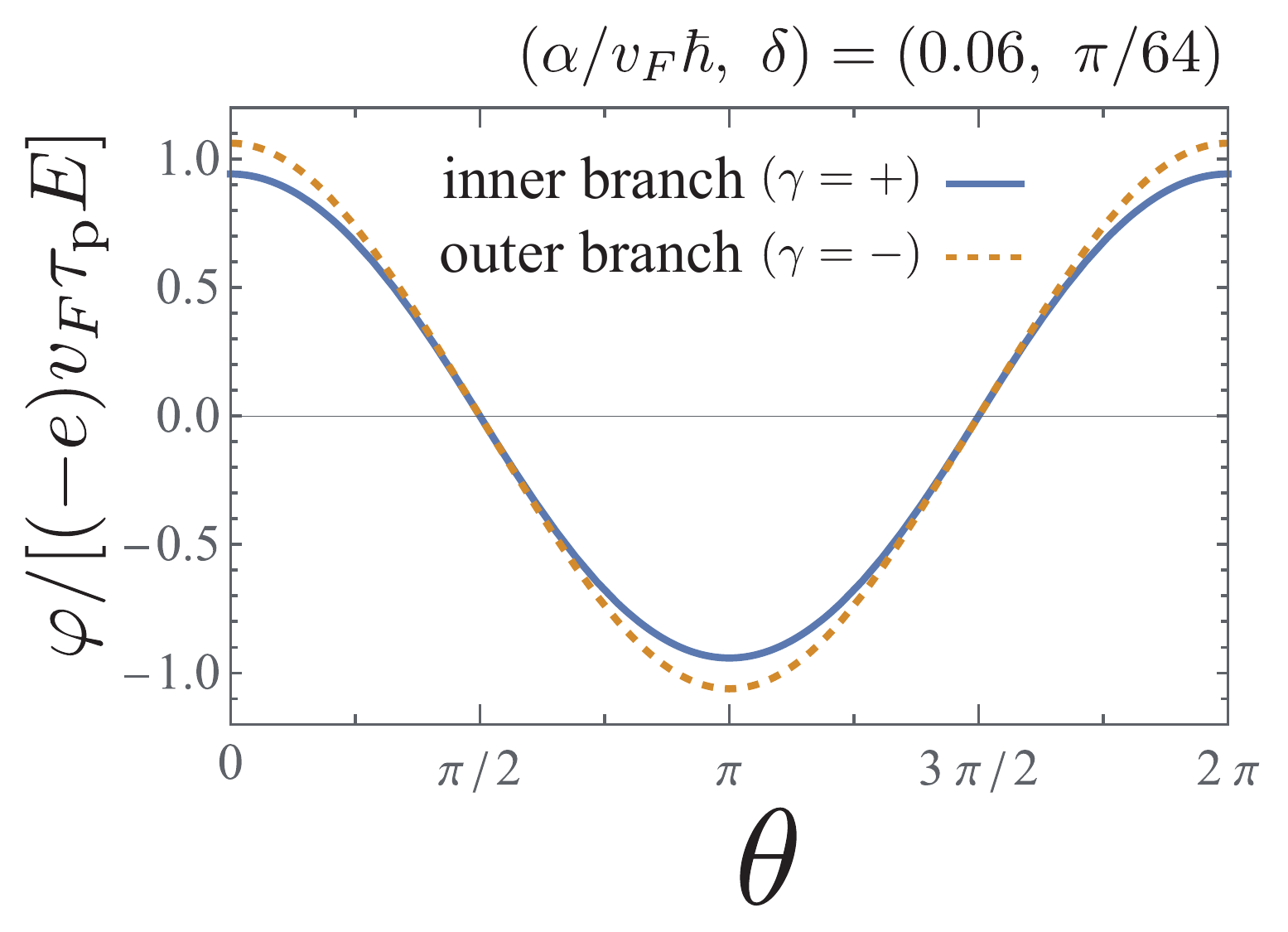}
 \caption{
 Deviation of distribution function of the 2D electron system, with no metals attached on,
 in the presence of constant electric field $E$.
 The deviation $\varphi$ is plotted around the Fermi contours with $0\leq \theta <2\pi$ and two bands $\gamma=\pm$.
 }
 \label{fig: Edelstein Mode}
\end{figure}

We find the analytical expression
for the deviation $\varphi(\theta,\gamma)$ that satisfies both the BTE~\eqref{eq: linear-algebra-inhomo} and 
charge neutrality condition~\eqref{eq: charge neutrality cond}, written as
\begin{subequations}
\label{eq: sol of BTE of EE in 2D} 
 \begin{align}
 \frac{\varphi(\theta, \gamma)}{(-e)Ev_\text{F}\tau_{\text{p}}}
& = \left[\sqrt{1+\tilde{\alpha}^2A^2(\theta)} - \gamma \tilde{\alpha }A(\theta)\right]\cos\theta \\
& = \tilde{k}(\theta,\gamma)\cos\theta ~.
 \end{align}
\end{subequations}
The deviation, plotted in Fig.~\ref{fig: Edelstein Mode},
is described as cosine-like curve in both inner and outer Fermi contours,
which indicates the shift of Fermi contours in the direction of the external electric field $(\theta = 0)$.

The 2D electric current density associated with that momentum shift is 
\begingroup
\allowdisplaybreaks
 \begin{align}
 &j^c_z = \frac{-e}{V} \sum_{\bm{k},\gamma}v_z(\bm{k},\gamma) \varphi (\bm{k},\gamma)\cdot 
\delta \bm{(}\epsilon (\bm{k},\gamma) -\epsilon_\text{F}\bm{)}\\
&= (-e)N_0 \langle v_z \varphi\rangle_{\text{2DFC}}\\
 &= (-e)N_0\cdot \frac{1}{2} \sum_{\gamma = \pm} \int^{2\pi}_0\frac{d\theta}{2\pi}
\left[
1-\frac{\gamma \tilde{\alpha}A(\theta)}{\sqrt{1+\tilde{\alpha}^2A^2(\theta) }}
\right]\nonumber\\ &\qquad \cdot 
v_\text{F}\left[ \tilde{k}(\theta,\gamma)\cos\theta +\gamma \tilde{\alpha}\cos\delta\cos\Theta (\theta)\right]\nonumber \\
&\qquad \cdot(-e)Ev_\text{F}\tau_{\text{p}}\tilde{k}(\theta,\gamma)\cos\theta\\
&= e^2 {v_\text{F}}^2N_0\tau_{\text{p}} E\cdot \frac{1}{2}(1+\tilde{\alpha}^2)
= \frac{n_\text{2D}e^2\tau_{\text{p}}}{m}E
\end{align}
\endgroup
with the particle number density 
\begin{equation}
 n_\text{2D} = \int^{\epsilon_\text{F}}_{-\infty} d\epsilon~N_0(\epsilon) = \frac{k_\text{F}^2}{2\pi}(1+\tilde{\alpha}^2)~.
\end{equation}
The linear response of spin density is also accompanied by 
the difference in the shift of the two Fermi contours:
\begingroup
\allowdisplaybreaks
\begin{align}
& s_z = \frac{1}{V}\sum_{\bm{k},\gamma} 
\Braket{\bm{k},\gamma|\sigma_z|\bm{k},\gamma}\varphi (\bm{k},\gamma)\cdot 
\delta \bm{(}\epsilon (\bm{k},\gamma) -\epsilon_\text{F}\bm{)}\\
 &= N_0\cdot \frac{1}{2} \sum_{\gamma = \pm} \int^{2\pi}_0\frac{d\theta}{2\pi}
\left[
1-\frac{\gamma \tilde{\alpha}A(\theta)}{\sqrt{1+\tilde{\alpha}^2A^2(\theta) }}
\right]\nonumber \\& \qquad \cdot \gamma \cos\Theta(\theta)
\cdot (-e)Ev_\text{F}\tau_{\text{p}}\tilde{k}(\theta,\gamma)\cos\theta\\
&= ev_\text{F} N_0\tau_{\text{p}}\tilde{\alpha}\cos\delta\cdot E\\
&= -2\cdot \tilde{\alpha}\cos\delta \cdot \frac{k_\text{F}^2}{2\pi}\cdot \frac{(-e)E\tau_{\text{p}}}{\hbar k_\text{F}}~.
\end{align}
\endgroup

\subsection{\label{subsec: Relaxation in time with spatially-homogeneous state}
Relaxation in time with spatially-homogeneous state}
In the main text, we have considered the relaxation case with the BTE and electron distribution $f = f(t,\bm{k}, \gamma)$,
which are provided as follows:
\begin{subequations}
 \label{eq: BTE for Relaxation}
 \begin{gather}
 \frac{\partial f}{\partial t} = \frac{df}{dt}\Big|_{\text{col}}~,\\
 f(t,\bm{k},\gamma) = f_0\bm{(}\epsilon (\bm{k},\gamma)\bm{)} + e^{-t/\tau} 
 \varphi_{\tau}(\bm{k},\gamma)\left(-\frac{\partial f_0 (\epsilon)}{\partial \epsilon}\right)~.
 \end{gather}
\end{subequations}
The time constant $\tau$ indicates the relaxation time,
while the deviation $\varphi_{\tau}$ characterize the non-equilibrium state.
The BTE after the substitution of this assumption results in an eigenvalue problem 
\begin{equation}
-\frac{\tau_{\text{p}}}{\tau} \varphi_{\tau}(\bm{k},\gamma)=\sum_{\bm{k}',\gamma'}
M_{\text{col}}(\bm{k},\gamma,\bm{k}',\gamma')\varphi_{\tau}(\bm{k}',\gamma')~.\label{eq: BTE for relaxation w tau}
\end{equation}

We now exactly solve this eigenvalue problem.
We first rewrite the relaxation matrix $M_{\text{col}}$ as an integral around the Fermi contours
based on Eqs.~\eqref{eq: collision integral as relaxation mat} and \eqref{eq: explicit collison integral}, which yields
\begin{widetext}
\begin{equation}
 -\frac{\tau_{\text{p}}}{\tau}\varphi (\theta, \gamma)
 = \sum_{\gamma'=\pm} \int^{2\pi}_0 \frac{d\theta'}{2\pi} 
\frac{1+\gamma\gamma'\cos[\Theta (\theta)-\Theta (\theta') ]}{2}
\left(
 1-\frac{\gamma' \tilde{\alpha} A(\theta')}{\sqrt{1+ \tilde{\alpha}^2A^2(\theta')}}
 \right)[\varphi(\theta',\gamma')-\varphi(\theta, \gamma)]~.
 \label{eq: relaxation theta integral}   
\end{equation}
\end{widetext}
Let us introduce two functions
\begin{align}
G(\theta) &= \frac{\varphi(\theta, +) + \varphi(\theta, -)}{2},
&F(\theta)& = \frac{\varphi(\theta, +) - \varphi(\theta, -)}{2}. 
\end{align}
The equation~\eqref{eq: relaxation theta integral} for $\gamma =\pm$ then yields simultaneous equations
\begingroup
\allowdisplaybreaks
\begin{subequations}
\label{eq: 1and2}
\begin{align}
&\left(1-\frac{\tau_{\text{p}}}{\tau}\right)
G(\theta) \nonumber\\
&=\int^{2\pi}_0\frac{d\theta'}{2\pi}
\left[G(\theta') -\frac{\tilde{\alpha} A(\theta')}{\sqrt{1+ \tilde{\alpha}^2A^2(\theta')}}F(\theta')\right],
\label{eq: 1st}\\
&\left(1-\frac{\tau_{\text{p}}}{\tau}\right)
F(\theta)\nonumber\\
&= \int^{2\pi}_0\frac{d\theta'}{2\pi}
\left[-\frac{\tilde{\alpha} A(\theta')}{\sqrt{1+ \tilde{\alpha}^2A^2(\theta')}}G(\theta')+ F(\theta')\right]
\nonumber\\
&\qquad \cdot \cos\left[\Theta (\theta)- \Theta (\theta')\right]~.
\label{eq: 2nd}
\end{align}
\end{subequations}
\endgroup

We first consider the case $\tau_{\text{p}}/\tau =1$.
There exist infinitely many solutions $(G(\theta), F(\theta))$ having that typical eigenvalue, and
we can write them down as
 \begin{subequations}
 \begin{align}
& G(\theta) -\frac{\tilde{\alpha} A(\theta)}{\sqrt{1+ \tilde{\alpha}^2A^2(\theta)}}F(\theta)\nonumber \\
&\quad = \frac{d \Theta (\theta)}{d \theta}
 \sum_{n=1}^{\infty}
 \left[ a_n\cos n\Theta (\theta) +b_n\sin n\Theta (\theta) 
 \right],\\
& -\frac{\tilde{\alpha} A(\theta)}{\sqrt{1+ \tilde{\alpha}^2A^2(\theta)}}G(\theta)+ F(\theta)\nonumber\\
 &\quad = \frac{d \Theta (\theta)}{d \theta}
 \left\{
 \frac{c_0}{2}+\sum_{n=2}^{\infty}\left[
 c_n \cos n\Theta (\theta) +d_n\sin n\Theta (\theta)\right]
 \right\}
 \end{align}
 \end{subequations}
for arbitrary constants $(a_n, b_n, c_n, d_n)$.

We then focus on the case $\tau_{\text{p}}/\tau \neq 1$.
The right-hand side of Eq.~\eqref{eq: 1st} is constant,
while the right-hand side of Eq.~\eqref{eq: 2nd} is 
a linear combination of $\cos\Theta (\theta)$ and $\sin\Theta (\theta)$. That leads to
\begin{equation}
G(\theta) = G_0, \qquad
F(\theta) = F_c \cos\Theta (\theta)+F_s \sin\Theta (\theta)
\end{equation}
with three constants $G_0$, $F_c$ and $F_s$.

The iterative substitution of $(G(\theta), F(\theta))$
into Eqs.~\eqref{eq: 1and2} gives 
\begingroup
\allowdisplaybreaks
 \begin{subequations}
 \begin{align}
&\qquad  \left(1-\frac{\tau_{\text{p}}}{\tau}\right)G_0 = G_0~,\\ 
& \left(1-\frac{\tau_{\text{p}}}{\tau}\right) \left[
 F_c \cos\Theta (\theta)+F_s \sin\Theta (\theta)\right]\nonumber\\
& =F_c  \cos\Theta (\theta)
 \int^{2\pi}_0\frac{d\theta'}{2\pi} \cos^2\Theta (\theta')\nonumber\\
&\qquad + F_s \sin\Theta (\theta)  \int^{2\pi}_0\frac{d\theta'}{2\pi} \sin^2\Theta (\theta')~.
 \end{align}
 \end{subequations}
\endgroup
We then obtain simpler eigenvalue problem 
\begin{equation}
\begin{pmatrix}
 0 & & \\
 & \frac{\tan\delta}{1+ \tan\delta}&  \\
 & & \frac{1}{1+ \tan\delta}
\end{pmatrix} 
\begin{pmatrix}
 G_0 \\F_c \\ F_s
\end{pmatrix}
= \frac{\tau_{\text{p}}}{\tau}
\begin{pmatrix}
 G_0 \\F_c \\ F_s
\end{pmatrix}\label{eq: simpler eig prob relaxation}
\end{equation}
that allows three eigenvalues
${\tau_{\text{p}}}/{\tau} = 0,~ \tan\delta/(1+\tan\delta),~ 1/(1+\tan\delta)$
corresponding to (i) a zero mode $\varphi(\theta, \gamma) = G_0 \equiv \text{constant}$,
(ii) a slow mode $\varphi(\theta, \gamma)= F_c \gamma \cos\Theta (\theta)\propto \gamma \cos\Theta (\theta)$ 
and (iii) another slow mode 
$\varphi(\theta, \gamma)=F_s\gamma \sin\Theta (\theta) \propto \gamma \sin\Theta (\theta)$, respectively.

In the main text, the modes with three different time constant
${\tau_{\text{p}}}/{\tau} = 1, \tan\delta/(1+\tan\delta), 1/(1+\tan\delta)$ are shown in Fig.~2(a), while
the zero mode with ${\tau_{\text{p}}}/{\tau} = 0$ is excluded from the figure
since this mode violates the charge-neutrality condition.
We now give the proof that the eigenmodes except for the zero mode follow the charge-neutrality condition.

We first multiply $\delta \bm{(}\epsilon (\bm{k},\gamma) -\epsilon_\text{F}\bm{)}$
on both sides of Eq.~\eqref{eq: BTE for relaxation w tau}
and sum up for states $(\bm{k},\gamma)$, which gives
\begin{align}
& -\frac{\tau_{\text{p}}}{\tau}\cdot
\frac{1}{V}\sum_{\bm{k},\gamma} \varphi_{\tau}(\bm{k},\gamma) \delta \bm{(}\epsilon (\bm{k},\gamma) -\epsilon_\text{F}\bm{)}\\
&= \frac{1}{V}\sum_{\bm{k},\gamma}\sum_{\bm{k}',\gamma'}
M_{\text{col}}(\bm{k},\gamma,\bm{k}',\gamma') \varphi_{\tau}(\bm{k}',\gamma')
\delta \bm{(}\epsilon (\bm{k},\gamma) -\epsilon_\text{F}\bm{)} \\
&= \frac{2\pi v_0^2n_{\text{imp}}}{\hbar V^2}\sum_{\bm{k},\gamma}\sum_{\bm{k}', \gamma'}
\Abs{\Braket{\bm{k}', \gamma'|\bm{k}, \gamma}}^2
[ \varphi_{\tau}(\bm{k}',\gamma') - \varphi_{\tau}(\bm{k},\gamma) ]\nonumber\\
&\qquad \cdot \delta\bm{ (}\epsilon (\bm{k}', \gamma') - \epsilon (\bm{k}, \gamma)\bm{) }
\delta \bm{(}\epsilon (\bm{k},\gamma) -\epsilon_\text{F}\bm{)} \\
&= 0~.\label{eq: relaxation charge neutral verification}
\end{align}
Here we used the definition of the collision term~\eqref{eq: col term from G.R.} in the middle, 
while at the last line we considered
a property that replacement of dummy variables $(\bm{k},\gamma)$ and $(\bm{k}',\gamma')$ changes the total sign.
The charge density 
\begin{equation}
 \rho = \frac{(-e)}{d_y} \frac{1}{V}\sum_{\bm{k},\gamma}
e^{-t/\tau} \varphi_{\tau} (\bm{k},\gamma)
\delta \bm{ (}\epsilon (\bm{k}, \gamma)-\epsilon_{\mathrm{F}} \bm{)}
\end{equation}
is thus automatically zero if ${\tau_{\text{p}}}/{\tau}\neq 0$.

\subsection{\label{subsec: Diffusion in space with temporally stationary state}
Diffusion in space with temporally stationary state}
In the main text, the diffusion case is described as both the BTE and Gauss' law
\begingroup
\allowdisplaybreaks
\begin{subequations}
\label{eq: both BTE and Gauss}
\begin{gather}
\bm{v}\cdot \frac{\partial f}{\partial \bm{r}} + \frac{(-e)\bm{E}_{\text{in}}}{\hbar}\cdot \frac{\partial f}{\partial \bm{k}}
= \frac{df}{dt}\Big|_{\text{col}}~, \\
\bm{\nabla}\cdot \bm{E}_{\text{in}} = \frac{-e}{\varepsilon_0 d_y}
\frac{1}{V}\sum_{\bm{k},\gamma}\left[
 f(\bm{r},\bm{k},\gamma) -f_0(\epsilon (\bm{k},\gamma))
\right]~.
\end{gather} 
\end{subequations}
\endgroup
with the distribution function $f(\bm{r}, \bm{k}, \gamma)$ and internal electric field $\bm{E}_{\text{in}}$
spatially decaying in the $+z$-direction with a diffusion length $\ell > 0$
\begingroup
\allowdisplaybreaks
\begin{subequations}
 \begin{align}
 f(\bm{r},\bm{k},\gamma) &= f_0\bm{(}\epsilon (\bm{k},\gamma)\bm{)} + e^{-z/\ell} \varphi_{\ell}(\bm{k},\gamma)
\left(-\frac{\partial f_0 (\epsilon)}{\partial \epsilon }\right)~,\\
  E_{\text{in}, z}&  =  \frac{\mathcal{E}_{\ell}}{(-e) v_\text{F} \tau_{\text{p}}} e^{-z/\ell}~.
 \end{align}
\end{subequations}
\endgroup
The equations~\eqref{eq: both BTE and Gauss} then reduce to a generalized eigenvalue equation:
\begin{widetext}
 \begin{equation}
 \left\{
 \begin{alignedat}{2}
 \sum_{\bm{k}',\gamma'}M_\text{col}(\bm{k},\gamma, \bm{k}',\gamma')\varphi_{\ell}(\bm{k}',\gamma')&
 + \frac{v_{z}(\bm{k},\gamma)}{v_\text{F}}\cdot \mathcal{E}_{\ell}
 & &= -\frac{v_\text{F}\tau_{\text{p}}}{\ell}\cdot \frac{v_{z}(\bm{k},\gamma)}{v_\text{F}} \varphi_{\ell}(\bm{k},\gamma)\\
 \sum_{\bm{k}',\gamma'}\frac{\delta\bm{(}\epsilon (\bm{k}',\gamma')- \epsilon_\text{F}\bm{)}}{VN_0/2} 
\varphi_{\ell}(\bm{k}',\gamma')
 & & &= -\frac{v_\text{F}\tau_{\text{p}}}{\ell}\cdot \eta  \mathcal{E}_{\ell}~.
 \end{alignedat}
 \right.
 \label{eq: BTE for diffusion in a matrix}
 \end{equation}
\end{widetext}
with unknown eigenvalues $-{v_\text{F}\tau_{\text{p}}}/{\ell}$ and
eigenvectors $\left(\{\varphi_{\ell} (\bm{k},\gamma)\}, \mathcal{E}_{\ell}\right)$.
Here we provide a ratio of screening length to diffusion length
\begin{equation}
 \eta^{-1} \equiv (v_\text{F} \tau_{\text{p}})^{2}\cdot 
\left(\frac{e^2N_0/2}{\varepsilon_0 d_y}\right)
\sim  (v_\text{F} \tau_{\text{p}}q_{\text{TF}})^{2}
\sim 10^{6}
\end{equation}
with $v_\text{F} \tau_{\text{p}}\sim \SI{e-8}{m}$ and Thomas--Fermi screening wave length $q_{\text{TF}}\sim \SI{e11}{m^{-1}}$.
Even if we set the parameter $\eta$ as $\eta = 10^{-4}, 10^{-6}, 10^{-8}$, the eigenvalue spectrum of long diffusion length 
has little changed and gives no strikingly difference in our results.

Let us mention on the numerical details.
The $(\bm{k},\gamma)$ summation in Eqs.~\eqref{eq: BTE for diffusion in a matrix}
is replaced with an integral with respect to $\theta$ and sum of two bands $\gamma = \pm$ around the Fermi contours.
We took mesh number $N_{\theta}= 358$ around both Fermi contours $\gamma = \pm$.

The eigenmodes of the BTE and Gauss' law with finite diffusion length $v_\text{F}\tau_\text{p}/\ell \neq 0$
do not induce charge current density.
That can be seen from the first line of Eqs.~\eqref{eq: BTE for diffusion in a matrix}.
We multiply $\delta \bm{(}\epsilon (\bm{k},\gamma) -\epsilon_\text{F}\bm{)}$ on the both sides 
and sum up for states $(\bm{k},\gamma)$, which gives
\begin{align}
&-\frac{v_\text{F}\tau_{\text{p}}}{\ell}\cdot \frac{1}{V}
\sum_{\bm{k},\gamma}
\frac{v_{z}(\bm{k},\gamma)}{v_\text{F}} \varphi_{\ell}(\bm{k},\gamma) 
\delta\bm{(}\epsilon(\bm{k},\gamma) -\epsilon_\text{F}\bm{)}\\
&=\frac{1}{V}
\sum_{\bm{k},\gamma}\sum_{\bm{k}',\gamma'}M_\text{col}(\bm{k},\gamma, \bm{k}',\gamma')\varphi_{\ell}(\bm{k}',\gamma')
\delta\bm{(}\epsilon(\bm{k},\gamma) -\epsilon_\text{F}\bm{)}\nonumber\\
&\quad +\frac{1}{V}
\sum_{\bm{k},\gamma}
 \frac{v_{z}(\bm{k},\gamma)}{v_\text{F}}\cdot \mathcal{E}_{\ell}
 \delta\bm{(}\epsilon(\bm{k},\gamma) -\epsilon_\text{F}\bm{)}\\
&=0~.
\end{align}
In the last line, 
we made use of the same technique as Eq.~\eqref{eq: relaxation charge neutral verification}
for the first term, 
as well as the absence of charge current in equilibrium $\langle v_z \rangle_\text{2DFC}=0$ for the second term.

\section{Microscopic Derivation of the Boltzmann equation}
In the main text, we have applied the Boltzmann transport equation (BTE) for the two-band system. 
We now follow its microscopic derivation starting from the Keldysh Green's function, 
paying attention to the validity of the BTE.
This section is organized as follows.
We first introduce quantum kinetic equation for density matrix in Sect.~\ref{subsec: Introduction of quantum kinetic equation},
which is an extension of the semiclassical BTE.
We then pick up the leading terms in both sides of the kinetic equation in the clean limit,
using order estimation (Sect.~\ref{subsec: Leading terms in both sides of the kinetic equation}).
We finally obtain transport equations for density matrix elements
and make some remarks on transport coefficients that are derived by the density matrix approach,
but cannot be derived by the semiclassical BTE approach (Sect.~\ref{subsec: Derived transport equations for density matrix elements}).
We set $\hbar = c =1$ in this section, for simplicity.

\subsection{\label{subsec: Introduction of quantum kinetic equation}
Introduction of quantum kinetic equation}
In the Keldysh formalism, quantum kinetic equation is given as follows~\cite{Rammer1986}:
\begin{equation}
 \left[
\begin{pmatrix}
 \underline{g}_0^{-1} -\underline{\Sigma}^{\mathrm{R}} & -\underline{\Sigma}^{\mathrm{K}}\\
0 & \underline{g}_0^{-1} -\underline{\Sigma}^{\mathrm{A}}
\end{pmatrix}
\overset{\otimes}{,}
\begin{pmatrix}
 \underline{g}^{\mathrm{R}} & \underline{g}^{\mathrm{K}}\\
0 & \underline{g}^{\mathrm{A}}
\end{pmatrix}
\right]_{-} = 0~.
\label{eq: Quantum kinetic equation}
\end{equation}
It is obtained by subtracting the left-Dyson equation from the right-Dyson equation.
Here underlined symbols, such as Green's functions 
$\underline{g}_0$, $\underline{g}^\text{R}$, $\underline{g}^\text{A}$, $\underline{g}^\text{K}$  
and self-energies
$\underline{\Sigma}^\text{R}$, $\underline{\Sigma}^\text{A}$, $\underline{\Sigma}^\text{K}$, 
are $2\times 2$ matrices reflecting the spin degrees of freedom.
The superscripts $\mathrm{R}, \mathrm{A}$ and $\mathrm{K}$ stand for retarded, advanced and Keldysh components of 
the Green's function and self-energy.
The convolution product $\otimes$ 
and (anti-)commutator between arbitrary two-point functions 
$\underline{M}(1,2)= {M}_{\sigma_1,\sigma_2}(t_1\bm{r}_1, t_2\bm{r}_2),\  
\underline{N}(1,2)=  {N}_{\sigma_1,\sigma_2}(t_1\bm{r}_1, t_2\bm{r}_2)$
are defined as 
\begin{multline}
  (\underline{M}\otimes \underline{N})(1,2) \\
\equiv \sum_{\sigma_3 = \uparrow, \downarrow}\int dt_3d\bm{r}_3~{M}_{\sigma_1,\sigma_3}(t_1\bm{r}_1, t_3\bm{r}_3)
{N}_{\sigma_3,\sigma_2}(t_3\bm{r}_3, t_2\bm{r}_2)     
\end{multline}
and $[\underline{M}~\overset{\otimes}{,}~ \underline{N}]_{\pm} 
\equiv \underline{M}\otimes \underline{N} \pm \underline{N}\otimes \underline{M}$.
A constant electric field $\bm{E} = -\partial_{\bm{r}} \phi -\partial_t \bm{A}$ is assumed to be applied onto the system.
Inverse Green's function in the absence of impurity scattering then takes the form
\begin{equation}
 \underline{g}_0^{-1}(1,2) = 
\left[i\frac{\partial}{\partial t_1} -q\phi (1)-\underline{H}(\bm{k}=\hat{\Pi})+\mu \right]
\delta (1,2)\underline{1}\label{eq: inv G before wigner transf}
\end{equation}
with the charge of the electron $q=-e$, chemical potential $\mu$ and 
kinetic momentum operator $\hat{\Pi}= -i\partial_{\bm{r}_1} -q\bm{A}(1)$.
The self-energy due to the impurity scattering 
potential~\eqref{eq: imp-scatt potential} is calculated as
\begin{equation}
 \underline{\Sigma}^{\text{X}}(1,2)
= n_{\text{imp}}v_0^2\cdot \delta (\bm{r}_1 -\bm{r}_2) \underline{g}^{\text{X}}(1,2)
\label{eq: Self-ene(1,2)}
\end{equation}
within the first Born approximation ($\text{X} = \mathrm{R},\mathrm{A}, \mathrm{K}$).

The Keldysh component of the kinetic equation~\eqref{eq: Quantum kinetic equation} 
is written
with the lesser component of the Green's function and self-energy 
\begin{align}
    \underline{g}^{<}&= -\frac{1}{2}(\underline{g}^{\mathrm{R}} - \underline{g}^{\mathrm{A}} - \underline{g}^{\mathrm{K}})~,
    &\underline{\Sigma}^{<}&= 
    -\frac{1}{2}(\underline{\Sigma}^{\mathrm{R}} - \underline{\Sigma}^{\mathrm{A}} - \underline{\Sigma}^{\mathrm{K}})
\end{align}
as follows: 
\begin{equation}
    [\underline{g}_0^{-1}~\overset{\otimes}{,}~\underline{g}^{<}]_{-}
    = \underline{\Sigma}^{\mathrm{R}}\otimes \underline{g}^{<}
    - \underline{g}^{<}\otimes\underline{\Sigma}^{\mathrm{A}}
    + \underline{\Sigma}^{\mathrm{<}}\otimes g^{\mathrm{A}}
    - \underline{g}^{\mathrm{R}}\otimes \underline{\Sigma}^{<}~.
\label{eq: kinetic Eq for G<(1,2)}
\end{equation}

We then apply the Wigner transform onto both sides of Eq.~\eqref{eq: kinetic Eq for G<(1,2)}, 
which replaces the relative coordinate $(t_1-t_2, \bm{r}_1 -\bm{r}_2)$ with frequency and wavevector $(\omega,\bm{k})$
and the convolution product~$\otimes$ with the Moyal product~$\star$.
More precisely, we make use of the 
gauge-invariant Wigner transform under the electromagnetic field~\cite{Levanda2001,Kita2001}, which is defined as
\begin{equation}
 \underline{M}(1,2)= e^{iI(1,2)}\frac{1}{V}\sum_{\bm{k}}\int \frac{d\omega}{2\pi}
\underline{M}(X,k)e^{ik^{\mu}x_{\mu}}
\end{equation}
for arbitrary two-point functions $\underline{M}(1,2)$.
In this transformation, we have used four-vectors
\begin{subequations}
 \begin{align}
 X^{\mu}&=\left(
 \frac{t_1+t_2}{2}, \frac{\bm{r}_1 +\bm{r}_2}{2}
 \right)=(T,\bm{R})~,\\
 x^{\mu}&= (t_1-t_2, \bm{r}_1 -\bm{r}_2) =(t,\bm{x})~,\\
 k^{\mu} &= (\omega,\bm{k})~,
 \end{align}
\end{subequations}
denoting the center-of-mass coordinate, relative coordinate and wavevector, respectively.
We have also used Einstein notation of summing over repeated indices $k^{\mu}x_{\mu}= \bm{k}\cdot \bm{x}-\omega t$.
The phase $I(1,2)$ is expressed as an integral along a linear path
\begin{equation}
 I(1,2)= q\int^{+ x^{\mu}/2}_{- x^{\mu}/2}A_{\lambda}(X^{\mu} +x'^{\mu})dx'^{\lambda}
\end{equation}
with $A^{\mu}=(\phi,\bm{A})$.
The Moyal product $\star$ after this Wigner transform
is obtained within the first order of the gradient expansion as~\cite{Onoda2006}
\begin{widetext}
 \begin{align}
 \underline{M}\star \underline{N} &=\underline{M}(X,k) \exp\left[
 \frac{i}{2}\left(
 \overleftarrow{\partial_{X^{\mu}}}~\overrightarrow{\partial_{k_{\mu}}}
 -\overleftarrow{\partial_{k^{\mu}}}~\overrightarrow{\partial_{X_{\mu}}}
 +qF^{\mu\nu}\overleftarrow{\partial_{k^{\mu}}}~\overrightarrow{\partial_{k^{\nu}}}
 \right)\right]\underline{N}(X,k)\\
 &= \underline{M}~\underline{N} + \frac{i}{2}
 \left[
 \frac{\partial \underline{M}}{\partial \bm{R}}\frac{\partial \underline{N}}{\partial \bm{k}}
 -\frac{\partial \underline{M}}{\partial T}\frac{\partial \underline{N}}{\partial \omega}
 -\frac{\partial \underline{M}}{\partial \bm{k}}\frac{\partial \underline{N}}{\partial \bm{R}}
 +\frac{\partial \underline{M}}{\partial \omega}\frac{\partial \underline{N}}{\partial T}
 -q\bm{E}\left(
 \frac{\partial \underline{M}}{\partial \bm{k}}\frac{\partial \underline{N}}{\partial \omega}
 -\frac{\partial \underline{M}}{\partial \omega}\frac{\partial \underline{N}}{\partial \bm{k}}
 \right)\right]
 + \mathcal{O}(\lambda^2)
 \end{align}
\end{widetext}
for arbitrary matrices $\underline{M}$ and $\underline{N}$.
Here $F^{\mu\nu} = \partial^{\mu}A^{\nu} - \partial^{\nu}A^{\mu}$ is the electromagnetic tensor.
We here introduced an expansion parameter
\begin{equation}
 \lambda \sim \frac{\hbar}{\epsilon_\text{F}\tau_{\text{p}}} \ll 1
\end{equation}
that characterizes the inhomogeneity of the system.
The inverse Green's function~\eqref{eq: inv G before wigner transf} and the self-energies~\eqref{eq: Self-ene(1,2)}
are also transformed as~\cite{Onoda2006}
\begin{subequations}
 \label{eq: self-ene (X,k)}  
 \begin{gather}
 \underline{g}_0^{-1}(\omega,\bm{k}) = \omega \underline{1}- \underline{H}(\bm{k}) +\mu \underline{1}~, \\
  \underline{\Sigma}^{\text{X}}(T, \bm{R},\omega) 
 = \frac{n_{\text{imp}}v_0^2}{V}\sum_{\bm{k}}\underline{g}^{\text{X}}(T,\bm{R},\omega,\bm{k})  
 \end{gather}
\end{subequations}
with $\text{X} = \mathrm{R},\mathrm{A},<$.

The kinetic equation~\eqref{eq: kinetic Eq for G<(1,2)} after the Wigner transform is thus expanded 
in the regime $\lambda \ll 1$ as
\begin{widetext}
 \begin{multline}
 i\left[\underline{H}(\bm{k}), -i\underline{g}^{<}\right]_{-} 
 + \frac{\partial (-i\underline{g}^{<})}{\partial T}
 +\frac{1}{2}\left[
 \underline{\bm{v}}(\bm{k}), \frac{\partial (-i\underline{g}^{<})}{\partial \bm{R}}
 \right]_{+}
 +q\bm{E}\frac{\partial (-i\underline{g}^{<})}{\partial \bm{k}}
 -\frac{\partial}{\partial \omega}
 \left(\frac{q\bm{E}}{2}\left[\underline{\bm{v}}, -i\underline{g}^{<}\right]_{+}\right)\\
 = -\underline{\Sigma}^{\mathrm{R}} \underline{g}^{<}
    + \underline{g}^{<}\underline{\Sigma}^{\mathrm{A}}
    - \underline{\Sigma}^{\mathrm{<}} \underline{g}^{\mathrm{A}}
    + \underline{g}^{\mathrm{R}} \underline{\Sigma}^{<}
 \label{eq: kinetic Eq for G<(X,k)}
 \end{multline}
\end{widetext}
with $\underline{g}^{<}(X,k) = \underline{g}^{<}(T,\bm{R},\omega, \bm{k})$ and velocity matrix
$\underline{\bm{v}}(\bm{k})= \partial_{\bm{k}} \underline{H}(\bm{k})$.
Here we take terms up to first order of the gradient expansion $\mathcal{O}(\lambda^1)$ and
neglect the spacetime derivatives of the self-energies (local approximation).

We further integrate out the $\omega$-dependence of the kinetic equation. 
In this integration, the lesser Green function provides (spin-)density matrix 
\begin{equation}
 \underline{f}(T,\bm{R},\bm{k})\equiv -i\int\frac{d\omega}{2\pi} \underline{g}^{<}(T,\bm{R},\omega, \bm{k})~.
\end{equation}
In equilibrium, the density matrix $\underline{f}(T,\bm{R},\bm{k})= \underline{f}_0(\bm{k})$ 
is diagonal in the band basis with its diagonal elements the Fermi distribution functions: 
\begin{equation}
\underline{\overline{f}}_0(\bm{k})
\equiv {U}(\bm{k})\underline{f}_0(\bm{k}){U}^{\dagger}(\bm{k}) 
=
\begin{bmatrix}
 f_0\bm{(}\epsilon (\bm{k},+)\bm{)} & 0\\
 0 & f_0\bm{(}\epsilon (\bm{k},-)\bm{)}
\end{bmatrix}
~.
\end{equation}
Here $U_{\gamma\sigma}(\bm{k}) = \braket{\bm{k},\gamma|\bm{k},\sigma}$
is an unitary matrix that diagonalize the Hamiltonian $\underline{H}(\bm{k})$.
The occupation probabilities of each band, i.e. distribution functions in the BTE,
are just diagonal elements of the density matrix in the band basis.

In a clean limit, we expect that the quantum kinetic equation is reduced to 
the semiclassical BTE with distribution functions for each band.
We thus apply the unitary transformation $U(\bm{k})$
to represent the kinetic equation~\eqref{eq: kinetic Eq for G<(X,k)} in the band basis
after the $\omega$-integration.
The matrix-form of the kinetic equation is then obtained as
follows~\cite{Dyakonov1984,Khaetskii2006,Shytov2006,Kailasvuori2009}:
\begin{subequations}
\label{eq: BTE as a Matrix Form and St[f] wo approx}
\begin{multline}
     \frac{\partial \underline{\overline{f}}}{\partial T}
     +\frac{1}{2}\left[\underline{\overline{\bm{v}}}, \frac{\partial \underline{\overline{f}}}{\partial \bm{R}}\right]_{+}
     +q\bm{E}\left[\frac{\partial}{\partial \bm{k}} -i\underline{\mathcal{A}}_{\bm{k}}, \underline{\overline{f}}\right]_{-}
     + i\left[\underline{\overline{H}}, \underline{\overline{f}}\right]_{-}\\
     = \text{St}[\underline{\overline{f}}]
    \label{eq: BTE as a Matrix Form} 
\end{multline}
 with collision integral
 \begin{equation}
     \text{St}[\underline{\overline{f}}]
     = \int \frac{d\omega}{2\pi} \left(
     -\underline{\overline{\Sigma}}^{\mathrm{R}} \underline{\overline{g}}^{<}
    + \underline{\overline{g}}^{<}\underline{\overline{\Sigma}}^{\mathrm{A}}
    - \underline{\overline{\Sigma}}^{\mathrm{<}} \underline{\overline{g}}^{\mathrm{A}}
    + \underline{\overline{g}}^{\mathrm{R}} \underline{\overline{\Sigma}}^{<}
     \right)~.
     \label{eq: St[f] wo approx}
 \end{equation}
\end{subequations}
Here each overlined matrix $\underline{\overline{M}}$ is a representation in the band basis 
$\underline{\overline{M}} =U \underline{M} U^{\dagger}$.
In Eq.~\eqref{eq: BTE as a Matrix Form}, the Hamiltonian in the band basis
$\underline{\overline{H}} = \diag [\epsilon(\bm{k}, +),~\epsilon(\bm{k}, -)]$
is diagonal, while
the velocity matrix $\underline{\overline{{v}}}_i$ and Berry connection $\underline{\mathcal{A}}_{\bm{k}}$, defined as
\begingroup
\allowdisplaybreaks
\begin{align}
& v_{i, \gamma\gamma'}(\bm{k}) \equiv \Braket{\bm{k},\gamma|\frac{\partial H(\bm{k})}{\partial k_{i}}|\bm{k},\gamma'}
\nonumber\\
&\quad = v_{i}(\bm{k},\gamma)\delta_{\gamma\gamma'}
+ i\left[\epsilon (\bm{k},\gamma)-\epsilon (\bm{k},\gamma')\right]\mathcal{A}_{k_{i}\gamma\gamma'}~,\\
&\underline{\mathcal{A}}_{\bm{k}}\equiv i{U}\partial_{\bm{k}}{U}^{\dagger}~,
\end{align}
\endgroup
have off-diagonal elements. Here $v_{i, \gamma\gamma'}(\bm{k})$ and $\mathcal{A}_{k_{i}\gamma\gamma'}$
are the $(\gamma,\gamma')$ element of the matrices $\underline{\overline{{v}}}_i$ and $\mathcal{A}_{k_{i}}$, respectively, and
$\bm{v}(\bm{k},\gamma)= \partial_{\bm{k}} \epsilon (\bm{k},\gamma)$ is the group velocity.
The density matrix in the band basis, i.e. $2\times 2$ distribution function,
also has off-diagonal elements in general:
\begin{equation}
\underline{\overline{f}}(T,\bm{R},\bm{k})=
\begin{bmatrix}
f_{++} & f_{+-}\\ f_{-+} & f_{--}
\end{bmatrix}
\end{equation}
with $f_{+-} = {f_{-+}}^*$.

\subsection{\label{subsec: Leading terms in both sides of the kinetic equation}
Leading terms in both sides of the kinetic equation}

\begin{figure}
    \centering
\includegraphics[width=0.6\columnwidth]{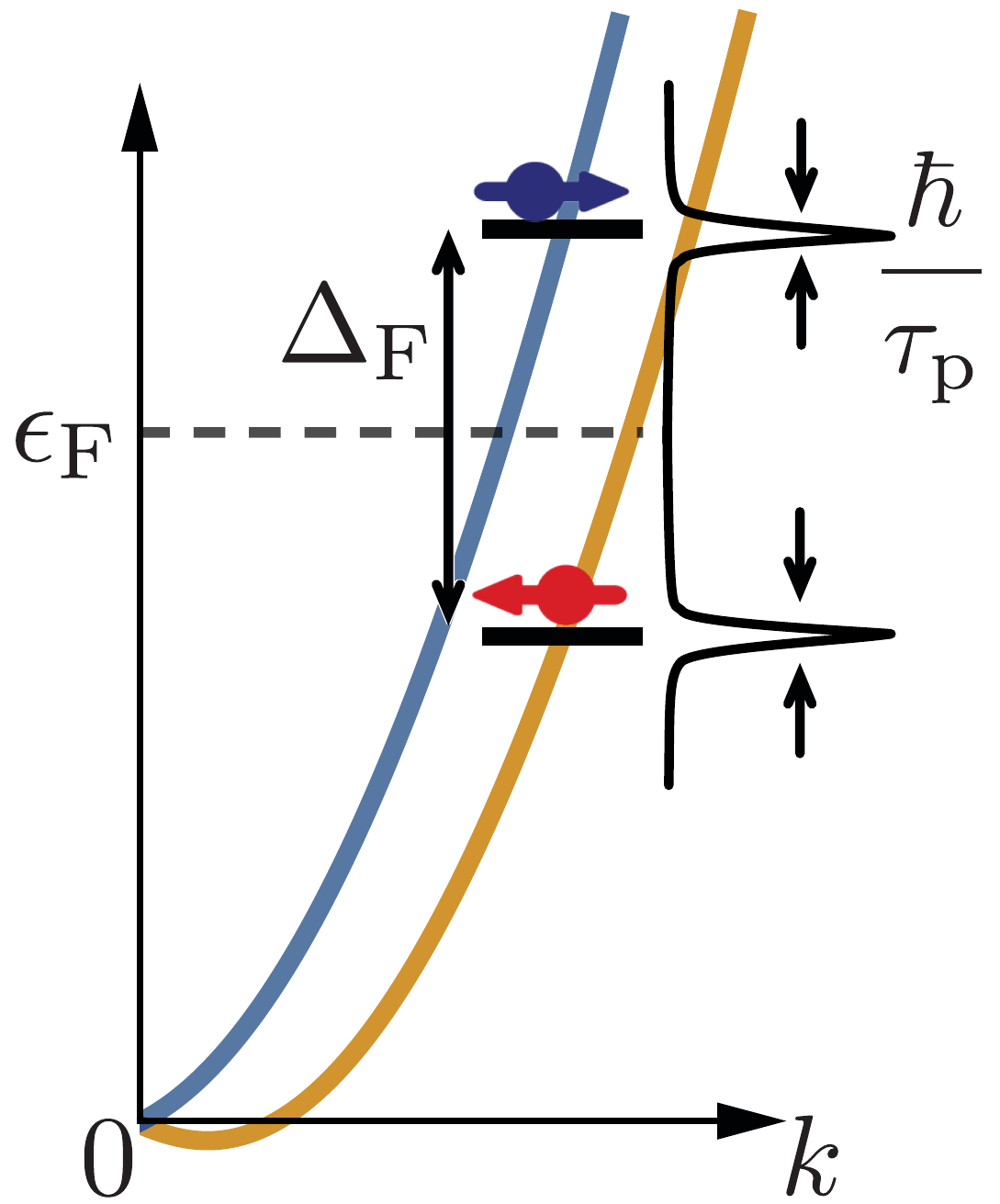}
    \caption{Spin-splitting two bands for 
    the case $\hbar/\tau_{\text{p}}\ll \Delta_\text{F} \ll \epsilon_\text{F}$. 
    Around the Fermi energy, 
    inverse of quasiparticle lifetime 
    $\hbar/\tau_{\text{p}}$ due to the impurity scattering
    is smaller than the spin splitting $\Delta_\text{F}$.
    }
    \label{fig: nearE_F Energyscales}
\end{figure}

In general, the off-diagonal elements of the density matrix $f_{+-}, f_{-+}$ cannot be neglected in the kinetic equation.
However, as we detail in this subsection, we can simplify the kinetic equation in a regime of energy scales (clean limit)
\begin{equation}
\frac{\hbar}{\tau_{\text{p}}} \ll \Delta_\text{F} \ll \epsilon_\text{F} 
\label{eq: nearE_F Energyscales}
\end{equation}
with Fermi energy $\epsilon_\text{F}$, spin-splitting energy gap around the Fermi energy $\Delta_\text{F} = \Delta (k_\text{F})$,
and typical impurity-scattering rate $\hbar/\tau_{\text{p}}= 2\pi n_{\text{imp}}v_0^2\cdot N_0/2$
(Fig.~\ref{fig: nearE_F Energyscales}).
In this region, the off-diagonal elements (interband contribution) $f_{+-}, f_{-+}$
are much smaller than the diagonal elements (intraband contribution) $f_{++}, f_{--}$, or more precisely,
\begin{equation}
\left|\frac{f_{-+}}{f_{\gamma\gamma}-f_0(\epsilon \bm{(}\bm{k},\gamma)\bm{)}}\right|,~
\left|\frac{f_{+-}}{f_{\gamma\gamma}-f_0(\epsilon \bm{(}\bm{k},\gamma)\bm{)}}\right|
\sim \frac{\hbar}{\Delta_\text{F}\tau_{\text{p}}} \ll 1
\label{eq: order diag and off-diag}
\end{equation}
holds for two bands $\gamma = \pm$.
That allows us to approximately separate the matrix-form of the kinetic equation
into two equations; one is a transport equation for the diagonal elements, and
the other is a equation that determines the leading terms of the off-diagonal elements by using the diagonal elements.
We will provide them in the next subsection.

Let us now closely examine both sides of
the matrix-formed kinetic equation~\eqref{eq: BTE as a Matrix Form}
in the clean limit~\eqref{eq: nearE_F Energyscales}
in order to see Eq.~\eqref{eq: order diag and off-diag}.
As for the left-hand side, the matrix elements take the following forms:
 \begin{subequations}
 \label{eq: LHS of BTE}
 \begin{align}
  (\text{LHS})_{++}
 &= \frac{\partial f_{++}}{\partial T}
 + \bm{v}(\bm{k},+)\cdot \frac{\partial f_{++}}{\partial \bm{R}}
 + q\bm{E}\cdot \frac{\partial f_{++}}{\partial \bm{k}}\nonumber\\
&\qquad + L^{(1)}_{++}[f_{-+}, f_{+-}]~, 
 \label{eq: LHS of BTE diag}\\
 (\text{LHS})_{-+} 
 & = -i \Delta(\bm{k})f_{-+}
 + L^{(1)}_{-+}[f_{++}, f_{--}]\nonumber\\
&\qquad + L^{(2)}_{-+}[f_{-+}]~.
 \label{eq: LHS of BTE off-diag}
 \end{align}
 \end{subequations}
 Here we introduced 
 the band splitting $\Delta(\bm{k}) = \epsilon (\bm{k},+) - \epsilon (\bm{k},-)$ and the following three terms
\begin{widetext}
 \begin{subequations}
 \begin{align}
 L^{(1)}_{++}[f_{-+}, f_{+-}]
 &= \frac{i}{2}\Delta(\bm{k})\left[
 \mathcal{A}_{\bm{k}+-}\frac{\partial f_{-+}}{\partial \bm{R}}
 -\frac{\partial f_{+-}}{\partial \bm{R}}\mathcal{A}_{\bm{k}-+}
 \right]  -iq\bm{E}\left[
 \mathcal{A}_{\bm{k}+-} f_{-+}
 -f_{+-}\mathcal{A}_{\bm{k}-+}
 \right] ~,\label{eq: anomal velo}\\
 L^{(1)}_{-+}[f_{++}, f_{--}]
 &=-i\mathcal{A}_{\bm{k}-+}\cdot \left[
 \frac{\Delta(\bm{k})}{2}\frac{\partial (f_{++}+ f_{--})
 }{\partial \bm{R}}
 +q\bm{E}(f_{++}-f_{--})
 \right]~,\\
 L^{(2)}_{-+}[f_{-+}]
 &=\frac{\partial f_{-+}}{\partial T}
 + \frac{\bm{v}(\bm{k},+)+\bm{v}(\bm{k},-)}{2}\cdot \frac{\partial f_{-+}}{\partial \bm{R}}
 + q\bm{E}\cdot \frac{\partial f_{-+}}{\partial \bm{k}} 
 + iq\bm{E}\cdot (\mathcal{A}_{\bm{k}++}-\mathcal{A}_{\bm{k}--})f_{-+}~.
 \end{align}
 \end{subequations}
\end{widetext}
The elements $ (\text{LHS})_{--}$ and $ (\text{LHS})_{+-}$ are also obtained
just as flipping all the signs of both the band index $\pm$ and 
$\Delta (\bm{k})= \epsilon (\bm{k}, +)- \epsilon (\bm{k}, -)$,
included in Eqs.~\eqref{eq: LHS of BTE}.

Let us assume a length (time) scale $\ell$ ($\tau$)
that characterizes slowly decaying component of the distribution function $\underline{\overline{f}}$,
satisfying a condition 
$qE/(\hbar k_\text{F})\ll 1/\tau \sim v_\text{F}/\ell \lesssim 1/\tau_{\text{p}}$.
Here $v_\text{F} = \hbar k_\text{F}/m = 2\epsilon_\text{F}/(\hbar k_\text{F})$ is the Fermi velocity in the absence of the SOC.
There is another inequality $\hbar/\tau_{\text{p}} \ll \Delta_\text{F}\ll \epsilon_\text{F}$ as we have mentioned.
These assumptions lead to the inequality around the Fermi energy
\begin{equation}
    L^{(2)}_{-+}[f_{-+}]
\sim
\frac{f_{-+}}{\tau_{\text{p}}}
= \frac{\hbar}{\Delta_\text{F}\tau_{\text{p}}}\cdot\frac{\Delta_\text{F} f_{-+}}{\hbar}
\ll\frac{\Delta(\bm{k})f_{-+}}{\hbar}~,
\end{equation}
which shows $L^{(2)}_{-+}$ is negligibly small in Eq.~\eqref{eq: LHS of BTE off-diag}.
We then compare
$(-, +)$ elements of both sides of the matrix-form of the kinetic equation~\eqref{eq: BTE as a Matrix Form and St[f] wo approx},
which yields
\begin{align}
\frac{-i\Delta(\bm{k})f_{-+}}{\hbar}&
&\sim & && L^{(1)}_{-+}[f_{++}, f_{--}],~ \text{St}[\underline{\overline{f}}]_{-+}\label{eq: evaluation of off-diag 1}\\
&&\sim&&&
\frac{\Delta_\text{F}}{\hbar k_\text{F}\ell}\left[f_{\gamma\gamma}-f_0(\epsilon \bm{(}\bm{k},\gamma)\bm{)}\right],\nonumber\\
&&&&& \frac{qE}{\hbar k_\text{F}}\left[f_{\gamma\gamma}-f_0\bm{(}\epsilon (\bm{k},\gamma)\bm{)}\right],\nonumber\\
&&&&& \frac{f_{\gamma\gamma}-f_0\bm{(}\epsilon (\bm{k},\gamma)\bm{)} }{\tau_{\text{p}}},\nonumber\\
&&&&&\frac{f_{-+}}{\tau_{\text{p}}},~\frac{f_{+-}}{\tau_{\text{p}}}~.
\label{eq: proof of off-diag is small}
\end{align}
Here we used $\underline{\mathcal{A}}_{\bm{k}}\sim 1/k_\text{F}$
around the Fermi energy, and evaluate the collision
integral by means of the relaxation time approximation.
The dominant term in Eq.~\eqref{eq: proof of off-diag is small} is 
$[f_{\gamma\gamma}-f_0(\epsilon \bm{(}\bm{k},\gamma)\bm{)}]/\tau_{\text{p}}$,
as can be seen by using the assumptions given.
Thus, we find the relation
\begin{equation}
    \frac{-i\Delta(\bm{k})f_{-+}}{\hbar}
    \sim 
 \frac{\hbar}{\Delta_\text{F}\tau_{\text{p}}}\cdot\frac{\Delta_\text{F} \left[f_{\gamma\gamma}-f_0(\epsilon \bm{(}\bm{k},\gamma)\bm{)}
\right]}{\hbar}~,
\end{equation}
which yields ${f}_{-+}/\left[f_{\gamma\gamma}-f_0(\epsilon \bm{(}\bm{k},\gamma)\bm{)}\right]\sim \hbar/(\Delta_\text{F}\tau_{\text{p}})\ll 1$
as a result.
The term $L^{(1)}_{++}$ in Eq.\eqref{eq: LHS of BTE diag}
is also shown to be smaller than the other terms in Eq.~\eqref{eq: LHS of BTE diag}, as can be seen from the following inequality:
\begingroup
\allowdisplaybreaks
\begin{align}
&L^{(1)}_{++}[f_{-+}, f_{+-}] \nonumber \\
&\sim\qquad
\frac{\Delta_\text{F}}{\hbar k_\text{F}\ell}f_{-+},~
\frac{qE}{\hbar k_\text{F}} f_{-+}\\
&\sim\qquad
\frac{\Delta_\text{F}}{\hbar k_\text{F}\ell}\cdot \frac{\hbar}{\Delta_\text{F}\tau_{\text{p}}}
\left[f_{\gamma\gamma}-f_0(\epsilon \bm{(}\bm{k},\gamma)\bm{)}\right]\\
&\sim\qquad
\frac{\hbar}{\epsilon_\text{F}\tau_{\text{p}}}\cdot\frac{v_\text{F}\tau_{\text{p}}}{\ell}\cdot \frac{f_{\gamma\gamma}-f_0(\epsilon \bm{(}\bm{k},\gamma)\bm{)}}{\tau_{\text{p}}}\\
&\ll \qquad
\frac{\partial f_{\gamma\gamma}}{\partial T}
+ \bm{v}(\bm{k},\gamma)\cdot \frac{\partial f_{\gamma\gamma}}{\partial \bm{R}}
+ \frac{q\bm{E}}{\hbar}\cdot \frac{\partial f_{\gamma\gamma}}{\partial \bm{k}}.
\end{align}
\endgroup
Here we used the assumption that the gradient expansion parameter is small enough, i.e.
$\lambda \sim \hbar/(\epsilon_\text{F}\tau_{\text{p}})\ll 1$.
We then neglect $L^{(1)}_{++}$ in Eq.~\eqref{eq: LHS of BTE diag}.

Therefore, the left-hand side of the BTE~\eqref{eq: BTE as a Matrix Form} is simplified as
\begin{widetext}
 \begin{multline}
 (\text{LHS})
 = 
 \begin{bmatrix}
 \displaystyle\frac{\partial f_{++}}{\partial T} & \displaystyle i\Delta(\bm{k}) f_{+-}\\
 \displaystyle -i\Delta(\bm{k}) f_{-+} & \displaystyle\frac{\partial f_{--}}{\partial T}
 \end{bmatrix}
 +
 \begin{bmatrix}
 \displaystyle\bm{v}_{++}(\bm{k})\frac{\partial f_{++}}{\partial \bm{R}} 
 & \displaystyle \bm{v}_{+-}(\bm{k})\frac{\partial (f_{++}+f_{--})/2}{\partial \bm{R}}\\
 \displaystyle\bm{v}_{-+}(\bm{k})\frac{\partial (f_{++}+f_{--})/2}{\partial \bm{R}} 
 & \displaystyle\bm{v}_{--}(\bm{k})\frac{\partial f_{--}}{\partial \bm{R}}
 \end{bmatrix}
 \\+q\bm{E}
 \begin{bmatrix}
 \displaystyle\frac{\partial f_{++}}{\partial \bm{k}} & \displaystyle i\mathcal{A}_{\bm{k}+-}(f_{++}-f_{--})\\
 \displaystyle -i\mathcal{A}_{\bm{k}-+}(f_{++}-f_{--}) & \displaystyle \frac{\partial f_{--}}{\partial \bm{k}}
 \end{bmatrix}
 ~.\label{eq: simplified LHS of BTE}
 \end{multline}
\end{widetext}

We now turn to approximate the right-hand side of the BTE, i.e. the collision integral~\eqref{eq: St[f] wo approx},
by replacing the retarded and advanced Green's functions with their expressions in equilibrium 
\begin{align}
&\underline{\overline{g}}^{\mathrm{R}}\simeq
\underline{\overline{g}}^{\mathrm{R}}_0\nonumber\\
&=
\begin{bmatrix}
 \omega -\epsilon (\bm{k},+) +\mu +i0+ & 0\\
 0 & \omega -\epsilon (\bm{k},-) +\mu +i0+
\end{bmatrix}^{-1},\\
&\underline{\overline{g}}^{\mathrm{A}}\simeq \underline{\overline{g}}^{\mathrm{R}\dagger}_0.
\end{align}
Here $i0+$ denotes an infinitesimal positive imaginary part.
We also neglect the off-diagonal elements of 
the lesser component $\underline{\overline{g}}^{<}$ and that of the density matrix $\underline{\overline{f}}$
since they are much smaller than the diagonal elements, as we have shown on the left-hand side of the BTE.

A conventional assumption $\underline{g}^{<}  =\underline{f}\underline{g}^{\mathrm{A}} -\underline{g}^{\mathrm{R}}\underline{f}$ 
then leads to the relation
\begin{equation}
 \underline{\overline{g}}^{<}
 \simeq (\underline{\overline{g}}^{\mathrm{A}}-\underline{\overline{g}}^{\mathrm{R}}) \underline{\overline{f}}
 = 2\pi i \cdot \delta \bm{(}\omega - \underline{\overline{H}}(\bm{k}) +\mu \bm{)} \underline{\overline{f}}~.
\end{equation}
The self-energy terms, given in Eq.~\eqref{eq: self-ene (X,k)}, are rewritten as 
\begin{equation}
\underline{\overline{\Sigma}}^{\text{X}}
= \frac{n_{\text{imp}}v_0^2}{V}
\sum_{\bm{k}'}\underline{U}(\bm{k})\underline{U}^{\dagger}(\bm{k}')
\underline{\overline{g}}^{\text{X}}(\bm{k}')\underline{U}(\bm{k}')\underline{U}^{\dagger}(\bm{k})
\end{equation}
with $\text{X} = \mathrm{R}, \mathrm{A}, <$.
The $(\gamma, \gamma')$  matrix element of $\underline{U}(\bm{k})\underline{U}^{\dagger}(\bm{k}')$
is also expressed as $\Braket{\bm{k},\gamma|\bm{k}',\gamma'}$.

In the following, we adopt the notation 
${g}^{\text{X}}(\bm{k},\gamma)= {g}^{\text{X}}_{\gamma\gamma}(T,\bm{R},\omega,\bm{k})$ and
$f(\bm{k},\gamma)= f_{\gamma\gamma}(T,\bm{R},\bm{k})$ for simplicity.
The collision integral~\eqref{eq: St[f] wo approx}
is then described by $f(\bm{k}, \pm)$ as 
 \begin{align}
 &\text{St}[\underline{\overline{f}}]_{\gamma\gamma'}
 = \frac{n_{\text{imp}}v_0^2}{V}\sum_{\bm{k}'', \gamma'' =\pm}
 \Braket{\bm{k},\gamma|\bm{k}'',\gamma''}\cdot \Braket{\bm{k}'',\gamma''|\bm{k},\gamma'}\nonumber\\
 &\cdot\int \frac{d\omega}{2\pi}\Big[
  -g^{\mathrm{R}}  (\bm{k}'',\gamma'') g^{<}(\bm{k},\gamma')
 + g^{<}(\bm{k},\gamma) g^{\mathrm{A}}  (\bm{k}'',\gamma'')\nonumber\\
&\qquad\qquad -g^{<}(\bm{k}'',\gamma'') g^{\mathrm{A}}  (\bm{k},\gamma')
 +g^{\mathrm{R}}  (\bm{k},\gamma) g^{<}(\bm{k}'',\gamma'')
 \Big]\\
 &=  \frac{2\pi n_{\text{imp}}v_0^2}{V}\sum_{\bm{k}'',\gamma''}
 \Braket{\bm{k},\gamma|\bm{k}'',\gamma''}\cdot \Braket{\bm{k}'',\gamma''|\bm{k},\gamma'}\nonumber\\
& \cdot \frac{1}{2\pi i}
 \Big[
 \frac{f(\bm{k}'',\gamma'') -f(\bm{k},,\gamma')}{\epsilon(\bm{k}'',\gamma'')  - \epsilon (\bm{k},\gamma')-i0+}\nonumber\\
&\qquad - \frac{f(\bm{k}'',\gamma'') -f(\bm{k},\gamma)}{\epsilon(\bm{k}'',\gamma'')  - \epsilon (\bm{k},\gamma)+i0+}
 \Big]~.
 \label{eq: simplified coll int}
 \end{align}
As for the diagonal elements, 
\begin{multline}
\text{St}[\underline{\overline{f}}]_{\gamma\gamma}
= \frac{2\pi n_{\text{imp}}v_0^2}{V}\sum_{\bm{k}',\gamma'}
\Abs{\Braket{\bm{k},\gamma|\bm{k}',\gamma'} }^2\\
\left[f(\bm{k}',\gamma') -f(\bm{k},\gamma)\right]
\delta \bm{(} \epsilon(\bm{k}',\gamma')  - \epsilon (\bm{k},\gamma)\bm{)} 
\end{multline}
holds, which is consistent with the result obtained from the Fermi's golden rule.

\subsection{\label{subsec: Derived transport equations for density matrix elements}
Derived transport equations for density matrix elements}

To summarize, the equations~\eqref{eq: simplified LHS of BTE} and~\eqref{eq: simplified coll int}
yield the BTE for diagonal distribution functions
\begingroup
\allowdisplaybreaks
 \begin{subequations}
 \label{eq: simplified BTE}
\begin{align}
& \frac{\partial f(\bm{k},\gamma)}{\partial T}
 +  \bm{v}(\bm{k},\gamma)  \frac{\partial f(\bm{k},\gamma)}{\partial \bm{R}}
 + q \bm{E} \frac{\partial f(\bm{k},\gamma)}{\partial \bm{k}}\nonumber\\
&= \frac{2\pi n_{\text{imp}}v_0^2}{V}\sum_{\bm{k}',\gamma'}
 \Abs{\Braket{\bm{k},\gamma|\bm{k}',\gamma'} }^2\nonumber\\
&\qquad \cdot\left[f(\bm{k}',\gamma') -f(\bm{k},\gamma)\right]
 \delta \bm{(} \epsilon(\bm{k}',\gamma')  - \epsilon (\bm{k},\gamma)\bm{)}   
\end{align}
 with $\gamma = \pm$. 
 There are other two equations that determine the leading terms of the off-diagonal distribution functions
\begin{widetext}
  \begin{multline}
 f_{-+}(\bm{k})=
 -\mathcal{A}_{\bm{k}-+}\cdot \left\{
 \frac{\partial [f(\bm{k},+)+ f(\bm{k},-)]/2}{\partial \bm{R}}
 +\frac{q\bm{E}}{\Delta (\bm{k})} [f(\bm{k},+)-f(\bm{k},-)]
 \right\}\\
 +\frac{n_{\text{imp}}v_0^2}{\Delta (\bm{k})V}\sum_{\bm{k}',\gamma'}
 \Braket{\bm{k},-|\bm{k}',\gamma'}\cdot \Braket{\bm{k}',\gamma'|\bm{k},+}
 \left[
 \frac{f(\bm{k}',\gamma') -f(\bm{k},+)}{\epsilon(\bm{k}',\gamma')  - \epsilon (\bm{k},+)-i0+}
 - \frac{f(\bm{k}',\gamma') -f(\bm{k},-)}{\epsilon(\bm{k}',\gamma')  - \epsilon (\bm{k},-)+i0+}
 \right]
 \label{eq: f-+ leading terms}
 \end{multline}
\end{widetext} 
and $f_{+-}=f^{*}_{-+}$.
 \end{subequations}
\endgroup
The equations~\eqref{eq: simplified BTE} are invariant 
under the U(1) gauge transformation $\ket{\bm{k},\gamma}\to \ket{\bm{k},\gamma}e^{-i\varphi (\bm{k},\gamma)}$.
Note that the equations~\eqref{eq: simplified BTE} are valid 
in the range of $\bm{k}$ that satisfies the condition~\eqref{eq: nearE_F Energyscales}, 
and in particular fail around the band-crossing point $\bm{k}=0$.
In the latter, we must seriously solve the matrix-form of kinetic equation.

The off-diagonal elements of the density matrix $f_{+-}, f_{-+}$
affect expectation values of the physical quantities, in general.
That is because density of a physical quantity $\hat{O}$ deviated from equilibrium per unit volume is calculated as
\begin{align}
& \langle \hat{O} \rangle
= \frac{1}{V}\Tr \sum_{\bm{k}} 
\Big\{
\begin{bmatrix}
 O_{++} & O_{+-}\\  O_{-+} & O_{--}
\end{bmatrix}\nonumber\\
&\quad \cdot 
\begin{bmatrix}
f_{++}-f_0(\epsilon\bm{(}\bm{k},+)\bm{)} & f_{+-}\\ f_{-+} & f_{--}-f_0(\epsilon\bm{(}\bm{k},-)\bm{)}
\end{bmatrix}
\Big\}\\
&= \frac{1}{V}\sum_{\bm{k},\gamma}
O_{\gamma\gamma} \left[
f(t,\bm{r}, \bm{k}, \gamma) -f_0(\epsilon\bm{(}\bm{k},\gamma)\bm{)}
\right]\nonumber\\
&\quad + \frac{1}{V}\sum_{\bm{k}}( O_{+-}f_{-+} + O_{-+}f_{+-}) 
\label{eq: exp val phys quantity}
\end{align}
with $O_{\gamma\gamma'}\equiv \Braket{\bm{k},\gamma|\hat{O}|\bm{k},\gamma'}$,
where $f_{-+}$ and $f_{+-}$ are included at the second term in the last line.
However, the condition~\eqref{eq: order diag and off-diag} enables us
to safely neglect its contribution in a leading order.
We thus omit the off-diagonal elements and use the semiclassical BTE in the main text.

There are, however, two exceptional cases that we must consider the contribution from $f_{-+}$ and $f_{+-}$; 
one case is when we calculate intrinsic contribution of physical quantity, which exists 
even when we replace the diagonal elements by the equilibrium distribution function
$f(\bm{k},\gamma)= f_0\bm{(}\epsilon (\bm{k},\gamma)\bm{)}$;
the other case is when we face a physical quantity whose intraband elements $O_{++}$, $O_{--}$
are much smaller than interband ones $O_{-+}$, $O_{+-}$, 
in particular when $O_{++} = O_{--}=0$.
The intrinsic anomalous Hall conductivity~\cite{Jungwirth2002,Onoda2002} corresponds to the former case, which
is first clarified by Kohn and Luttinger~\cite{Kohn1957}.
The spin Hall effect in the Rashba 2D system corresponds to the latter case.
The density matrix approach reproduces the well-known result~\cite{Inoue2004}
that the intrinsic spin Hall conductivity is exactly canceled
by the contribution from the impurity scattering in Rashba two-dimensional system~\cite{Khaetskii2006}.

\section{Transmission rate across the interface with arbitrary spin polarization}
At Sect.~III~A in the main text, we have omitted the derivation of the transmission rate 
across the interface $\frac{df}{dt}\Big|_{\text{int}}$ and $\frac{dF}{dt}\Big|_{\text{int}}$.
In this subsection, we detail their derivation based on the Keldysh Green's function, as a complement.
We also consider the case when the 3D metal has an arbitrary spin-polarization direction in the following formalism,
though it is fixed in the $z$-direction in the main text.

Before we start with the Green's function approach, 
we see how to incorporate
spin degree of freedom into electron distribution in the 3D nonmagnetic metal attached on the surface of the chiral metal.
The 3D metal is considered to be isotropic in spin space.
The electron distribution is thus expressed with a spin density matrix
\begin{equation}
 \underline{F}(\bm{q})= \underline{F}^{\dagger}(\bm{q})= 
\begin{pmatrix}
 F_{\uparrow\uparrow}(\bm{q}) & F_{\uparrow\downarrow}(\bm{q})\\
 F_{\downarrow\uparrow}(\bm{q}) &  F_{\downarrow\downarrow}(\bm{q})
\end{pmatrix}
= F_0(\bm{q})\underline{1}+ \bm{F}_1(\bm{q})\cdot \underline{\bm{\sigma}}~,\label{eq: spin density matrix expression}
\end{equation}
where every component of the spin is treated on an equal footing.

The spin-polarization direction $\hat{\bm{\mu}}\equiv {\bm{F}_1}/{|\bm{F}_1|}$
is usually set to be a constant vector, which is determined by boundary conditions for the 3D metals.
As the 3D metal is attached with the surface, the spin density induced at the 2D system
defines the polarization direction $\hat{\bm{\mu}}$ in the 3D metal.
That constraint is expressed as a condition
\begin{equation}
 \hat{\bm{\mu}} = \frac{\langle \bm{S}(\bm{k},\gamma) \varphi(\bm{k},\gamma)\rangle_{\text{2DFC}}}
{\left|\langle \bm{S}(\bm{k},\gamma) \varphi(\bm{k},\gamma)\rangle_{\text{2DFC}}\right|}
\label{eq: 3D spin pol dir is 2D spin pol dir}
\end{equation}
with spin vector represented on each band $\bm{S}(\bm{k},\gamma)\equiv \braket{\bm{k},\gamma|\bm{\sigma}|\bm{k},\gamma}$.
The average over the 2D Fermi contours $\langle \cdots \rangle_\text{2DFC}$ is defined at Eq.~\eqref{eq: def average over 2DFC}.
Spin current injected from the other surface $y=L$ also affects $\hat{\bm{\mu}}$.
We, however, assume that the spin current is polarized in the same direction 
specified by Eq.~\eqref{eq: 3D spin pol dir is 2D spin pol dir}, for convenience.
In the main text, we take $\hat{\bm{\mu}}$ in the $z$-direction
and adopt $F(\bm{q}, \sigma) \equiv F_{\sigma\sigma}(\bm{q})$ 
as the 3D distribution function for spin states $\sigma =\uparrow,\downarrow$ polarized in the $z$-direction,
which is favored when electric field is applied in $z$-direction at the 2D system,
or when spin current polarized in $z$-direction is injected from  the 3D metal.
In this subsection, on the other hand, we make no assumptions on the polarization direction $\hat{\bm{\mu}}$
other than the constraint~\eqref{eq: 3D spin pol dir is 2D spin pol dir}.

The kinetic equation for the spin density matrix $\underline{F}(\bm{q})$ is microscopically given by the Keldysh Green's function.
Let us first define the Green's function in each metal attached on the interface.
In the surface of the chiral metal, the Green's function is defined as
\begin{equation}
  g_{\sigma_1\sigma_2}(\bm{k}; t^{C}_1, t^{C}_2)
\equiv -i \langle T_C\hat{S}_C~\hat{c}_{\bm{k}\sigma_1}(t^C_1)~
\hat{c}^{\dagger}_{\bm{k}\sigma_2}(t^C_2) \rangle~.
\label{eq: Keldysh g in 2D}
\end{equation}
Here we set $\hbar =1$ and introduced fermion annihilation (creation) operator $c_{\bm{k},\sigma}$ ($c^{\dagger}_{\bm{k},\sigma}$)
that specifies one-particle state with the wavevector $\bm{k}$ and $z$-component of spin $\sigma =\uparrow, \downarrow$.
The annihilation operator $c_{\bm{k},\sigma}$ is related to another annihilation operator 
$a_{\bm{k},\gamma}$ that specifies the state $(\bm{k},\gamma)$ in the band basis as 
\begin{equation}
 \begin{pmatrix}
  c_{\bm{k},\uparrow} \\ c_{\bm{k},\downarrow}
 \end{pmatrix}
=
\begin{pmatrix}
 \cos\frac{\Theta(\bm{k})}{2} & \sin\frac{\Theta(\bm{k})}{2} \\
\sin\frac{\Theta(\bm{k})}{2} & -\cos\frac{\Theta(\bm{k})}{2}
\end{pmatrix}
\begin{bmatrix}
 a_{\bm{k},+} \\ a_{\bm{k},-}
\end{bmatrix}
= U^{\dagger}(\bm{k})
\begin{bmatrix}
 a_{\bm{k},+} \\ a_{\bm{k},-}
\end{bmatrix}
\end{equation}
with an unitary matrix $U_{\gamma\sigma}(\bm{k}) = \braket{\bm{k},\gamma|\bm{k},\sigma}$.
The Hamiltonian of the 2D system is diagonalized by the latter operators as
$\hat{H}_{\text{2D}}=\sum_{\bm{k},\gamma}\epsilon (\bm{k},\gamma) \hat{a}^{\dagger}_{\bm{k},\gamma}\hat{a}_{\bm{k},\gamma}$.

On the other hand, the Green's function of the 3D metal is defined as
\begin{equation}
G_{\sigma_1\sigma_2}(\bm{q}; t^{C}_1, t^{C}_2) \equiv -i
\langle T_C\hat{S}_C~\hat{d}_{\bm{q}\sigma_1}(t^C_1)~\hat{d}^{\dagger}_{\bm{q}\sigma_2}(t^C_2) \rangle
\label{eq: Keldysh G in 3D}
\end{equation}
with fermion annihilation operator $\hat{d}_{\bm{q}\sigma_1}$ that specifies one-particle state with $(\bm{q},\sigma)$.
The Hamiltonian of the 3D system is expressed as
$\hat{H}^{\text{(n)}}
= \sum_{\bm{q},\sigma}\epsilon^{\text{(n)}} (|\bm{q}|) \hat{d}^{\dagger}_{\bm{q},\sigma}\hat{d}_{\bm{q},\sigma}$~.

The 2D and 3D systems are in contact at the interface $y=0$ plane where a phenomenological tunneling Hamiltonian
\begin{equation}
 \hat{H}_{\rm T}
=\sum_{\bm{k},\bm{q}}\sum_\sigma\left(T_{\bm{k}\bm{q}}c_{\bm{k},\sigma}^\dagger d_{\bm{q},\sigma}
+T^{*}_{\bm{k}\bm{q}}d_{\bm{q},\sigma}^\dagger c_{\bm{k},\sigma} \right)
\end{equation} 
allows spin-independent transmission across the interface.
We take $T_{\bm{k}\bm{q}} = T$ according to the main text.

We then consider $\hat{H}_{\rm T}$ as a perturbation to the 2D and 3D systems $\hat{H}_{\text{2D}} + \hat{H}^{\text{(n)}}$.
The interaction representation of the operators $c_{\bm{k},\sigma}$
and $d_{\bm{k},\sigma}$ are denoted as
$c_{\bm{k},\sigma}(t^C)$ and $d_{\bm{k},\sigma}(t^C)$ in Eqs.~\eqref{eq: Keldysh g in 2D} and \eqref{eq: Keldysh G in 3D},  
respectively.
We have also introduced an operator so-called scattering matrix
\begin{equation}
 \hat{S}_C = T_C \exp\left[ -i\int_C \hat{H}_{\text{T}}(t^C)~dt^C \right]
\end{equation}
with the Keldysh time ordering operator $T_C$.

As a lowest order approximation, the diagram approach shows that 
\begin{widetext}
 \begin{equation}
   g_{\sigma_1\sigma_2}(\bm{k}; t^{C}_1, t^{C}_2)
= g^{(0)}_{\sigma_1\sigma_2}(\bm{k}; t^{C}_1, t^{C}_2)
+ |T|^2 \int dt^C_3\int dt^C_4 \sum_{\bm{q}, \sigma_3, \sigma_4}
g^{(0)}_{\sigma_1\sigma_3}(\bm{k}; t^{C}_1, t^{C}_3)
G^{(0)}_{\sigma_3\sigma_4}(\bm{q}; t^{C}_3, t^{C}_4)
g^{(0)}_{\sigma_4\sigma_2}(\bm{k}; t^{C}_4, t^{C}_2). 
 \end{equation}
\end{widetext}
The self-energy for the 2D system is then obtained as
\begin{equation}
 \Sigma_{\sigma_1\sigma_2}(t^C_1, t^C_2)
\equiv |T|^2\sum_{\bm{q}} G^{(0)}_{\sigma_1\sigma_2}(\bm{q}; t^{C}_1, t^{C}_2)~.
\end{equation}
The self-energy for the 3D system is also given as
\begin{equation}
 \Sigma^{\text{(n)}}_{\sigma_1\sigma_2}(t^C_1, t^C_2)
\equiv |T|^2\sum_{\bm{k}} g^{(0)}_{\sigma_1\sigma_2}(\bm{k}; t^{C}_1, t^{C}_2)~.
\end{equation}
There are other contributions to the self-energy; impurity scattering in the 2D metal, given at Eq.~\eqref{eq: Self-ene(1,2)}
and Eq.~\eqref{eq: self-ene (X,k)}, and spin-dependent scattering in the 3D metal.
These contributions are considered later.

The Boltzmann collision integral represents the contribution from the self-energy,
as we have seen in the \eqref{eq: St[f] wo approx}.
In the 2D system, the additional collision term is expressed as a matrix form
\begin{align}
 & \underline{\frac{d f(\bm{k})}{d t}\Big|_{\text{int}}}
= \int \frac{d\omega}{2\pi} \left(
-\underline{\Sigma}^{\text{R}}\underline{g}^{<} + \underline{g}^{<}\underline{\Sigma}^{\text{A}}
-\underline{\Sigma}^{<}\underline{g}^{\text{A}} + \underline{g}^{\text{R}}\underline{\Sigma}^{<}
\right)\\
 &= |T|^2\sum_{\bm{q}} \int \frac{d\omega}{2\pi}\Big[
-\underline{G}^{\text{R}}(\bm{q}\omega) \underline{g}^{<}(\bm{k}\omega)
+\underline{g}^{<}(\bm{k}\omega) \underline{G}^{\text{A}}(\bm{q}\omega)
\nonumber\\
&\qquad -\underline{G}^{<}(\bm{q}\omega) \underline{g}^{\text{A}}(\bm{k}\omega)
+\underline{g}^{\text{R}}(\bm{k}\omega) \underline{G}^{<}(\bm{q}\omega)
\Big]~,\label{eq: dfdt matrix form}
\end{align}
while in the 3D system that is expressed as
\begin{multline}
 \underline{\frac{d F(\bm{q})}{d t}\Big|_{\text{int}}}\\
 = |T|^2\sum_{\bm{k}} \int \frac{d\omega}{2\pi}\Big[-\underline{g}^{\text{R}}(\bm{k}\omega) \underline{G}^{<}(\bm{q}\omega)
+\underline{G}^{<}(\bm{q}\omega) \underline{g}^{\text{A}}(\bm{k}\omega)\\
-\underline{g}^{<}(\bm{k}\omega) \underline{G}^{\text{A}}(\bm{q}\omega)
+\underline{G}^{\text{R}}(\bm{q}\omega) \underline{g}^{<}(\bm{k}\omega)
\Big]. 
\end{multline}
We approximate retarded and advanced Green's functions $\underline{g}^{\text{R}}$, 
$\underline{g}^{\text{A}}$, $\underline{G}^{\text{R}}$ and $\underline{G}^{\text{A}}$
shown in the equations above with their expressions in equilibrium: 
\begin{widetext}
\begin{subequations}
 \begin{align}
 \underline{g}^{\text{R}}(\bm{k}\omega) &= {U}^{\dagger}(\bm{k})
 \begin{bmatrix}
 \omega -\epsilon (\bm{k},+) + \mu + i0+ & 0\\
 0 & \omega -\epsilon (\bm{k},-) + \mu + i0+
 \end{bmatrix}^{-1}
 {U}(\bm{k})~, 
&\underline{g}^{\text{A}}(\bm{k}\omega)&= \left[\underline{g}^{\text{R}}(\bm{k}\omega) \right]^{\dagger}~, \\
 \underline{G}^{\text{R}}(\bm{q}\omega)
 &= \frac{1}{\omega -\epsilon^{\text{(n)}} (|\bm{q}|) + \mu + i0+} \underline{1}~,
 &\underline{G}^{\text{A}}(\bm{q}\omega)& 
 =\left[\underline{G}^{\text{R}}(\bm{q}\omega)\right]^{\dagger}~.
 \end{align}
\end{subequations}
We also assume that the lesser Green's functions in the 2D and 3D systems are described as follows:
 \begin{subequations}
 \begin{align}
 \underline{g}^{<}(\bm{k}\omega)
 &= 2\pi i {U}^{\dagger}(\bm{k})
 \begin{bmatrix}
 f(\bm{k},+)\delta \bm{(}\omega -\epsilon (\bm{k},+)+\mu\bm{)} & 0\\
 0 & f(\bm{k},-)\delta \bm{(}\omega -\epsilon (\bm{k},-)+\mu\bm{)} 
 \end{bmatrix}
 {U}(\bm{k})~,\label{eq: lesser G in 2D w distribution func}\\
 \underline{ G}^{<}(\bm{q}\omega)
 &= 2\pi i
 \begin{pmatrix}
 F_{\uparrow\uparrow}(\bm{q}) & F_{\uparrow\downarrow}(\bm{q})\\
 F_{\downarrow\uparrow}(\bm{q}) & F_{\downarrow\downarrow}(\bm{q})
 \end{pmatrix}
 \cdot \delta \bm{(}\omega -\epsilon^{\text{(n)}} (|\bm{q}|)+\mu\bm{)}
 = 2\pi i \underline{F}(\bm{q}) 
 \cdot \delta \bm{(}\omega -\epsilon^{\text{(n)}} (|\bm{q}|)+\mu\bm{)}
 ~.
 \end{align}
 \end{subequations}
\end{widetext}
Here $\underline{F}(\bm{q})$ is the $2\times 2$ spin density matrix in Eq.~\eqref{eq: spin density matrix expression}.
Indeed, electron distribution in the 2D system is also described as a $2\times 2$ density matrix with off-diagonal elements
in general.
However, as we have evaluated in Eq.~\eqref{eq: evaluation of off-diag 1}, the off-diagonal elements $f_{-+}$, $f_{+-}$ 
are of the order of the collision term divided by spin-splitting band gap $\Delta_\text{F}$.
Here the collision term due to the impurity scattering 
is of the order $[f(\bm{k},\gamma) -f_0\bm{(}\epsilon(\bm{k},\gamma)\bm{)}]/\tau_{\text{p}}$,
while the collision term due to the tunneling across the interface~\eqref{eq: dfdt matrix form} is of the order 
$[F(\bm{q},\sigma) -f(\bm{k},\gamma)]/\tau_{\text{t}}$, where
\begin{equation}
 \frac{1}{\tau_{\text{t}}} = \frac{2\pi |T|^2 V^{\text{(n)}} N^{\text{(n)}} /2}{\hbar}
\end{equation}
is the transmission rate across the interface, given in the main text.
It follows that the off-diagonal elements $f_{-+}$, $f_{+-}$ are safely neglected 
in the clean limit $\hbar/(\Delta_\text{F} \tau_\text{p})\ll 1$ and in the low tunneling rate limit $\hbar/(\Delta_\text{F} \tau_\text{t})\ll 1$.
We assume these conditions in the following.

The matrix elements of additional collision term to the 3D metal are now calculated as follows:
\begin{multline}
\left(\frac{dF(\bm{q})}{dt}\Big|_\text{int}\right)_{\sigma\sigma'}\\
= \frac{2\pi |T|^2}{\hbar}
\sum_{\bm{k},\gamma} \Big[
 \braket{\sigma|\bm{k},\gamma} f(\bm{k},\gamma)\cdot
\delta \bm{(}\epsilon^{\text{(n)}}(\bm{q}) -\epsilon (\bm{k},\gamma)\bm{)}
\braket{\bm{k},\gamma |\sigma'}\\
- \frac{1}{2\pi i}\sum_{\sigma''} F_{\sigma\sigma''}\braket{\sigma''|\bm{k},\gamma }
\frac{1}{\epsilon^{\text{(n)}}(\bm{q}) -\epsilon (\bm{k},\gamma) -i0+}
\braket{\bm{k},\gamma |\sigma'}\\
- \braket{\sigma|\bm{k},\gamma}\frac{1}{\epsilon^{\text{(n)}}(\bm{q}) -\epsilon (\bm{k},\gamma) +i0+}
\braket{\bm{k},\gamma |\sigma''} F_{\sigma''\sigma'}
\Big]
\end{multline}

In the same way, the diagonal elements of additional collision term to the 2D metal in the band basis
$\displaystyle U(\bm{k})\underline{\frac{df(\bm{k})}{dt}\Big|_\text{int}}U^{\dagger}(\bm{k})$ are calculated as 
\begin{align}
&\frac{df(\bm{k},\gamma)}{dt}\Big|_\text{int}\equiv
\left(\frac{df(\bm{k})}{dt}\Big|_\text{int}\right)_{\gamma\gamma}\\
&= \frac{2\pi |T|^2}{\hbar}
\sum_{\bm{q}} \left[\braket{\bm{k},\gamma|\underline{F}(\bm{q})|\bm{k},\gamma} - f(\bm{k},\gamma)\right]\nonumber\\
&\qquad\qquad\qquad\qquad
\cdot \delta \bm{(}\epsilon^{\text{(n)}}(\bm{q}) -\epsilon (\bm{k},\gamma)\bm{)}\label{eq: dfdt definition}   
\end{align}
where we defined
\begin{equation}
 \braket{\bm{k},\gamma|\underline{F}(\bm{q})|\bm{k},\gamma}
\equiv \sum_{\sigma\sigma'}\braket{\bm{k},\gamma|\sigma} F_{\sigma\sigma'}(\bm{q})\braket{\sigma'|\bm{k},\gamma}.
\end{equation}

The Boltzmann transport equation (BTE) in the 3D nonmagnetic metal in a stationary case is then expressed as
\begin{equation}
 v_y^{\text{(n)}}\cdot \underline{\frac{\partial F}{\partial y}}
= \underline{\frac{d F}{d t}\Big|_{\text{int}}} +  \underline{\frac{d F}{d t}\Big|_{\text{col}}}\label{eq: BTE in the 3D metal in a matirx form}
\end{equation}
On the right-hand side, the first term gives transmissions at the interface $y=0$,
while the second term describes both spin-conserving and spin-flip scattering process in the bulk.
Let us assume the spin density matrix $\underline{F}$ in the bulk as a solution of the BTE~\eqref{eq: BTE in the 3D metal in a matirx form} to be 
\begin{subequations}
 \label{eq: const spin polarization approx}  
 \begin{align}
& \underline{F} (y, \bm{q}) 
 = F_0(y, \bm{q})\underline{1} + F_1(y, \bm{q}) (\hat{\bm{\mu}}\cdot \underline{\bm{\sigma}})\\
& = \frac{\underline{1} +\hat{\bm{\mu}}\cdot \underline{\bm{\sigma}}}{2}
 \left[F_0(y, \bm{q}) + F_1(y, \bm{q})\right]\nonumber\\
&\qquad + \frac{\underline{1} -\hat{\bm{\mu}}\cdot \underline{\bm{\sigma}}}{2}
 \left[F_0(y, \bm{q}) - F_1(y, \bm{q})\right]
 \end{align}
\end{subequations}
with $\hat{\bm{\mu}}$ satisfying Eq.~\eqref{eq: 3D spin pol dir is 2D spin pol dir}.
The diagonal elements of $\underline{F}$ are $F_0 \pm F_1$,
which give distribution function in the 3D metal.

It is important to check if the collision term $\underline{\frac{d F}{d t}\Big|_{\text{int}}}$ is also
spin polarized parallel to $\hat{\bm{\mu}}$;
otherwise, the spin polarization of the density matrix $\underline{F}(y,\bm{q})$ is deviated from the direction $\hat{\bm{\mu}}$,
according to the BTE~\eqref{eq: BTE in the 3D metal in a matirx form}.
In other words, 
the condition if the collision term $\underline{\frac{d F}{d t}\Big|_{\text{int}}}$ is in the direction $\hat{\bm{\mu}}$
serve as a sufficient condition for our assumption~\eqref{eq: const spin polarization approx}.
The matrix $\frac{d F}{d t}\Big|_{\text{int}}$ is first decomposed into charge and spin sectors as
\begin{multline}
 \underline{\frac{d F (y,\bm{q})}{d t}\Big|_{\text{int}}}
= \frac{1}{2}\Big\{
\Tr \left[\underline{\frac{dF(\bm{q})}{dt}\Big|_\text{int}}\right] \underline{1}\\
+ \sum_{\hat{\bm{n}} = \hat{\bm{\mu}}, \hat{\bm{\nu}}, \hat{\bm{\lambda}}}
\Tr \left[(\hat{\bm{n}}\cdot \underline{\bm{\sigma}})
\underline{\frac{dF(\bm{q})}{dt}\Big|_\text{int}}\right] (\hat{\bm{n}}\cdot \underline{\bm{\sigma}})
\Big\}\label{eq: charge spin sectors decomposed dFdt} 
\end{multline}
where $\hat{\bm{\nu}},~\hat{\bm{\lambda}}$ are unit vectors normal to $\hat{\bm{\mu}}$, satisfying 
$\hat{\bm{\mu}}\times \hat{\bm{\nu}} = \hat{\bm{\lambda}}$.
The charge sector after the substitution of the assumption~\eqref{eq: const spin polarization approx}
is calculated as
\begin{align}
& \Tr \left[\underline{\frac{dF(\bm{q})}{dt}\Big|_\text{int}}\right]\nonumber\\
&= \frac{2\pi |T|^2}{\hbar}\sum_{\bm{k},\gamma} 
\Big[
\sum_\sigma |\braket{\sigma|\bm{k},\gamma}|^2 f(\bm{k},\gamma)
\delta \bm{(}\epsilon^{\text{(n)}}(\bm{q}) -\epsilon (\bm{k},\gamma)\bm{)}\nonumber\\
&\qquad -\frac{1}{2\pi i}\sum_{\sigma,\sigma''}
\left(
\frac{F_{\sigma\sigma''}\braket{\sigma''|\bm{k},\gamma }\braket{\bm{k},\gamma |\sigma}
}{\epsilon^{\text{(n)}}(\bm{q}) -\epsilon (\bm{k},\gamma) -i0+}
- \text{c.c.}
\right)\Big]\\
 &=  \frac{2\pi |T|^2}{\hbar}\sum_{\bm{k},\gamma} \left\{
f(\bm{k},\gamma) - \left[
F_0(\bm{q}) + F_1(\bm{q}) \cdot\braket{\bm{k},\gamma|\hat{\bm{\mu}}\cdot \underline{\bm{\sigma}}|\bm{k},\gamma}
\right]\right\} \nonumber\\
&\qquad \cdot\delta \bm{(}\epsilon^{\text{(n)}}(\bm{q}) -\epsilon (\bm{k},\gamma)\bm{)}\\
 & =  \frac{2\pi |T|^2}{\hbar}\sum_{\bm{k},\gamma} \left[f(\bm{k},\gamma) - F_0(\bm{q}) \right] 
\delta \bm{(}\epsilon^{\text{(n)}}(\bm{q}) -\epsilon (\bm{k},\gamma)\bm{)}~.
\label{eq: charge transmission rate}
\end{align}
Here $\text{c.c.}$ stands for complex conjugate.
In the last line, we used no spin polarization in equilibrium at the 2D metal
$\left\langle\braket{\bm{k},\gamma|\bm{\sigma}|\bm{k},\gamma}\right\rangle_{\text{2DFC}} =0$.
The spin sectors polarized in $\hat{\bm{\mu}}$ and $\hat{\bm{\nu}}$ are also given as follows:
\begin{align}
& \Tr \left[(\hat{\bm{\mu}}\cdot \underline{\bm{\sigma}}) \underline{\frac{dF(\bm{q})}{dt}\Big|_\text{int}}\right]\nonumber\\
&= \frac{2\pi |T|^2}{\hbar}\sum_{\bm{k},\gamma} 
\Big[
\braket{\bm{k},\gamma|\hat{\bm{\mu}}\cdot \underline{\bm{\sigma}}|\bm{k},\gamma} f(\bm{k},\gamma)
\delta \bm{(}\epsilon^{\text{(n)}}(\bm{q}) -\epsilon (\bm{k},\gamma)\bm{)}\nonumber\\
&\qquad -\frac{1}{2\pi i}\left(
\frac{\braket{\bm{k},\gamma|(\hat{\bm{\mu}}\cdot \underline{\bm{\sigma}}) \underline{F}|\bm{k},\gamma}}
{\epsilon^{\text{(n)}}(\bm{q}) -\epsilon (\bm{k},\gamma) -i0+}
- \text{c.c.}
\right)\Big]\\
 & = \frac{2\pi |T|^2}{\hbar}\sum_{\bm{k},\gamma} \left[
\braket{\bm{k},\gamma|\hat{\bm{\mu}}\cdot \underline{\bm{\sigma}}|\bm{k},\gamma}
f(\bm{k},\gamma) -  F_1(\bm{q})\right]\nonumber\\
&\qquad \cdot \delta \bm{(}\epsilon^{\text{(n)}}(\bm{q}) -\epsilon (\bm{k},\gamma)\bm{)}~,
\label{eq: spin transmission rate}
\end{align}
\begin{align}
& \Tr \left[(\hat{\bm{\nu}}\cdot \underline{\bm{\sigma}}) \underline{\frac{dF(\bm{q})}{dt}\Big|_\text{int}}\right]\\
&= \frac{2\pi |T|^2}{\hbar}\sum_{\bm{k},\gamma} \Big\{
\braket{\bm{k},\gamma|\hat{\bm{\nu}}\cdot \underline{\bm{\sigma}}|\bm{k},\gamma}
f(\bm{k},\gamma) \cdot\delta \bm{(}\epsilon^{\text{(n)}}(\bm{q}) -\epsilon (\bm{k},\gamma)\bm{)}\nonumber\\
&\qquad \qquad 
- \frac{\mathscr{P}}{\pi}\cdot \frac{\braket{\bm{k},\gamma|\hat{\bm{\lambda}}\cdot \underline{\bm{\sigma}}|\bm{k},\gamma}}
{\epsilon^{\text{(n)}}(\bm{q}) -\epsilon (\bm{k},\gamma)}
F_1(\bm{q}) \Big\}~.\label{eq: dFdt normal to bulk pol} 
\end{align}
Here we used relations 
$\braket{\bm{k},\gamma|(\hat{\bm{\mu}}\cdot \underline{\bm{\sigma}}) \underline{F}|\bm{k},\gamma}
= \braket{\bm{k},\gamma|(\hat{\bm{\mu}}\cdot \underline{\bm{\sigma}}) 
\left(F_0(\bm{q})\underline{1} + F_1(\bm{q}) (\hat{\bm{\mu}}\cdot \underline{\bm{\sigma}})\right)|\bm{k},\gamma}
= F_0(\bm{q})  \braket{\bm{k},\gamma|(\hat{\bm{\mu}}\cdot \underline{\bm{\sigma}}) |\bm{k},\gamma} + F_1(\bm{q})$
and $(\hat{\bm{\mu}}\cdot \underline{\bm{\sigma}})(\hat{\bm{\nu}}\cdot \underline{\bm{\sigma}})
= i\hat{\bm{\lambda}}\cdot \underline{\bm{\sigma}}$.
The principle value of the integral is denoted as $\mathscr{P}$.
The spin sector polarized in $\hat{\bm{\nu}}$, shown in Eq.~\eqref{eq: dFdt normal to bulk pol}, becomes
exactly zero. The first term vanishes due to the constraint~\eqref{eq: 3D spin pol dir is 2D spin pol dir},
where spin polarization at the 2D metal is set to be in the direction $\hat{\bm{\mu}}$, normal to $\hat{\bm{\nu}}$.
As for the second term, since the 2D metal itself has time reversal symmetry,
the sum over the states in the 2D metal $\ket{\bm{k},\gamma}$ with energy $\epsilon (\bm{k},\gamma)$
is equivalent to sum over states $\ket{\tilde{\bm{k}},\tilde{\gamma}}\equiv
\Theta\ket{\bm{k},\gamma}$ with energy $\epsilon (\tilde{\bm{k}},\tilde{\gamma}) = \epsilon (\bm{k},\gamma)$,
where $\Theta$ denotes
time-reversal anti-unitary operator for spin-$1/2$ system, satisfying $\Theta^2 =-1$
and $\underline{\sigma}\Theta = -\Theta \underline{\sigma}$.
It follows that
\begin{align}
&\mathscr{P}\sum_{\bm{k},\gamma}
\frac{\braket{\bm{k},\gamma|\hat{\bm{\lambda}}\cdot \underline{\bm{\sigma}}|\bm{k},\gamma}}
{\epsilon^{\text{(n)}}(\bm{q}) -\epsilon (\bm{k},\gamma)}
= \mathscr{P}\sum_{\tilde{\bm{k}},\tilde{\gamma}}
\frac{\braket{\tilde{\bm{k}},\tilde{\gamma}|\hat{\bm{\lambda}}\cdot \underline{\bm{\sigma}}|\tilde{\bm{k}},\tilde{\gamma}}}
{\epsilon^{\text{(n)}}(\bm{q}) -\epsilon (\tilde{\bm{k}}, \tilde{\gamma})}\nonumber\\
&\qquad  =  \mathscr{P}\sum_{{\bm{k}},{\gamma}}
\frac{\braket{\Theta {\bm{k}},{\gamma}|\hat{\bm{\lambda}}\cdot \underline{\bm{\sigma}}|\Theta {\bm{k}},{\gamma}}}
{\epsilon^{\text{(n)}}(\bm{q}) -\epsilon ({\bm{k}}, {\gamma})}\nonumber\\
&\qquad = - \mathscr{P}\sum_{\bm{k},\gamma}
\frac{\braket{\bm{k},\gamma|\hat{\bm{\lambda}}\cdot \underline{\bm{\sigma}}|\bm{k},\gamma}}
{\epsilon^{\text{(n)}}(\bm{q}) -\epsilon (\bm{k},\gamma)}\\
 &= 0 ~.
\end{align}
The spin sector polarized in the $\hat{\bm{\lambda}}$-direction is also zero 
in the same way.
It is thus proved that the collision term due to the interface transmission is polarized in the $\hat{\bm{\mu}}$-direction,
and that the assumption~\eqref{eq: const spin polarization approx} is consistent 
with the BTE of the 3D metal~\eqref{eq: BTE in the 3D metal in a matirx form}.

The collision term $\displaystyle \underline{\frac{dF(y=0,\bm{q})}{dt}\Big|_\text{int}}$, decomposed in 
Eq.~\eqref{eq: charge spin sectors decomposed dFdt}, is thus expressed
as a linear combination of the unit matrix $\underline{1}$ and $\hat{\bm{\mu}}\cdot \underline{\bm{\sigma}}$.
That is diagonalized as follows:
\begin{widetext}
 \begin{multline}
 \underline{\frac{dF(y=0,\bm{q})}{dt}\Big|_\text{int}}
 =\frac{\underline{1} + \hat{\bm{\mu}}\cdot \underline{\bm{\sigma}}}{2}\cdot
 \frac{2\pi |T|^2}{\hbar}\sum_{\bm{k},\gamma} \left[
 \frac{1 + \hat{\bm{\mu}}\cdot\bm{S}(\bm{k},\gamma)}{2}f(\bm{k},\gamma)
 - \frac{1}{2}(F_0(\bm{q})+ F_1(\bm{q}))
 \right]\cdot \delta \bm{(}\epsilon^{\text{(n)}}(\bm{q}) -\epsilon (\bm{k},\gamma)\bm{)}\\
 + \frac{\underline{1} - \hat{\bm{\mu}}\cdot \underline{\bm{\sigma}}}{2}\cdot
 \frac{2\pi |T|^2}{\hbar}\sum_{\bm{k},\gamma} \left[
 \frac{1 - \hat{\bm{\mu}}\cdot\bm{S}(\bm{k},\gamma)}{2}f(\bm{k},\gamma)
 - \frac{1}{2}(F_0(\bm{q})- F_1(\bm{q}))
 \right]\cdot \delta \bm{(}\epsilon^{\text{(n)}}(\bm{q}) -\epsilon (\bm{k},\gamma)\bm{)}~.
 \end{multline}
 The collision term to the 2D metal for each band
 $\displaystyle {\frac{df(\bm{k},\gamma)}{dt}\Big|_\text{int}}$
 is also written by substituting Eq.~\eqref{eq: const spin polarization approx} 
 for Eq.~\eqref{eq: dfdt definition} as
 \begin{align}
 & \frac{df(\bm{k},\gamma)}{dt}\Big|_\text{int}
 = \frac{2\pi |T|^2}{\hbar}
 \sum_{\bm{q}} \left[
 F_0(\bm{q})+ \hat{\bm{\mu}}\cdot \bm{S}(\bm{k},\gamma) F_1(\bm{q})  - f(\bm{k},\gamma)\right]
 \cdot \delta \bm{(}\epsilon^{\text{(n)}}(\bm{q}) -\epsilon (\bm{k},\gamma)\bm{)}  \label{eq: dfdt int F0F1}\\
 &= \frac{2\pi |T|^2}{\hbar}
 \sum_{\bm{q}} \left[
 \frac{1 + \hat{\bm{\mu}}\cdot\bm{S}(\bm{k},\gamma)}{2}\cdot (F_0(\bm{q})+ F_1(\bm{q}))
 + \frac{1 -\hat{\bm{\mu}}\cdot\bm{S}(\bm{k},\gamma)}{2}\cdot (F_0(\bm{q})- F_1(\bm{q}))
 - f(\bm{k},\gamma)\right]
 \cdot \delta \bm{(}\epsilon^{\text{(n)}}(\bm{q}) -\epsilon (\bm{k},\gamma)\bm{)} 
 \end{align}
\end{widetext}

If we take $\hat{\bm{\mu}} = \hat{\bm{z}}$ as the main text,
the relations $\displaystyle |\braket{\bm{k},\gamma|\sigma = \uparrow}|^2 = \frac{1 + \hat{\bm{\mu}}\cdot\bm{S}(\bm{k},\gamma)}{2}
= \frac{1 + \cos\Theta (\bm{k})}{2}$ and
$\displaystyle |\braket{\bm{k},\gamma|\sigma = \downarrow}|^2 = \frac{1- \hat{\bm{\mu}}\cdot\bm{S}(\bm{k},\gamma)}{2}
= \frac{1 + \cos\Theta (\bm{k})}{2}$ follow.
We can also take
$F(y,\bm{q},\sigma) = F_0 (y,\bm{q}) + \sigma F_1 (y,\bm{q})$ as distribution function in the 3D metal,
which gives the expression for the transmission rates
\begin{subequations}
 \begin{align}
& \frac{d f(\bm{k},\gamma)}{d t}\Big|_{\rm int}\nonumber\\
&=\frac{2\pi|T|^2}{\hbar}\sum_{\bm{q},\sigma}|\langle \bm{k},\gamma|\sigma\rangle|^2
\left[F(y=0,\bm{q},\sigma)-f(\bm{k},\gamma)\right] \nonumber\\
&\qquad \cdot\delta\bm{(}\epsilon(\bm{k},\gamma)-\epsilon^{\rm (n)}(\bm{q})\bm{)},\\
& \frac{d F(y=0,\bm{q},\sigma)}{d t}\Big|_{\rm int} \nonumber\\
&= \frac{2\pi|T|^2}{\hbar}\sum_{\bm{k},\gamma}|\langle \bm{k},\gamma|\sigma\rangle|^2
\left[f(\bm{k},\gamma)-F(\bm{q},\sigma)\right]\nonumber\\
&\cdot \delta\bm{(}\epsilon(\bm{k},\gamma)-\epsilon^{\rm (n)}(\bm{q})\bm{)}
 \end{align}
\end{subequations}
as we provide in the main text.

\section{Boltzmann equation for the 3D metal and the Valet--Fert solution}
In this section, we review the derivation of 
the electron distribution function in a 3D nonmagnetic metal with spin-degenerate bands, along with the work of 
Valet and Fert~\cite{Valet1993}.

As we have shown in the last section, the BTE in the 3D metal is expressed as $2\times 2$ matrices in general.
However, under appropriate boundary conditions,
the BTE and its solution, spin density matrix, are polarized in a direction $\hat{\bm{\mu}}$,
which allows us to take two distribution functions corresponding to the spin in the $\pm \hat{\bm{\mu}}$ directions.
In this subsection, we assume the case $\hat{\bm{\mu}} = \hat{\bm{z}}$, and 
take $F(y, \bm{q},\sigma = \uparrow, \downarrow)$ as distribution function.

In the bulk, the collision term is due to the spin-dependent scattering
$\displaystyle \underline{\frac{dF}{dt}\Big|_{\text{col}}}$ in the BTE~\eqref{eq: BTE in the 3D metal in a matirx form}.
When there are spin-conserving impurity scattering and much weaker spin-flip impurity scattering, 
the distribution function $F(y,\bm{q},\sigma)$ in a steady state follows
\begin{multline}
v_y\frac{\partial F(y,\bm{q},\sigma)}{\partial y}\\
=-\int \frac{d\bm{q}'}{(2\pi)^3} \sum_{\sigma'=\uparrow, \downarrow}
W(\bm{q}\sigma,\bm{q}'\sigma') 
\left[F(y,\bm{q},\sigma) - F(y,\bm{q}',\sigma')\right] \label{eq: ValetFert BTE} 
\end{multline}
with the scattering rates by impurities
\begin{subequations}
\begin{equation}
 W(\bm{q},\sigma,\bm{q}',\sigma)
= P_\text{s}(\bm{q}\cdot \bm{q}'/|\bm{q}|^2)\cdot \delta\bm{(}\epsilon^{\rm (n)}(\bm{q}')-\epsilon^{\rm (n)}(\bm{q})\bm{)} 
\end{equation}
for spin-conserving process, and
\begin{equation}
W(\bm{q},\sigma,\bm{q}',-\sigma)
= P_{\text{sf}}(\bm{q}\cdot \bm{q}'/|\bm{q}|^2)\cdot \delta\bm{(}\epsilon^{\rm (n)}(\bm{q}')-\epsilon^{\rm (n)}(\bm{q})\bm{)} 
\end{equation} 
\end{subequations}
for spin-flip process. 
Here $-\sigma$ represents $\downarrow (\uparrow)$ when $\sigma=\uparrow (\downarrow)$, and $P_{\text{sf}}\ll P_\text{s}$ holds.
Valet and Fert have also considered a driving term due to external electric field,
but we here focus on the case in the absence of the electric fields.

We then separate the distribution function to the isotropic part $\mu (y,\sigma)$ and anisotropic part $g(y, \bm{q},\sigma)$
with respect to the momentum $\bm{q}$ as
\begin{multline}
F(y,\bm{q},\sigma)\\
= f_0 \bm{(}\epsilon^{\text{(n)}}(\bm{q}) \bm{)} 
+ \frac{\partial f_0 (\epsilon)}{\partial \epsilon^{\text{(n)}}(\bm{q})}
\left[\mu_0 -\mu (y,\sigma) + g(y, \bm{q},\sigma)\right]\label{eq: VF distribution fuc} 
\end{multline}
with $\mu_0$ equilibrium chemical potential.
For simplicity, we approximate the scattering rates and anisotropic part of the distribution function as
\begin{subequations}
 \begin{align}
 P_\text{s}(\bm{q}\cdot \bm{q}'/|\bm{q}|^2)
 &\simeq P_\text{s}^{\text{(0)}} + P_\text{s}^{\text{(1)}}\cdot \frac{\bm{q}\cdot \bm{q}'}{|\bm{q}|^2}~,\\
 P_\text{sf}(\bm{q}\cdot \bm{q}'/|\bm{q}|^2)
 &\simeq P_\text{sf}^{\text{(0)}}~,\\
 g(y, \bm{q},\sigma)&\simeq g^{\text{(1)}}(y,\sigma)\cdot \frac{q_y}{|\bm{q}|}~,
 \end{align}
\end{subequations}
and omit higher multipole terms of them.
The BTE~\eqref{eq: ValetFert BTE} is then expressed as
\begin{multline}
v_y\left[\frac{\partial g(y, \bm{q},\sigma)}{\partial y} -\frac{\partial \mu (y,\sigma)}{\partial y}\right] \\
= \frac{\mu (y, \sigma) - \mu (y, -\sigma)}{\tau_{\text{sf}}}
- \left(\frac{1}{\tau_{\text{s}}}+\frac{1}{\tau_{\text{sf}}}\right)g(y, \bm{q},\sigma) 
\end{multline}
around the Fermi surface $\epsilon_\text{F} = \epsilon^{\text{(n)}}(\bm{q})$.
There are spin-conserving scattering rate $1/\tau_{\text{s}}$
and spin-flip scattering rate $1/\tau_{\text{sf}}$, defined as
\begin{align}
\frac{1}{\tau_{\text{s}}(\mu_0)}
&\equiv\int \frac{d\bm{q}'}{(2\pi)^3}P_\text{s} (\bm{q}\cdot \bm{q}'/q_\text{F}^2)
\left(1 - \frac{\bm{q}\cdot \bm{q}'}{q_\text{F}^2}\right)\nonumber\\
&\qquad \cdot \delta\bm{(}\epsilon^{\rm (n)}(\bm{q}')-\mu_0\bm{)}~,\\
\frac{1}{\tau_{\text{sf}}(\mu_0)} &\equiv \int \frac{d\bm{q}'}{(2\pi)^3}P_\text{sf} (\bm{q}\cdot \bm{q}'/q_\text{F}^2)
\cdot \delta\bm{(}\epsilon^{\rm (n)}(\bm{q}')-\mu_0\bm{)}.
\end{align}
It is easy to see $\tau_{\text{s}}\ll \tau_{\text{sf}}$.
This BTE is arranged to be a dimensionless one
\begin{multline}
 \cos^2\vartheta \frac{\partial g^{\text{(1)}}(y,\sigma)}{\partial (y/\lambda^{\text{(n)}})} 
- \frac{1}{6}\left(\frac{\lambda^{\text{(n)}}}{\ell_{\text{sf}}}
\right)^2\left[\mu (y, \sigma) - \mu (y, -\sigma)\right] \\
=  \cos\vartheta 
\left[
\frac{\partial \mu (y, \sigma)}{\partial (y/\lambda^{\text{(n)}})}
-g^{\text{(1)}}(y,\sigma)
\right]
\label{eq: BTE VF dimensionless} 
\end{multline}
with electron mean free path $\lambda^{\text{(n)}}$ and spin diffusion length $\ell_{\text{sf}}$ such that
\begin{align}
 \lambda^{\text{(n)}} &\equiv v^{\text{(n)}}_\text{F} \left(\frac{1}{\tau_\text{s}} + \frac{1}{\tau_\text{sf}}\right)^{-1},
&\ell_{\text{sf}} &\equiv \sqrt{\frac{1}{6}v^{\text{(n)}}_\text{F}\lambda^{\text{(n)}}\tau_{\text{sf}}}\label{eq: def ele mean free path and spin diff length}
\end{align}
and $\cos\vartheta \equiv q_y/q^{\text{(n)}}_\text{F} = v_y/v^{\text{(n)}}_\text{F}$.

As we have neglected higher multipoles in $\bm{q}$, i.e. Legendre polynomials $P_m(\cos\vartheta)$ with $m\geq 2$,
the factor $\cos^2\vartheta = \frac{1}{3}P_0(\cos\vartheta) + \frac{2}{3}P_2(\cos\vartheta)$
in Eq.~\eqref{eq: BTE VF dimensionless} is approximated as $1/3$.
Equation~\eqref{eq: BTE VF dimensionless} is then divided into isotropic part 
and anisotropic part
\begin{subequations}
 \label{eq: VF SDL pre}
 \begin{align}
 \frac{\partial g^{\text{(1)}}(y,\sigma)}{\partial (y/\lambda^{\text{(n)}})}
 &= \frac{1}{2}\left(\frac{\lambda^{\text{(n)}}}{\ell_{\text{sf}}}\right)^2\left[\mu (y, \sigma) - \mu (y, -\sigma)\right],\\
 g^{\text{(1)}}(y,\sigma) 
 &=\frac{\partial \mu (y, \sigma)}{\partial (y/\lambda^{\text{(n)}})}~.
\end{align}
\end{subequations}
Here $g^{\text{(1)}}$ contributes to 
both electric current density and spin current density that flow in $y$-direction;
\begin{align}
 &J^{\text{(n)}}_c(y) = -e\int \frac{d\bm{q}}{(2\pi)^3} 
\left[F(y,\bm{q},\uparrow) + F(y,\bm{q},\downarrow)\right]
v_y(\bm{q})\\
&= -ev^{\text{(n)}}_\text{F} \int \frac{d\bm{q}}{(2\pi)^3} 
\frac{\partial f_0 (\epsilon)}{\partial \epsilon^{\text{(n)}}(\bm{q})}
\left[g(y, \bm{q},\uparrow) +g(y, \bm{q},\downarrow)\right]  \frac{q_y}{k^{\text{(n)}}_\text{F}}\\
&= ev^{\text{(n)}}_\text{F} \int \frac{d\bm{q}}{(2\pi)^3} 
\left[g^{\text{(1)}}(y,\uparrow) + g^{\text{(1)}}(y,\downarrow) \right]  \cos^2\vartheta\nonumber\\
&\qquad \cdot \delta \bm{(} \epsilon^{\text{(n)}}(\bm{q}) -\mu_0\bm{)}\\
&= ev^{\text{(n)}}_\text{F}\cdot \frac{N^{\text{(n)}}(\mu_0)}{2}\cdot \frac{1}{3}
\left[g^{\text{(1)}}(y,\uparrow) +g^{\text{(1)}}(y,\downarrow) \right]\\
&= \frac{\sigma^{\text{(n)}}}{2e\lambda^{\text{(n)}}}
\left[g^{\text{(1)}}(y,\uparrow) + g^{\text{(1)}}(y,\downarrow) \right]~,
\end{align}
and
\begin{align}
J^{\text{(n)}}_\text{s}(y)
&= \int \frac{d\bm{q}}{(2\pi)^3} 
\left[F(y,\bm{q},\uparrow) - F(y,\bm{q},\downarrow)\right] v_y(\bm{q})\\
&= -\frac{\sigma^{\text{(n)}}}{2e^2\lambda^{\text{(n)}}}
\left[g^{\text{(1)}}(y,\uparrow) -g^{\text{(1)}}(y,\downarrow) \right]  
\end{align}
with particle number density $n$ and electric conductivity in the 3D metal $\sigma^{\text{(n)}}$ given as
\begin{gather}
n = 2\cdot \frac{4\pi}{3}\left(\frac{k^{\text{(n)}}_\text{F}}{2\pi}\right)^3
= \frac{2}{3}\mu_0 N^{\text{(n)}}(\mu_0)~,\label{eq: particle number density n}\\
\sigma^{\text{(n)}}\equiv \frac{ne^2 (1/\tau_\text{s}+ 1/\tau_{\text{sf}})^{-1}}{m}~.\label{eq: 3D electric conductivity} 
\end{gather}

Equations~\eqref{eq: VF SDL pre} then yield the expression for diffusive spin current and spin diffusion equation~\cite{Valet1993}
\begin{align}
J^{\text{(n)}}_\text{s}(y)&= - \frac{\sigma^{\text{(n)}}}{2e^2}\frac{\partial \mu^{\text{(n)}}_\text{s}(y)}{\partial y}~,
& \frac{\partial^2 \mu^{\text{(n)}}_\text{s}(y) }{\partial y^2} &= \frac{1}{\ell_{\text{sf}}^2} \mu_\text{s}^{\text{(n)}}(y)
\end{align}
with spin accumulation $\mu^{\text{(n)}}_\text{s} (y)\equiv \mu (y, \uparrow)- \mu (y, \downarrow)$,
as well as ordinary diffusive electric current and charge conservation law in a steady state
\begin{align}
 J^{\text{(n)}}_c(y)&= - \frac{\sigma^{\text{(n)}}}{(-e)}\frac{\partial \bar{\mu}(y)}{\partial y}~,
& \frac{\partial J^{\text{(n)}}_c(y) }{\partial y} &= 0
\end{align}
with net chemical potential $\bar{\mu}\equiv \left[\mu (y, \uparrow)+\mu (y, \downarrow)\right]/2$.
It follows that the spin accumulation and spin current are expressed as
\begin{subequations}
 \label{eq: js-n-y-edelstein}
 \begin{align}
 \mu^{\text{(n)}}_\text{s}(y) &= 2(A e^{-y/\ell_{\rm sf}}+B e^{y/\ell_{\rm sf}})~,\\
 J_{\rm s}^{\rm (n)}(y)&=\frac{\sigma^{\rm (n)}}{e^2\ell_{\rm sf}}\left(A e^{-y/\ell_{\rm sf}}
 -B e^{y/\ell_{\rm sf}}\right)~,
 \end{align}
\end{subequations}
where two coefficients $A, B$ will be determined by boundary conditions.
We assume that the junction system we consider is bounded in the $y$-direction,
which leads to the absence of electric current through the 3D nonmagnetic metal
$J^\text{(n)}_c (y) = 0$.
The net chemical potential $\bar{\mu}$ is then spatially constant,
and $g^{\text{(1)}}(y,\uparrow) + g^{\text{(1)}}(y,\downarrow) =0 $ holds.
Therefore, the distribution function in the 3D metal~\eqref{eq: VF distribution fuc} is described as
\begin{widetext}
 \begin{align}
 &F(y,\bm{q},\sigma)
 = f_0 \bm{(}\epsilon^{\text{(n)}}(\bm{q}) \bm{)} 
 -\frac{\partial f_0\bm{(}\epsilon^{\text{(n)}}(|\bm{q}|)\bm{)}}{\partial \epsilon}
 \left\{\bar{\mu}- \mu_0 + \frac{\sigma}{2}\left[
 \mu^{\rm (n)}_{\rm s}(y)
 + \frac{2e^2\lambda^{\rm (n)}}{\sigma^{\rm (n)}}\frac{q_y}{q}
 J_{\rm s}^{\rm (n)}(y)\right]
 \right\}~,\\
 &= f_0 \bm{(}\epsilon^{\text{(n)}}(\bm{q}) \bm{)} 
 - \frac{\partial f_0\bm{(}\epsilon^{\text{(n)}}(|\bm{q}|)\bm{)}}{\partial \epsilon}
 \left[\bar{\mu}-\mu_0
 +\sigma A \left(1+\frac{\lambda^{\rm (n)}}{\ell_{\rm sf}}\frac{q_y}{q}\right)\exp\left(-\frac{y}{\ell_{\rm sf}}\right)
 +\sigma B \left(1-\frac{\lambda^{\rm (n)}}{\ell_{\rm sf}}\frac{q_y}{q}\right)\exp\left(\frac{y}{\ell_{\rm sf}}\right)\right]~.
 \end{align}
\end{widetext}
As compared with Eq.~\eqref{eq: const spin polarization approx}
where $F(y,\bm{q},\sigma) = F_0 (y,\bm{q}) + \sigma F_1 (y,\bm{q})$ follows, 
we obtain the expressions
\begin{subequations}
\label{eq: F0 F1 determined by ValetFert}
 \begin{align}
 F_0 (y,\bm{q})&= f_0 \bm{(}\epsilon^{\text{(n)}}(\bm{q}) \bm{)} 
 + (\bar{\mu}-\mu_0) \delta \bm{(}\epsilon^{\text{(n)}}(\bm{q}) -\mu_0\bm{)} \\
 F_1 (y,\bm{q})&= 
 \frac{1}{2}
 \left[\mu^{\rm (n)}_{\rm s}(y)
 + \frac{2e^2\lambda^{\rm (n)}}{\sigma^{\rm (n)}}\frac{q_y}{q}
 J_{\rm s}^{\rm (n)}(y)\right]\nonumber\\
&\qquad \cdot \delta \bm{(}\epsilon^{\text{(n)}}(\bm{q}) -\mu_0\bm{)}~.
\end{align}
\end{subequations}
The result above holds for arbitrary spin polarization direction $\hat{\bm{\mu}}$.

\section{Derivation of the effective Boltzmann equation at the interface}
At Sect.~III~A in the main text, we omit the derivation of the effective Boltzmann equation
at the interface with the additional relaxation matrix $M_\text{int}$ and driving term $b_{\text{IEE}}$.
As a complement, we follow the derivation in this section.
In Sect.~\ref{subsec: Elimination of parameters in the Valet--Fert solution}, we review
how the boundary conditions determine parameters included in
the 3D electron distribution derived by Valet and Fert~\eqref{eq: F0 F1 determined by ValetFert}.
In Sect.~\ref{subsec: The Boltzmann equation for the 2D electron distribution function},
the effective BTE for the 2D electron is formulated by using the 3D electron distribution $F_0$ and $F_1$.

\subsection{\label{subsec: Elimination of parameters in the Valet--Fert solution}
Elimination of parameters in the Valet--Fert solution}
In accordance with the Keldysh formalism,
we will express the three parameters $\bar{\mu}$, $A$ and $B$ as the functionals of the distribution function 
of the 2D metal by the three conditions: 
(i) the absence of charge current through the interface,
(ii) the continuity of spin current at the interface,
(iii) the boundary condition on the other side of the 3D metal at $y=L$.  
The deviation of the 2D distribution function is given as
\begin{equation}
f(\bm{k},\gamma) = f_0\bm{(}\epsilon (\bm{k},\gamma)\bm{)}
+ \varphi (\bm{k},\gamma)\cdot \delta \bm{(}\epsilon (\bm{k},\gamma) -\mu_0\bm{)}~.
\end{equation}

\subsubsection*{(i) Condition of the absence of charge current through the interface}
The number of electrons tunneling through the unit area of the interface 
from the 2D system to 3D system per unit time is written as
\begin{align}
& \frac{1}{V} \sum_{\bm{q}} \Tr \left[\underline{\frac{dF(\bm{q})}{dt}\Big|_\text{int}}\right]\nonumber\\
&=  \frac{2\pi |T|^2}{\hbar V}
\sum_{\bm{q},\bm{k},\gamma} \left[f(\bm{k},\gamma) - F_0(\bm{q}) \right] 
\delta \bm{(}\epsilon^{\text{(n)}}(\bm{q}) -\epsilon (\bm{k},\gamma)\bm{)}\\
 &=  \frac{2\pi |T|^2}{\hbar V}
\sum_{\bm{q},\bm{k},\gamma} \left[\varphi (\bm{k},\gamma) - (\bar{\mu}-\mu_0)\right] \nonumber\\
&\qquad \delta \bm{(}\epsilon (\bm{k},\gamma) -\mu_0\bm{)}
\delta \bm{(}\epsilon^{\text{(n)}}(\bm{q}) -\mu_0\bm{)}\\
 &= \frac{2\pi |T|^2}{\hbar}V^{\text{(n)}}N^{\text{(n)}} N_0
\cdot \left[\left\langle \varphi (\bm{k},\gamma)\right\rangle_{\text{2DFC}} - (\bar{\mu}-\mu_0)\right]
\end{align}
where we made use of Eq.~\eqref{eq: charge transmission rate}.
The condition of no charge current tunneling through the interface thus becomes
\begin{equation}
\bar{\mu}-\mu_0=\langle \varphi(\bm{k},\gamma)\rangle_{\rm 2DFC}~.
\label{eq: mubar}
\end{equation}

\subsubsection*{ (ii) Condition of continuity of spin current at the interface}
Similarly, the $\hat{\bm{\mu}}$-component spin entering from the 2D system to 3D system per unit time per unit area of the interface 
is written as
\begin{align}
& \frac{1}{V} \sum_{\bm{q}}\Tr \left[(\hat{\bm{\mu}}\cdot \underline{\bm{\sigma}}) 
\underline{\frac{dF(\bm{q})}{dt}\Big|_\text{int}}\right]\nonumber\\
& = \frac{2\pi |T|^2}{\hbar V}\sum_{\bm{q}, \bm{k},\gamma} \left[
\braket{\bm{k},\gamma|\hat{\bm{\mu}}\cdot \underline{\bm{\sigma}}|\bm{k},\gamma}
f(\bm{k},\gamma) -  F_1(\bm{q})\right]\nonumber\\
&\qquad \delta \bm{(}\epsilon^{\text{(n)}}(\bm{q}) -\epsilon (\bm{k},\gamma)\bm{)}\\
 &=  \frac{2\pi |T|^2}{\hbar V}\sum_{\bm{q}, \bm{k},\gamma} \Big\{
\braket{\bm{k},\gamma|\hat{\bm{\mu}}\cdot \underline{\bm{\sigma}}|\bm{k},\gamma}
\varphi (\bm{k},\gamma)\nonumber\\
&\quad -  \left[\mu^{\rm (n)}_{\rm s}(y=0)
+ \frac{2e^2\lambda^{\rm (n)}}{\sigma^{\rm (n)}}\frac{q_y}{q}
J_{\rm s}^{\rm (n)}(y=0)\right]
\Big\}\nonumber\\
& \cdot \delta \bm{(}\epsilon (\bm{k},\gamma) -\mu_0\bm{)}\cdot
\delta \bm{(}\epsilon^{\text{(n)}}(\bm{q}) -\mu_0 \bm{)}\\
 &=  \frac{2\pi |T|^2}{\hbar}V^{\text{(n)}}N^{\text{(n)}} N_0\nonumber\\
&\quad \cdot \left[\hat{\bm{\mu}}\cdot
\left\langle \bm{S} (\bm{k},\gamma)\varphi (\bm{k},\gamma)\right\rangle_{\text{2DFC}} 
- \frac{\mu^{\rm (n)}_{\rm s}(0)}{2}\right]
\label{eq: spin-density-from2to3}
\end{align}
where we made use of Eq.~\eqref{eq: spin transmission rate}.
Here the anisotropic term proportional to $q_y$ vanishes in the $\bm{q}$ summation.
The quantity~\eqref{eq: spin-density-from2to3} is equal to $J^{\rm (n)}_{\rm s}(y=0)$
polarized in the $\hat{\bm{\mu}}$ direction, which provides a condition
\begin{multline}
\frac{2\pi |T|^2}{\hbar}V^{\text{(n)}}N^{\text{(n)}} N_0
\cdot \left[\hat{\bm{\mu}}\cdot
\left\langle \bm{S} (\bm{k},\gamma)\varphi (\bm{k},\gamma)\right\rangle_{\text{2DFC}} 
- \frac{\mu^{\rm (n)}_{\rm s}(0)}{2}\right]\\
= J^{\rm (n)}_{\rm s}(0)~. 
\label{eq: A-BvsA+B} 
\end{multline}
This equality indicates that the term
$2 \hat{\bm{\mu}}\cdot\left\langle \bm{S} (\bm{k},\gamma)\varphi (\bm{k},\gamma)\right\rangle_{\text{2DFC}}$
serves as a spin accumulation at the 2D metal.
Here spin accumulation and spin current density at the interface is given as
$\mu^{\rm (n)}_{\rm s}(0) = 2(A+B)$ and $J^{\rm (n)}_{\rm s}(0) = (N_0/\tau_\text{3D})\cdot (A-B)$, where
we introduced a rate that measures spin diffusion in the 3D metal
\begin{equation}
\frac{1}{\tau_{\text{3D}}} \equiv \frac{\sigma^{\rm (n)}}{e^2 \ell_{\rm sf}N_0}
= \sqrt{\frac{2}{3\left(1+ \tau_\text{sf}/\tau_\text{s}\right)}}\cdot 
\frac{N^{\text{(n)}}(\mu_0) v^\text{(n)}_\text{F}}{N_0(\mu_0)}~.
\label{eq: tau3D explanation}
\end{equation}
The most right-hand side of Eq.~\eqref{eq: tau3D explanation}
is obtained by substituting Eqs.~\eqref{eq: def ele mean free path and spin diff length},
\eqref{eq: particle number density n},
\eqref{eq: 3D electric conductivity}
and by assuming $\mu_0 \simeq m (v^\text{(n)}_\text{F})^2/2$.
As $\tau_\text{sf}/\tau_\text{s}\gg 1$ when spin diffusion equation holds,
we find a relation $\tau_{\text{3D}} \propto \sqrt{\tau_\text{sf}/\tau_\text{s}}$,
which we have mentioned in Sect.~IIIB in the main text.

The equality~\eqref{eq: A-BvsA+B} is thus expressed as 
\begin{equation}
 \hat{\bm{\mu}}\cdot
\left\langle \bm{S} (\bm{k},\gamma)\varphi (\bm{k},\gamma)\right\rangle_{\text{2DFC}} 
- (A+B) = \frac{\tau_\text{t}}{\tau_\text{3D}}(A-B)~,
\label{eq: second boudary cond}
\end{equation}
which gives one of the conditions that $A$ and $B$ have to satisfy. 

\subsubsection*{(iii) The boundary condition at 
\texorpdfstring{$y=L$}{y=L}}
In the same way as the main text, we can express the boundary condition as  
$J_{\rm s}^{\rm (n)}(L) = 0,~J_{\rm s}^{\rm ext}$, in other words,
\begin{multline}
(A-B)\cosh \frac{L}{\ell_\text{sf}} - (A+B)\sinh \frac{L}{\ell_\text{sf}}\\
 = \frac{\tau_\text{3D}}{N_0}J_{\rm s}^{\rm (n)}(L) = 0,~\frac{\tau_\text{3D}J_{\rm s}^{\rm ext}}{N_0}
\label{eq:  bc-JcL} 
\end{multline}
The relation~\eqref{eq:  bc-JcL} is the other condition for $A$ and $B$. 

We thus obtain the parameters $A$ and $B$ from Eqs.~\eqref{eq: second boudary cond} and \eqref{eq:  bc-JcL} as 
\begin{multline}
\begin{pmatrix}
 A+B \\ A-B
\end{pmatrix} 
\\= {2K}
\begin{pmatrix}
 \cosh \frac{L}{\ell_\text{sf}} & -\frac{\tau_\text{t}}{\tau_\text{3D}} \\
-\sinh \frac{L}{\ell_\text{sf}} & 1
\end{pmatrix}
\begin{pmatrix}
\hat{\bm{\mu}}\cdot
\left\langle \bm{S} (\bm{k},\gamma)\varphi (\bm{k},\gamma)\right\rangle_{\text{2DFC}} \\
-\tau_\text{3D}J^\text{ext}_\text{s} /N_0
\end{pmatrix} 
\end{multline}
with $2K \equiv [\cosh L/\ell_\text{sf} + (\tau_\text{t}/\tau_\text{3D})\sinh L/\ell_\text{sf}]^{-1}$.

\subsection{\label{subsec: The Boltzmann equation for the 2D electron distribution function}
The Boltzmann equation for the 2D electron distribution function}

The Boltzmann collision integral due to the interface transmission into or out of the 2D metal for each band
$\displaystyle {\frac{df(\bm{k},\gamma)}{dt}\Big|_\text{int}}$, shown in Eq.~\eqref{eq: dfdt int F0F1},
is then derived as
\begin{align}
 & \frac{df(\bm{k},\gamma)}{dt}\Big|_\text{int}\nonumber\\
&= \frac{2}{\tau_\text{t}V^{\text{(n)}} N^\text{(n)}}
\sum_{\bm{q}} \left[
F_0(\bm{q})+ \hat{\bm{\mu}}\cdot \bm{S}(\bm{k},\gamma) F_1(\bm{q})  - f(\bm{k},\gamma)\right]\nonumber\\
&\qquad \cdot \delta \bm{(}\epsilon^{\text{(n)}}(\bm{q}) -\epsilon (\bm{k},\gamma)\bm{)}  \\
 &=  \frac{2}{\tau_\text{t}V^{\text{(n)}} N^\text{(n)}}
\sum_{\bm{q}} \Big\{(\bar{\mu}-\mu_0) - \varphi (\bm{k},\gamma) \nonumber\\
&+ \frac{1}{2}
\left[\mu^{\rm (n)}_{\rm s}(y=0)
+ \frac{2e^2\lambda^{\rm (n)}}{\sigma^{\rm (n)}}\frac{q_y}{q}
J_{\rm s}^{\rm (n)}(y=0)\right]\cdot [\hat{\bm{\mu}}\cdot \bm{S}(\bm{k},\gamma)]
\Big\}\nonumber\\
&\qquad \qquad\cdot \delta \bm{(}\epsilon (\bm{k},\gamma) -\mu_0\bm{)}
\delta \bm{(}\epsilon^{\text{(n)}}(\bm{q}) -\mu_0\bm{)}\\
 &= \frac{1}{\tau_\text{t}}\left\{
(\bar{\mu}-\mu_0) + (A+B)[\hat{\bm{\mu}}\cdot \bm{S}(\bm{k},\gamma)] -\varphi (\bm{k},\gamma)
\right\} \nonumber\\
&\qquad \cdot \delta \bm{(}\epsilon (\bm{k},\gamma) -\mu_0\bm{)}
\end{align}

The BTE in the 2D metal with applied electric field $\bm{E}$
\begin{equation}
 (-e)\bm{E}\cdot \bm{v}(\bm{k},\gamma)\frac{\partial f_0(\bm{k},\gamma)}{\partial \epsilon(\bm{k})}
=\frac{d f}{d t}\Big|_{\rm col}+\frac{d f}{d t}\Big|_{\rm int}
\end{equation}
with the impurity collision term $\frac{d f}{d t}\Big|_{\rm col}$ given in Eq.~\eqref{eq: collision integral as relaxation mat}
thus yields the following equation around the Fermi contours:
\begin{widetext}
\begin{multline}
 e\tau_\text{p} \bm{E}\cdot \bm{v}(\bm{k},\gamma) + 
 \frac{2K\tau_{\rm p}}{N_0}\cdot [J_{\rm s}^{\rm ext}\hat{\bm{\mu}}\cdot \bm{S}(\bm{k},\gamma)]\\
 = \sum_{\bm{k}',\gamma'}
 \Big(
 M_{\rm col}(\bm{k},\gamma,\bm{k}',\gamma')
 + \frac{\tau_\text{p}}{\tau_\text{t}}
 \Big\{ -\delta_{\bm{k},\bm{k}'}\delta_{\gamma,\gamma'}
 +\frac{\delta\bm{(}\epsilon(\bm{k},\gamma)-
 \epsilon(\bm{k}',\gamma')
 \bm{)}}{N_0V}
\left[1+K'\left(\hat{\bm{\mu}}\cdot \bm{S}(\bm{k},\gamma)\right)
\cdot 
\left(\hat{\bm{\mu}}\cdot \bm{S}(\bm{k}',\gamma')\right)
\right] \Big\}
 \Big)\varphi (\bm{k}',\gamma')
 \label{eq: eff BTE in 2D pre}
\end{multline}
\end{widetext}
with $K'\equiv 2K \cosh L/\ell_{\rm sf}$.

Before applying the electric field or spin current injection, the 2D metal itself has no net spin polarization.
It is thus unnatural that the relaxation matrix on the right-hand side of Eq.~\eqref{eq: eff BTE in 2D pre}
contains the information on spin-polarization direction in the 3D metal $\hat{\bm{\mu}}$.
We can exclude $\hat{\bm{\mu}}$ from the expression 
by using the self-consistent condition~\eqref{eq: 3D spin pol dir is 2D spin pol dir}
to the non-equilibrium spin polarization in the 2D metal as follows:
\begin{align}
& \frac{1}{N_0V}\sum_{\bm{k}',\gamma'} 
[\hat{\bm{\mu}}\cdot \bm{S}(\bm{k},\gamma)]\cdot [\hat{\bm{\mu}}\cdot \bm{S}(\bm{k}',\gamma')]
\varphi (\bm{k}',\gamma') \nonumber\\
&\qquad \qquad \qquad\cdot \delta\bm{(}\epsilon(\bm{k},\gamma)-
\epsilon(\bm{k}',\gamma')\bm{)}\nonumber\\
&= \frac{1}{N_0V}\sum_{\bm{k}',\gamma'} \sum_{\hat{\bm{n}} = \hat{\bm{\mu}}, \hat{\bm{\nu}}, \hat{\bm{\lambda}}}
[\hat{\bm{n}}\cdot \bm{S}(\bm{k},\gamma)]\cdot [\hat{\bm{n}}\cdot \bm{S}(\bm{k}',\gamma')]
\varphi (\bm{k}',\gamma') \nonumber\\
&\qquad\qquad \cdot \delta\bm{(}\epsilon(\bm{k},\gamma)-
\epsilon(\bm{k}',\gamma')\bm{)}~, 
\end{align}
where $\sum_{\hat{\bm{n}} = \hat{\bm{\mu}}, \hat{\bm{\nu}}, \hat{\bm{\lambda}}}
[\hat{\bm{n}}\cdot \bm{S}(\bm{k},\gamma)]\cdot [\hat{\bm{n}}\cdot \bm{S}(\bm{k}',\gamma')]
=  \bm{S}(\bm{k},\gamma)\cdot\bm{S}(\bm{k}',\gamma')$ holds.
The right-hand side of Eq.~\eqref{eq: eff BTE in 2D pre}
is thus expressed as $\sum_{\bm{k},\gamma} M_{\text{tot}} (\bm{k},\gamma,\bm{k}',\gamma')$
with $M_\text{tot} = M_\text{col} + M_\text{int}$, as is given in the main text.

\section{Analytical expressions of electron distributions for Edelstein effect and its inverse}
In this section, we provide analytical expression for the deviation of the 2D electron distribution 
$\varphi = \varphi_\text{EE}$ and $\varphi = \varphi_\text{IEE}$
that are solutions of
the effective Boltzmann transport equation at the interface for the EE and IEE, respectively.

\subsection{Direct Edelstein effect}
Let us see the linear equation
\begin{equation}
 b_\text{EE}(\bm{k},\gamma) =  \sum_{\bm{k}', \gamma'} 
M_{\text{tot}}(\bm{k},\gamma, \bm{k}',\gamma') \varphi (\bm{k}',\gamma') \label{eq: EE}
\end{equation}
with an inhomogeneous term
\begin{equation}
 b_\text{EE}(\bm{k},\gamma) = e\tau_{\text{p}} E v_z(\bm{k},\gamma)~.
\end{equation}
We find that the solution of that equation is expressed as a linear combination of
$v_z (\bm{k}, \gamma)$ and $S_z (\bm{k},\gamma) = \gamma \cos\Theta (\theta)$:
\begin{widetext}
 \begin{align}
 &\frac{\varphi_\text{EE}(\theta, \gamma)}{(-e)Ev_\text{F}\tau_{\text{p}}} 
 = \frac{v_z(\theta, \gamma)/v_\text{F}}{1+ \tau_{\text{p}}/\tau_\text{t}}
 - \frac{\frac{\tilde{\alpha}\sin\delta}{1+\tan\delta}\left(1 
 + \frac{\tau_{\text{p}}}{\tau_\text{t}}K'\right)\cdot \gamma\cos\Theta (\theta)}
 {\left(1 + \frac{\tau_{\text{p}}}{\tau_\text{t}}\right)\left[
 \frac{\tan\delta}{1 + \tan\delta} + \frac{\tau_{\text{p}}}{\tau_\text{t}}\left(
 1-\frac{K'}{1+\tan\delta}\right) \right]}\\
 &=  \frac{\tau_\text{a}\tan\delta}{\tau_{\text{p}}}\Big[
 \left(\sqrt{1 + \tilde{\alpha}^2A^2(\theta)} -\gamma \tilde{\alpha}A(\theta) \right)\cos\theta
 + \gamma \tilde{\alpha} \cos\delta \cos\Theta (\theta)\cdot 
 \frac{1 +\tan\delta}{1 + \tau_\text{b}/\tau_\text{a}}\Big]~.
 \end{align}
\end{widetext}
That solution also satisfies the charge neutrality condition
$\left\langle \varphi_\text{EE} \right\rangle_{\text{2DFC}} = 0$.

In a low transmission limit $\tau_\text{p}/\tau_\text{t}\to 0$,
the expression above becomes the distribution function of the Edelstein effect 
in the surface without transmission~\eqref{eq: sol of BTE of EE in 2D}.

\subsection{Inverse Edelstein effect}
We next consider the BTE
\begin{multline}
b_\text{IEE}(\bm{k},\gamma)\\
=\sum_{\bm{k}',\gamma'}\left[
M_\text{col}(\bm{k},\gamma,\bm{k}',\gamma')+M_{\rm int}(\bm{k},\gamma,\bm{k}',\gamma')\right]\varphi (\bm{k}',\gamma') 
\end{multline}
with an inhomogeneous term
\begin{equation}
b_\text{IEE}(\bm{k},\gamma)
={2K\tau_{\rm p} \gamma\cos\Theta(\bm{k})J_{\rm s}^{\rm ext}}/{N_0}~.
\end{equation}

We find that the solution of that equation is proportional to $S_z(\bm{k},\gamma) = \gamma\cos\Theta (\theta)$.
Indeed, as we have seen at the relaxation process in the surface 
(see Eqs.~\eqref{eq: BTE for relaxation w tau}--\eqref{eq: simpler eig prob relaxation}),
$\varphi (\theta, \gamma) \propto \gamma\cos\Theta (\theta)$ satisfies
\begin{equation}
\sum_{\bm{k}',\gamma'}
M_\text{col}(\bm{k},\gamma,\bm{k}',\gamma')\varphi (\bm{k}',\gamma')
= -\frac{\tan\delta}{1 + \tan\delta} \varphi (\bm{k},\gamma)~.
\end{equation}
In addition, we can show that $\varphi (\theta, \gamma) \propto \gamma\cos\Theta (\theta)$ satisfies
\begin{equation}
\sum_{\bm{k}',\gamma'}
M_\text{int}(\bm{k},\gamma,\bm{k}',\gamma')\varphi (\bm{k}',\gamma')
= \frac{\tau_{\text{p}}}{\tau_\text{t}}\left(\frac{K'}{1+\tan\delta}-1 \right)  \varphi (\bm{k},\gamma)~.
\end{equation}
In other words, $S_z(\bm{k},\gamma) = \gamma\cos\Theta (\theta)$ is an eigenvector of
the relaxation matrix $M_\text{tot} = M_\text{col}+ M_\text{int}$.

By comparison with both sides, we thus obtain the solution $\varphi_\text{IEE}(\theta, \gamma) = C\cdot \gamma\cos\Theta (\theta)$
with 
\begin{align}
 C &= \frac{2K \tau_{\text{p}}J^{\text{ext}}_\text{s} }{ N_0 \left[
\frac{-\tan\delta }{1+\tan\delta} + \frac{\tau_{\text{p}}}{\tau_\text{t}}\left(
\frac{K'}{1+\tan\delta}-1 \right)\right] } \\
 &=  - \frac{\tau_{\text{p}} J^{\text{ext}}_\text{s}}{N_0}\cdot 
\frac{1 + \tan\delta}{\tan\delta ( 1 + \tau_{\text{p}}/\tau_\text{t})}\nonumber\\
&\quad \cdot\left\{
\cosh\frac{L}{\ell_{\text{sf}}} + \left[
\frac{\tau_\text{t}}{\tau_\text{3D}} + \frac{\left(1/\tau_\text{p} + 1/\tau_\text{t}\right)^{-1}}{\tau_\text{3D} \tan\delta}
\right]\sinh \frac{L}{\ell_{\text{sf}}} \right\}^{-1}\\
 &= -\frac{\tau_\text{3D} J^\text{ext}_\text{s}}{N_0\sinh L/\ell_\text{sf}}
\cdot \frac{1 + \tan\delta}{1 + \tau_\text{b}/\tau_\text{a}}\cdot \gamma \cos\Theta (\bm{k})~.
\end{align}
That solution also satisfies the charge neutrality condition
$\left\langle \varphi_\text{IEE} \right\rangle_{\text{2DFC}} = 0$.

\section{Dependence on the Thickness of the three-dimensional metal}

\begin{figure*}
 \centering
\includegraphics[width=0.9\textwidth]{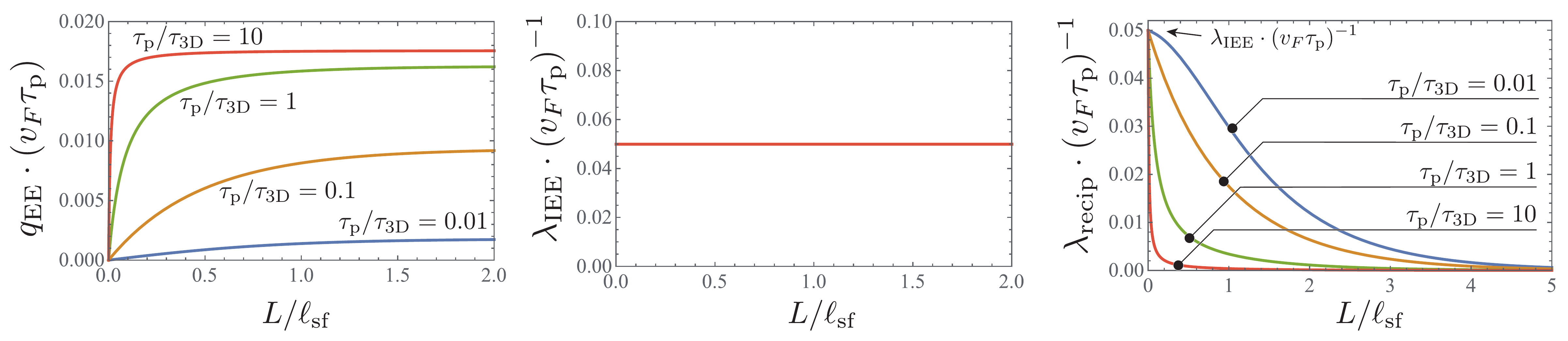}
\caption{Thickness dependence of the charge current--spin current interconversion ratios $q_\text{EE}$ (left),
$\lambda_\text{IEE}$ (middle), and reciprocity ratio $\lambda_\text{recip}$ (right).
The parameters are chosen as $(\alpha/v_\text{F}\hbar, \delta, \tau_\text{p}/\tau_\text{t})= (0.1, \pi/64, 1)$.
}
\label{fig: L dep}
\end{figure*}

In the main text, we stress that the finite thickness of the nonmagnetic metal attached on the chiral metal surface
$L$ is considered in our formalism.
Figures~\ref{fig: L dep} illustrate how the charge current--spin current interconversion ratios $q_\text{EE}$
and $\lambda_\text{IEE}$, and reciprocity ratio $\lambda_\text{recip}$ are dependent on $L$.
As we have described in the main text,
thickness dependence of these quantities are roughly estimated as
\begin{gather}
 q_\text{EE}\sim (\tau_\text{t} + \tau_\text{3D}\coth L/\ell_\text{sf})^{-1},\quad
\lambda_\text{IEE}\sim (1/\tau_\text{p} + 1/\tau_\text{t})^{-1},\\
\lambda_\text{recip}\sim 
\tau_\text{3D}/(\tau_\text{t}\sinh L/\ell_\text{sf} + \tau_\text{3D}\cosh L/\ell_\text{sf}), 
\end{gather}
which coincides with the behavior of each figure.

\begin{figure*}
\centering
\includegraphics[width=0.8\textwidth]{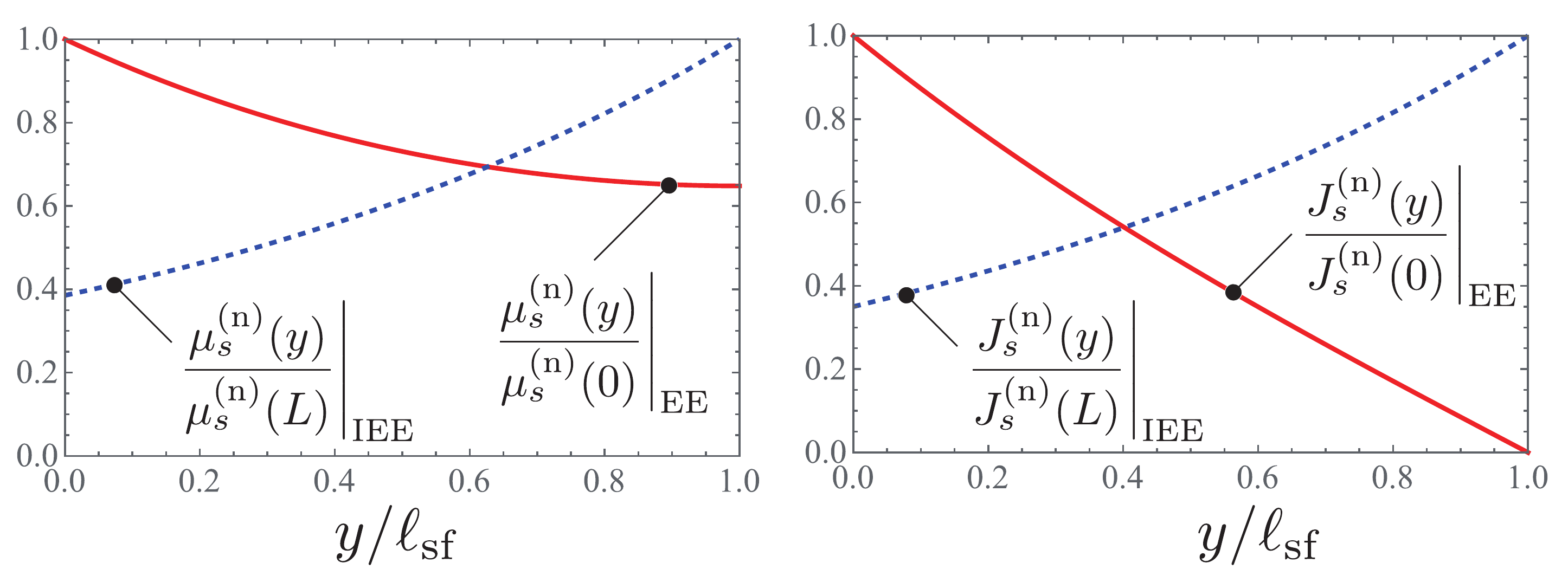}
\caption{Spatial distribution of spin accumulation and spin current density in the three-dimensional 
nonmagnetic metal.
The parameters are chosen as $(\delta, \tau_\text{p}/\tau_\text{t},\tau_\text{p}/\tau_\text{3D}, 
L/\ell_{\text{sf}})= (\pi/64, 1, 0.1, 1)$.
}
\label{fig: spin acc spin cur in 3D metal}
\end{figure*}

Before closing this section, we focus on 
the spin accumulation and spin current density spread in the 3D metal 
as a function of $y$, the direction of which is normal to the interface:
\begingroup
\allowdisplaybreaks
\begin{subequations}
 \begin{align}
 \left.\frac{\mu^{(n)}_\text{s}(y)}{\mu^{(n)}_\text{s}(0)}\right|_{\text{EE}}
 &= \frac{\cosh (L-y)/\ell_{\text{sf}}}{\cosh L/\ell_{\text{sf}}},\\
 \left.\frac{J^{(n)}_\text{s}(y)}{J^{(n)}_\text{s}(0)}\right|_{\text{EE}}&
 = \frac{\sinh (L-y)/\ell_{\text{sf}}}{\sinh L/\ell_{\text{sf}}}, \\
 \left.\frac{\mu^{(n)}_\text{s}(y)}{\mu^{(n)}_\text{s}(0)}\right|_{\text{IEE}}
 &= \cosh y/\ell_{\text{sf}} + \frac{\tau_\text{3D}}{\tau_\text{t}+\tau_\text{a}}\sinh y/\ell_{\text{sf}},\\
 \left.\frac{J^{(n)}_\text{s}(y)}{J^{(n)}_\text{s}(0)}\right|_{\text{IEE}}&
 = \cosh y/\ell_{\text{sf}} + \frac{\tau_\text{t}+ \tau_\text{a}}{\tau_\text{3D}}\sinh y/\ell_{\text{sf}}. 
\end{align}
\end{subequations}
\endgroup
That spatial distribution of spin is depicted in Figs.~\ref{fig: spin acc spin cur in 3D metal}.


\begin{thebibliography}{52}%
\makeatletter
\providecommand \@ifxundefined [1]{%
 \@ifx{#1\undefined}
}%
\providecommand \@ifnum [1]{%
 \ifnum #1\expandafter \@firstoftwo
 \else \expandafter \@secondoftwo
 \fi
}%
\providecommand \@ifx [1]{%
 \ifx #1\expandafter \@firstoftwo
 \else \expandafter \@secondoftwo
 \fi
}%
\providecommand \natexlab [1]{#1}%
\providecommand \enquote  [1]{``#1''}%
\providecommand \bibnamefont  [1]{#1}%
\providecommand \bibfnamefont [1]{#1}%
\providecommand \citenamefont [1]{#1}%
\providecommand \href@noop [0]{\@secondoftwo}%
\providecommand \href [0]{\begingroup \@sanitize@url \@href}%
\providecommand \@href[1]{\@@startlink{#1}\@@href}%
\providecommand \@@href[1]{\endgroup#1\@@endlink}%
\providecommand \@sanitize@url [0]{\catcode `\\12\catcode `\$12\catcode
  `\&12\catcode `\#12\catcode `\^12\catcode `\_12\catcode `\%12\relax}%
\providecommand \@@startlink[1]{}%
\providecommand \@@endlink[0]{}%
\providecommand \url  [0]{\begingroup\@sanitize@url \@url }%
\providecommand \@url [1]{\endgroup\@href {#1}{\urlprefix }}%
\providecommand \urlprefix  [0]{URL }%
\providecommand \Eprint [0]{\href }%
\providecommand \doibase [0]{https://doi.org/}%
\providecommand \selectlanguage [0]{\@gobble}%
\providecommand \bibinfo  [0]{\@secondoftwo}%
\providecommand \bibfield  [0]{\@secondoftwo}%
\providecommand \translation [1]{[#1]}%
\providecommand \BibitemOpen [0]{}%
\providecommand \bibitemStop [0]{}%
\providecommand \bibitemNoStop [0]{.\EOS\space}%
\providecommand \EOS [0]{\spacefactor3000\relax}%
\providecommand \BibitemShut  [1]{\csname bibitem#1\endcsname}%
\let\auto@bib@innerbib\@empty
\bibitem [{\citenamefont {Edelstein}(1990)}]{Edelstein1990}%
  \BibitemOpen
  \bibfield  {author} {\bibinfo {author} {\bibfnamefont {V.~M.}\ \bibnamefont
  {Edelstein}},\ }\bibfield  {title} {\bibinfo {title} {{Spin polarization of
  conduction electrons induced by electric current in two-dimensional
  asymmetric electron systems}},\ }\href
  {https://doi.org/10.1016/0038-1098(90)90963-C} {\bibfield  {journal}
  {\bibinfo  {journal} {Solid State Commun.}\ }\textbf {\bibinfo {volume}
  {73}},\ \bibinfo {pages} {233} (\bibinfo {year} {1990})}\BibitemShut
  {NoStop}%
\bibitem [{\citenamefont {Aronov}\ and\ \citenamefont {{Yu B.
  Lyanda-Geller}}(1989)}]{Aronov1989}%
  \BibitemOpen
  \bibfield  {author} {\bibinfo {author} {\bibfnamefont {A.~G.}\ \bibnamefont
  {Aronov}}\ and\ \bibinfo {author} {\bibnamefont {{Yu B. Lyanda-Geller}}},\
  }\bibfield  {title} {\bibinfo {title} {Nuclear electric resonance and
  orientation of carrier spins by an electric field},\ }\href@noop {}
  {\bibfield  {journal} {\bibinfo  {journal} {Zh. Eksp. Teor. Fiz.}\ }\textbf
  {\bibinfo {volume} {50}},\ \bibinfo {pages} {398} (\bibinfo {year} {1989})},\
  \bibinfo {note} {[JETP Lett. \textbf{50}, {431} ({1989})]}\BibitemShut
  {NoStop}%
\bibitem [{\citenamefont {Kato}\ \emph {et~al.}(2004)\citenamefont {Kato},
  \citenamefont {Myers}, \citenamefont {Gossard},\ and\ \citenamefont
  {Awschalom}}]{Kato2004a}%
  \BibitemOpen
  \bibfield  {author} {\bibinfo {author} {\bibfnamefont {Y.~K.}\ \bibnamefont
  {Kato}}, \bibinfo {author} {\bibfnamefont {R.~C.}\ \bibnamefont {Myers}},
  \bibinfo {author} {\bibfnamefont {A.~C.}\ \bibnamefont {Gossard}},\ and\
  \bibinfo {author} {\bibfnamefont {D.~D.}\ \bibnamefont {Awschalom}},\
  }\bibfield  {title} {\bibinfo {title} {{Current-induced spin polarization in
  strained semiconductors}},\ }\href
  {https://doi.org/10.1103/PhysRevLett.93.176601} {\bibfield  {journal}
  {\bibinfo  {journal} {Phys. Rev. Lett.}\ }\textbf {\bibinfo {volume} {93}},\
  \bibinfo {pages} {176601} (\bibinfo {year} {2004})}\BibitemShut {NoStop}%
\bibitem [{\citenamefont {Ganichev}\ \emph {et~al.}(2002)\citenamefont
  {Ganichev}, \citenamefont {Ivchenko}, \citenamefont {Bel'Kov}, \citenamefont
  {Tarasenko}, \citenamefont {Sollinger}, \citenamefont {Weiss}, \citenamefont
  {Wegscheider},\ and\ \citenamefont {Prettl}}]{Ganichev2002}%
  \BibitemOpen
  \bibfield  {author} {\bibinfo {author} {\bibfnamefont {S.~D.}\ \bibnamefont
  {Ganichev}}, \bibinfo {author} {\bibfnamefont {E.~L.}\ \bibnamefont
  {Ivchenko}}, \bibinfo {author} {\bibfnamefont {V.~V.}\ \bibnamefont
  {Bel'Kov}}, \bibinfo {author} {\bibfnamefont {S.~A.}\ \bibnamefont
  {Tarasenko}}, \bibinfo {author} {\bibfnamefont {M.}~\bibnamefont
  {Sollinger}}, \bibinfo {author} {\bibfnamefont {D.}~\bibnamefont {Weiss}},
  \bibinfo {author} {\bibfnamefont {W.}~\bibnamefont {Wegscheider}},\ and\
  \bibinfo {author} {\bibfnamefont {W.}~\bibnamefont {Prettl}},\ }\bibfield
  {title} {\bibinfo {title} {Spin-galvanic effect},\ }\href
  {https://doi.org/10.1038/417153a} {\bibfield  {journal} {\bibinfo  {journal}
  {Nature (London)}\ }\textbf {\bibinfo {volume} {417}},\ \bibinfo {pages}
  {153} (\bibinfo {year} {2002})}\BibitemShut {NoStop}%
\bibitem [{\citenamefont {{Ka Shen}}\ \emph {et~al.}(2014)\citenamefont {{Ka
  Shen}}, \citenamefont {Vignale},\ and\ \citenamefont {Raimondi}}]{Shen2014}%
  \BibitemOpen
  \bibfield  {author} {\bibinfo {author} {\bibnamefont {{Ka Shen}}}, \bibinfo
  {author} {\bibfnamefont {G.}~\bibnamefont {Vignale}},\ and\ \bibinfo {author}
  {\bibfnamefont {R.}~\bibnamefont {Raimondi}},\ }\bibfield  {title} {\bibinfo
  {title} {{Microscopic theory of the inverse Edelstein effect}},\ }\href
  {https://doi.org/10.1103/PhysRevLett.112.096601} {\bibfield  {journal}
  {\bibinfo  {journal} {Phys. Rev. Lett.}\ }\textbf {\bibinfo {volume} {112}},\
  \bibinfo {pages} {096601} (\bibinfo {year} {2014})}\BibitemShut {NoStop}%
\bibitem [{\citenamefont {Silsbee}(2004)}]{Silsbee2004}%
  \BibitemOpen
  \bibfield  {author} {\bibinfo {author} {\bibfnamefont {R.~H.}\ \bibnamefont
  {Silsbee}},\ }\bibfield  {title} {\bibinfo {title} {{Spin-orbit induced
  coupling of charge current and spin polarization}},\ }\href
  {https://doi.org/10.1088/0953-8984/16/7/R02} {\bibfield  {journal} {\bibinfo
  {journal} {J. Phys.: Condens. Matter}\ }\textbf {\bibinfo {volume} {16}},\
  \bibinfo {pages} {R179} (\bibinfo {year} {2004})}\BibitemShut {NoStop}%
\bibitem [{\citenamefont {Gambardella}\ and\ \citenamefont
  {Miron}(2011)}]{Gambardella2011}%
  \BibitemOpen
  \bibfield  {author} {\bibinfo {author} {\bibfnamefont {P.}~\bibnamefont
  {Gambardella}}\ and\ \bibinfo {author} {\bibfnamefont {I.~M.}\ \bibnamefont
  {Miron}},\ }\bibfield  {title} {\bibinfo {title} {{Current-induced spin-orbit
  torques}},\ }\href {https://doi.org/10.1098/rsta.2010.0336} {\bibfield
  {journal} {\bibinfo  {journal} {Philos. Trans. R. Soc. A}\ }\textbf {\bibinfo
  {volume} {369}},\ \bibinfo {pages} {3175} (\bibinfo {year}
  {2011})}\BibitemShut {NoStop}%
\bibitem [{\citenamefont {Zhang}\ and\ \citenamefont {Fert}(2016)}]{Zhang2016}%
  \BibitemOpen
  \bibfield  {author} {\bibinfo {author} {\bibfnamefont {S.}~\bibnamefont
  {Zhang}}\ and\ \bibinfo {author} {\bibfnamefont {A.}~\bibnamefont {Fert}},\
  }\bibfield  {title} {\bibinfo {title} {{Conversion between spin and charge
  currents with topological insulators}},\ }\href
  {https://doi.org/10.1103/PhysRevB.94.184423} {\bibfield  {journal} {\bibinfo
  {journal} {Phys. Rev. B}\ }\textbf {\bibinfo {volume} {94}},\ \bibinfo
  {pages} {184423} (\bibinfo {year} {2016})}\BibitemShut {NoStop}%
\bibitem [{\citenamefont {Dey}\ \emph {et~al.}(2018)\citenamefont {Dey},
  \citenamefont {Prasad}, \citenamefont {Register},\ and\ \citenamefont
  {Banerjee}}]{Dey2018}%
  \BibitemOpen
  \bibfield  {author} {\bibinfo {author} {\bibfnamefont {R.}~\bibnamefont
  {Dey}}, \bibinfo {author} {\bibfnamefont {N.}~\bibnamefont {Prasad}},
  \bibinfo {author} {\bibfnamefont {L.~F.}\ \bibnamefont {Register}},\ and\
  \bibinfo {author} {\bibfnamefont {S.~K.}\ \bibnamefont {Banerjee}},\
  }\bibfield  {title} {\bibinfo {title} {{Conversion of spin current into
  charge current in a topological insulator: Role of the interface}},\ }\href
  {https://doi.org/10.1103/PhysRevB.97.174406} {\bibfield  {journal} {\bibinfo
  {journal} {Phys. Rev. B}\ }\textbf {\bibinfo {volume} {97}},\ \bibinfo
  {pages} {174406} (\bibinfo {year} {2018})}\BibitemShut {NoStop}%
\bibitem [{\citenamefont {Rojas-S\'anchez}\ \emph {et~al.}(2013)\citenamefont
  {Rojas-S\'anchez}, \citenamefont {Vila}, \citenamefont {Desfonds},
  \citenamefont {Gambarelli}, \citenamefont {Attan{\'{e}}}, \citenamefont {{De
  Teresa}}, \citenamefont {Mag{\'{e}}n},\ and\ \citenamefont
  {Fert}}]{Sanchez2013}%
  \BibitemOpen
  \bibfield  {author} {\bibinfo {author} {\bibfnamefont {J.~C.}\ \bibnamefont
  {Rojas-S\'anchez}}, \bibinfo {author} {\bibfnamefont {L.}~\bibnamefont
  {Vila}}, \bibinfo {author} {\bibfnamefont {G.}~\bibnamefont {Desfonds}},
  \bibinfo {author} {\bibfnamefont {S.}~\bibnamefont {Gambarelli}}, \bibinfo
  {author} {\bibfnamefont {J.~P.}\ \bibnamefont {Attan{\'{e}}}}, \bibinfo
  {author} {\bibfnamefont {J.~M.}\ \bibnamefont {{De Teresa}}}, \bibinfo
  {author} {\bibfnamefont {C.}~\bibnamefont {Mag{\'{e}}n}},\ and\ \bibinfo
  {author} {\bibfnamefont {A.}~\bibnamefont {Fert}},\ }\bibfield  {title}
  {\bibinfo {title} {{Spin-to-charge conversion using Rashba coupling at the
  interface between non-magnetic materials}},\ }\href
  {https://doi.org/10.1038/ncomms3944} {\bibfield  {journal} {\bibinfo
  {journal} {Nat. Commun.}\ }\textbf {\bibinfo {volume} {4}},\ \bibinfo {pages}
  {2944} (\bibinfo {year} {2013})}\BibitemShut {NoStop}%
\bibitem [{\citenamefont {Zhang}\ \emph {et~al.}(2015)\citenamefont {Zhang},
  \citenamefont {Yamamoto}, \citenamefont {Gu}, \citenamefont {Li},
  \citenamefont {Maekawa}, \citenamefont {Fukaya},\ and\ \citenamefont
  {Kawasuso}}]{Zhang2015a}%
  \BibitemOpen
  \bibfield  {author} {\bibinfo {author} {\bibfnamefont {H.~J.}\ \bibnamefont
  {Zhang}}, \bibinfo {author} {\bibfnamefont {S.}~\bibnamefont {Yamamoto}},
  \bibinfo {author} {\bibfnamefont {B.}~\bibnamefont {Gu}}, \bibinfo {author}
  {\bibfnamefont {H.}~\bibnamefont {Li}}, \bibinfo {author} {\bibfnamefont
  {M.}~\bibnamefont {Maekawa}}, \bibinfo {author} {\bibfnamefont
  {Y.}~\bibnamefont {Fukaya}},\ and\ \bibinfo {author} {\bibfnamefont
  {A.}~\bibnamefont {Kawasuso}},\ }\bibfield  {title} {\bibinfo {title}
  {{Charge-to-spin conversion and spin diffusion in Bi/Ag bilayers observed by
  spin-polarized positron beam}},\ }\href
  {https://doi.org/10.1103/PhysRevLett.114.166602} {\bibfield  {journal}
  {\bibinfo  {journal} {Phys. Rev. Lett.}\ }\textbf {\bibinfo {volume} {114}},\
  \bibinfo {pages} {166602} (\bibinfo {year} {2015})}\BibitemShut {NoStop}%
\bibitem [{\citenamefont {Lesne}\ \emph {et~al.}(2016)\citenamefont {Lesne},
  \citenamefont {{Yu Fu}}, \citenamefont {Oyarzun}, \citenamefont
  {Rojas-S{\'a}nchez}, \citenamefont {Vaz}, \citenamefont {Naganuma},
  \citenamefont {Sicoli}, \citenamefont {Attan{\'e}}, \citenamefont {Jamet},
  \citenamefont {Jacquet}, \citenamefont {George}, \citenamefont
  {Barth{\'e}l{\'e}my}, \citenamefont {Jaffr{\`e}s}, \citenamefont {Fert},
  \citenamefont {Bibes},\ and\ \citenamefont {Vila}}]{Lesne2016}%
  \BibitemOpen
  \bibfield  {author} {\bibinfo {author} {\bibfnamefont {E.}~\bibnamefont
  {Lesne}}, \bibinfo {author} {\bibnamefont {{Yu Fu}}}, \bibinfo {author}
  {\bibfnamefont {S.}~\bibnamefont {Oyarzun}}, \bibinfo {author} {\bibfnamefont
  {J.~C.}\ \bibnamefont {Rojas-S{\'a}nchez}}, \bibinfo {author} {\bibfnamefont
  {D.~C.}\ \bibnamefont {Vaz}}, \bibinfo {author} {\bibfnamefont
  {H.}~\bibnamefont {Naganuma}}, \bibinfo {author} {\bibfnamefont
  {G.}~\bibnamefont {Sicoli}}, \bibinfo {author} {\bibfnamefont {J.-P.}\
  \bibnamefont {Attan{\'e}}}, \bibinfo {author} {\bibfnamefont
  {M.}~\bibnamefont {Jamet}}, \bibinfo {author} {\bibfnamefont
  {E.}~\bibnamefont {Jacquet}}, \bibinfo {author} {\bibfnamefont {J.-M.}\
  \bibnamefont {George}}, \bibinfo {author} {\bibfnamefont {A.}~\bibnamefont
  {Barth{\'e}l{\'e}my}}, \bibinfo {author} {\bibfnamefont {H.}~\bibnamefont
  {Jaffr{\`e}s}}, \bibinfo {author} {\bibfnamefont {A.}~\bibnamefont {Fert}},
  \bibinfo {author} {\bibfnamefont {M.}~\bibnamefont {Bibes}},\ and\ \bibinfo
  {author} {\bibfnamefont {L.}~\bibnamefont {Vila}},\ }\bibfield  {title}
  {\bibinfo {title} {{Highly efficient and tunable spin-to-charge conversion
  through Rashba coupling at oxide interfaces}},\ }\href
  {https://doi.org/10.1038/nmat4726} {\bibfield  {journal} {\bibinfo  {journal}
  {Nat. Mater.}\ }\textbf {\bibinfo {volume} {15}},\ \bibinfo {pages} {1261}
  (\bibinfo {year} {2016})}\BibitemShut {NoStop}%
\bibitem [{\citenamefont {Shiomi}\ \emph {et~al.}(2014)\citenamefont {Shiomi},
  \citenamefont {Nomura}, \citenamefont {Kajiwara}, \citenamefont {Eto},
  \citenamefont {Novak}, \citenamefont {Segawa}, \citenamefont {Ando},\ and\
  \citenamefont {Saitoh}}]{Shiomi2014}%
  \BibitemOpen
  \bibfield  {author} {\bibinfo {author} {\bibfnamefont {Y.}~\bibnamefont
  {Shiomi}}, \bibinfo {author} {\bibfnamefont {K.}~\bibnamefont {Nomura}},
  \bibinfo {author} {\bibfnamefont {Y.}~\bibnamefont {Kajiwara}}, \bibinfo
  {author} {\bibfnamefont {K.}~\bibnamefont {Eto}}, \bibinfo {author}
  {\bibfnamefont {M.}~\bibnamefont {Novak}}, \bibinfo {author} {\bibfnamefont
  {K.}~\bibnamefont {Segawa}}, \bibinfo {author} {\bibfnamefont
  {Y.}~\bibnamefont {Ando}},\ and\ \bibinfo {author} {\bibfnamefont
  {E.}~\bibnamefont {Saitoh}},\ }\bibfield  {title} {\bibinfo {title}
  {{Spin-electricity conversion induced by spin injection into topological
  insulators}},\ }\href {https://doi.org/10.1103/PhysRevLett.113.196601}
  {\bibfield  {journal} {\bibinfo  {journal} {Phys. Rev. Lett.}\ }\textbf
  {\bibinfo {volume} {113}},\ \bibinfo {pages} {196601} (\bibinfo {year}
  {2014})}\BibitemShut {NoStop}%
\bibitem [{\citenamefont {Rojas-S\'anchez}\ \emph {et~al.}(2016)\citenamefont
  {Rojas-S\'anchez}, \citenamefont {Oyarz\'un}, \citenamefont {Fu},
  \citenamefont {Marty}, \citenamefont {Vergnaud}, \citenamefont {Gambarelli},
  \citenamefont {Vila}, \citenamefont {Jamet}, \citenamefont {Ohtsubo},
  \citenamefont {Taleb-Ibrahimi}, \citenamefont {Le~F\`evre}, \citenamefont
  {Bertran}, \citenamefont {Reyren}, \citenamefont {George},\ and\
  \citenamefont {Fert}}]{Sanchez2016}%
  \BibitemOpen
  \bibfield  {author} {\bibinfo {author} {\bibfnamefont {J.-C.}\ \bibnamefont
  {Rojas-S\'anchez}}, \bibinfo {author} {\bibfnamefont {S.}~\bibnamefont
  {Oyarz\'un}}, \bibinfo {author} {\bibfnamefont {Y.}~\bibnamefont {Fu}},
  \bibinfo {author} {\bibfnamefont {A.}~\bibnamefont {Marty}}, \bibinfo
  {author} {\bibfnamefont {C.}~\bibnamefont {Vergnaud}}, \bibinfo {author}
  {\bibfnamefont {S.}~\bibnamefont {Gambarelli}}, \bibinfo {author}
  {\bibfnamefont {L.}~\bibnamefont {Vila}}, \bibinfo {author} {\bibfnamefont
  {M.}~\bibnamefont {Jamet}}, \bibinfo {author} {\bibfnamefont
  {Y.}~\bibnamefont {Ohtsubo}}, \bibinfo {author} {\bibfnamefont
  {A.}~\bibnamefont {Taleb-Ibrahimi}}, \bibinfo {author} {\bibfnamefont
  {P.}~\bibnamefont {Le~F\`evre}}, \bibinfo {author} {\bibfnamefont
  {F.}~\bibnamefont {Bertran}}, \bibinfo {author} {\bibfnamefont
  {N.}~\bibnamefont {Reyren}}, \bibinfo {author} {\bibfnamefont {J.-M.}\
  \bibnamefont {George}},\ and\ \bibinfo {author} {\bibfnamefont
  {A.}~\bibnamefont {Fert}},\ }\bibfield  {title} {\bibinfo {title} {{Spin to
  charge conversion at room temperature by spin pumping into a new type of
  topological insulator: $\ensuremath{\alpha}$-Sn films}},\ }\href
  {https://doi.org/10.1103/PhysRevLett.116.096602} {\bibfield  {journal}
  {\bibinfo  {journal} {Phys. Rev. Lett.}\ }\textbf {\bibinfo {volume} {116}},\
  \bibinfo {pages} {096602} (\bibinfo {year} {2016})}\BibitemShut {NoStop}%
\bibitem [{\citenamefont {Inui}\ \emph {et~al.}(2020)\citenamefont {Inui},
  \citenamefont {Aoki}, \citenamefont {Nishiue}, \citenamefont {Shiota},
  \citenamefont {Kousaka}, \citenamefont {Shishido}, \citenamefont {Hirobe},
  \citenamefont {Suda}, \citenamefont {Ohe}, \citenamefont {Kishine},
  \citenamefont {Yamamoto},\ and\ \citenamefont {Togawa}}]{Inui2020}%
  \BibitemOpen
  \bibfield  {author} {\bibinfo {author} {\bibfnamefont {A.}~\bibnamefont
  {Inui}}, \bibinfo {author} {\bibfnamefont {R.}~\bibnamefont {Aoki}}, \bibinfo
  {author} {\bibfnamefont {Y.}~\bibnamefont {Nishiue}}, \bibinfo {author}
  {\bibfnamefont {K.}~\bibnamefont {Shiota}}, \bibinfo {author} {\bibfnamefont
  {Y.}~\bibnamefont {Kousaka}}, \bibinfo {author} {\bibfnamefont
  {H.}~\bibnamefont {Shishido}}, \bibinfo {author} {\bibfnamefont
  {D.}~\bibnamefont {Hirobe}}, \bibinfo {author} {\bibfnamefont
  {M.}~\bibnamefont {Suda}}, \bibinfo {author} {\bibfnamefont {J.-i.}\
  \bibnamefont {Ohe}}, \bibinfo {author} {\bibfnamefont {J.-i.}\ \bibnamefont
  {Kishine}}, \bibinfo {author} {\bibfnamefont {H.~M.}\ \bibnamefont
  {Yamamoto}},\ and\ \bibinfo {author} {\bibfnamefont {Y.}~\bibnamefont
  {Togawa}},\ }\bibfield  {title} {\bibinfo {title} {{Chirality-induced
  spin-polarized state of a chiral crystal CrNb$_3$S$_6$}},\ }\href
  {https://doi.org/10.1103/PhysRevLett.124.166602} {\bibfield  {journal}
  {\bibinfo  {journal} {Phys. Rev. Lett.}\ }\textbf {\bibinfo {volume} {124}},\
  \bibinfo {pages} {166602} (\bibinfo {year} {2020})}\BibitemShut {NoStop}%
\bibitem [{\citenamefont {Nabei}\ \emph {et~al.}(2020)\citenamefont {Nabei},
  \citenamefont {Hirobe}, \citenamefont {Shimamoto}, \citenamefont {Shiota},
  \citenamefont {Inui}, \citenamefont {Kousaka}, \citenamefont {Togawa},\ and\
  \citenamefont {Yamamoto}}]{Nabei2020}%
  \BibitemOpen
  \bibfield  {author} {\bibinfo {author} {\bibfnamefont {Y.}~\bibnamefont
  {Nabei}}, \bibinfo {author} {\bibfnamefont {D.}~\bibnamefont {Hirobe}},
  \bibinfo {author} {\bibfnamefont {Y.}~\bibnamefont {Shimamoto}}, \bibinfo
  {author} {\bibfnamefont {K.}~\bibnamefont {Shiota}}, \bibinfo {author}
  {\bibfnamefont {A.}~\bibnamefont {Inui}}, \bibinfo {author} {\bibfnamefont
  {Y.}~\bibnamefont {Kousaka}}, \bibinfo {author} {\bibfnamefont
  {Y.}~\bibnamefont {Togawa}},\ and\ \bibinfo {author} {\bibfnamefont {H.~M.}\
  \bibnamefont {Yamamoto}},\ }\bibfield  {title} {\bibinfo {title}
  {{Current-induced bulk magnetization of a chiral crystal CrNb$_3$S$_6$}},\
  }\href {https://doi.org/10.1063/5.0017882} {\bibfield  {journal} {\bibinfo
  {journal} {Appl. Phys. Lett.}\ }\textbf {\bibinfo {volume} {117}},\ \bibinfo
  {pages} {052408} (\bibinfo {year} {2020})}\BibitemShut {NoStop}%
\bibitem [{\citenamefont {Shiota}\ \emph {et~al.}(2021)\citenamefont {Shiota},
  \citenamefont {Inui}, \citenamefont {Hosaka}, \citenamefont {Amano},
  \citenamefont {\ifmmode~\bar{O}\else \={O}\fi{}nuki}, \citenamefont {Hedo},
  \citenamefont {Nakama}, \citenamefont {Hirobe}, \citenamefont {Ohe},
  \citenamefont {Kishine}, \citenamefont {Yamamoto}, \citenamefont {Shishido},\
  and\ \citenamefont {Togawa}}]{Shiota2021}%
  \BibitemOpen
  \bibfield  {author} {\bibinfo {author} {\bibfnamefont {K.}~\bibnamefont
  {Shiota}}, \bibinfo {author} {\bibfnamefont {A.}~\bibnamefont {Inui}},
  \bibinfo {author} {\bibfnamefont {Y.}~\bibnamefont {Hosaka}}, \bibinfo
  {author} {\bibfnamefont {R.}~\bibnamefont {Amano}}, \bibinfo {author}
  {\bibfnamefont {Y.}~\bibnamefont {\ifmmode~\bar{O}\else \={O}\fi{}nuki}},
  \bibinfo {author} {\bibfnamefont {M.}~\bibnamefont {Hedo}}, \bibinfo {author}
  {\bibfnamefont {T.}~\bibnamefont {Nakama}}, \bibinfo {author} {\bibfnamefont
  {D.}~\bibnamefont {Hirobe}}, \bibinfo {author} {\bibfnamefont {J.-i.}\
  \bibnamefont {Ohe}}, \bibinfo {author} {\bibfnamefont {J.-i.}\ \bibnamefont
  {Kishine}}, \bibinfo {author} {\bibfnamefont {H.~M.}\ \bibnamefont
  {Yamamoto}}, \bibinfo {author} {\bibfnamefont {H.}~\bibnamefont {Shishido}},\
  and\ \bibinfo {author} {\bibfnamefont {Y.}~\bibnamefont {Togawa}},\
  }\bibfield  {title} {\bibinfo {title} {{Chirality-induced spin polarization
  over macroscopic distances in chiral disilicide crystals}},\ }\href
  {https://doi.org/10.1103/PhysRevLett.127.126602} {\bibfield  {journal}
  {\bibinfo  {journal} {Phys. Rev. Lett.}\ }\textbf {\bibinfo {volume} {127}},\
  \bibinfo {pages} {126602} (\bibinfo {year} {2021})}\BibitemShut {NoStop}%
\bibitem [{\citenamefont {Shishido}\ \emph {et~al.}(2021)\citenamefont
  {Shishido}, \citenamefont {Sakai}, \citenamefont {Hosaka},\ and\
  \citenamefont {Togawa}}]{Shishido2021}%
  \BibitemOpen
  \bibfield  {author} {\bibinfo {author} {\bibfnamefont {H.}~\bibnamefont
  {Shishido}}, \bibinfo {author} {\bibfnamefont {R.}~\bibnamefont {Sakai}},
  \bibinfo {author} {\bibfnamefont {Y.}~\bibnamefont {Hosaka}},\ and\ \bibinfo
  {author} {\bibfnamefont {Y.}~\bibnamefont {Togawa}},\ }\bibfield  {title}
  {\bibinfo {title} {{Detection of chirality-induced spin polarization over
  millimeters in polycrystalline bulk samples of chiral disilicides NbSi$_2$
  and TaSi$_2$}},\ }\href {https://doi.org/10.1063/5.0074293} {\bibfield
  {journal} {\bibinfo  {journal} {Appl. Phys. Lett.}\ }\textbf {\bibinfo
  {volume} {119}},\ \bibinfo {pages} {182403} (\bibinfo {year}
  {2021})}\BibitemShut {NoStop}%
\bibitem [{Note1()}]{Note1}%
  \BibitemOpen
  \bibinfo {note} {Note that the observed spin polarization unique to the
  chiral metals cannot be explained as a linear spin Hall effect, as described
  in Ref.~\cite {Roy2022}. The spin Hall conductivity relates spin current and
  electric field, which are odd under spatial inversion. It follows that the
  spin Hall conductivity itself is independent of whether the spatial inversion
  is included in the point group or not; in particular, no linear spin Hall
  effect is unique to chiral crystals. \par More precisely, in the point group
  622 without magnetic orders, spin Hall conductivity vanishes when the
  electric field and spin polarization direction are parallel~\cite
  {Seemann2015}. The spin Hall effect is thus unrelated to the observed spin
  polarization that is parallel to the applied electric field.}\BibitemShut
  {Stop}%
\bibitem [{\citenamefont {Furukawa}\ \emph {et~al.}(2021)\citenamefont
  {Furukawa}, \citenamefont {Watanabe}, \citenamefont {Ogasawara},
  \citenamefont {Kobayashi},\ and\ \citenamefont {Itou}}]{Furukawa2021}%
  \BibitemOpen
  \bibfield  {author} {\bibinfo {author} {\bibfnamefont {T.}~\bibnamefont
  {Furukawa}}, \bibinfo {author} {\bibfnamefont {Y.}~\bibnamefont {Watanabe}},
  \bibinfo {author} {\bibfnamefont {N.}~\bibnamefont {Ogasawara}}, \bibinfo
  {author} {\bibfnamefont {K.}~\bibnamefont {Kobayashi}},\ and\ \bibinfo
  {author} {\bibfnamefont {T.}~\bibnamefont {Itou}},\ }\bibfield  {title}
  {\bibinfo {title} {{Current-induced magnetization caused by crystal chirality
  in nonmagnetic elemental tellurium}},\ }\href
  {https://doi.org/10.1103/physrevresearch.3.023111} {\bibfield  {journal}
  {\bibinfo  {journal} {Phys. Rev. Res.}\ }\textbf {\bibinfo {volume} {3}},\
  \bibinfo {pages} {023111} (\bibinfo {year} {2021})}\BibitemShut {NoStop}%
\bibitem [{\citenamefont {Yoda}\ \emph {et~al.}(2015)\citenamefont {Yoda},
  \citenamefont {Yokoyama},\ and\ \citenamefont {Murakami}}]{Yoda2015}%
  \BibitemOpen
  \bibfield  {author} {\bibinfo {author} {\bibfnamefont {T.}~\bibnamefont
  {Yoda}}, \bibinfo {author} {\bibfnamefont {T.}~\bibnamefont {Yokoyama}},\
  and\ \bibinfo {author} {\bibfnamefont {S.}~\bibnamefont {Murakami}},\
  }\bibfield  {title} {\bibinfo {title} {{Current-induced orbital and spin
  magnetizations in crystals with helical structure}},\ }\href
  {https://doi.org/10.1038/srep12024} {\bibfield  {journal} {\bibinfo
  {journal} {Sci. Rep.}\ }\textbf {\bibinfo {volume} {5}},\ \bibinfo {pages}
  {12024} (\bibinfo {year} {2015})}\BibitemShut {NoStop}%
\bibitem [{\citenamefont {Frigeri}(2005)}]{Frigeri2005}%
  \BibitemOpen
  \bibfield  {author} {\bibinfo {author} {\bibfnamefont {P.~A.}\ \bibnamefont
  {Frigeri}},\ }\emph {\bibinfo {title} {Superconductivity in crystals without
  an inversion center}},\ \href@noop {} {Ph.D. thesis},\ \bibinfo  {school}
  {ETH-Z{\"u}rich} (\bibinfo {year} {2005})\BibitemShut {NoStop}%
\bibitem [{\citenamefont {Furukawa}\ \emph {et~al.}(2017)\citenamefont
  {Furukawa}, \citenamefont {Shimokawa}, \citenamefont {Kobayashi},\ and\
  \citenamefont {Itou}}]{Furukawa2017}%
  \BibitemOpen
  \bibfield  {author} {\bibinfo {author} {\bibfnamefont {T.}~\bibnamefont
  {Furukawa}}, \bibinfo {author} {\bibfnamefont {Y.}~\bibnamefont {Shimokawa}},
  \bibinfo {author} {\bibfnamefont {K.}~\bibnamefont {Kobayashi}},\ and\
  \bibinfo {author} {\bibfnamefont {T.}~\bibnamefont {Itou}},\ }\bibfield
  {title} {\bibinfo {title} {{Observation of current-induced bulk magnetization
  in elemental tellurium}},\ }\href
  {https://doi.org/10.1038/s41467-017-01093-3} {\bibfield  {journal} {\bibinfo
  {journal} {Nat. Commun.}\ }\textbf {\bibinfo {volume} {8}},\ \bibinfo {pages}
  {954} (\bibinfo {year} {2017})}\BibitemShut {NoStop}%
\bibitem [{\citenamefont {Tatara}(2022)}]{Tatara2022a}%
  \BibitemOpen
  \bibfield  {author} {\bibinfo {author} {\bibfnamefont {G.}~\bibnamefont
  {Tatara}},\ }\bibfield  {title} {\bibinfo {title} {{Nonlocality of
  electrically-induced spin accumulation in chiral metals}},\ }\href
  {https://doi.org/10.7566/JPSJ.91.073701} {\bibfield  {journal} {\bibinfo
  {journal} {J. Phys. Soc. Jpn.}\ }\textbf {\bibinfo {volume} {91}},\ \bibinfo
  {pages} {073701} (\bibinfo {year} {2022})}\BibitemShut {NoStop}%
\bibitem [{\citenamefont {Roy}\ \emph {et~al.}(2022)\citenamefont {Roy},
  \citenamefont {Cerasoli}, \citenamefont {Jayaraj}, \citenamefont {Tenzin},
  \citenamefont {Nardelli},\ and\ \citenamefont
  {S{\l}awi{\'{n}}ska}}]{Roy2022}%
  \BibitemOpen
  \bibfield  {author} {\bibinfo {author} {\bibfnamefont {A.}~\bibnamefont
  {Roy}}, \bibinfo {author} {\bibfnamefont {F.~T.}\ \bibnamefont {Cerasoli}},
  \bibinfo {author} {\bibfnamefont {A.}~\bibnamefont {Jayaraj}}, \bibinfo
  {author} {\bibfnamefont {K.}~\bibnamefont {Tenzin}}, \bibinfo {author}
  {\bibfnamefont {M.~B.}\ \bibnamefont {Nardelli}},\ and\ \bibinfo {author}
  {\bibfnamefont {J.}~\bibnamefont {S{\l}awi{\'{n}}ska}},\ }\bibfield  {title}
  {\bibinfo {title} {{Long-range current-induced spin accumulation in chiral
  crystals}},\ }\href {https://doi.org/10.1038/s41524-022-00931-3} {\bibfield
  {journal} {\bibinfo  {journal} {npj Comput. Mater.}\ }\textbf {\bibinfo
  {volume} {8}},\ \bibinfo {pages} {243} (\bibinfo {year} {2022})}\BibitemShut
  {NoStop}%
\bibitem [{\citenamefont {Valet}\ and\ \citenamefont {Fert}(1993)}]{Valet1993}%
  \BibitemOpen
  \bibfield  {author} {\bibinfo {author} {\bibfnamefont {T.}~\bibnamefont
  {Valet}}\ and\ \bibinfo {author} {\bibfnamefont {A.}~\bibnamefont {Fert}},\
  }\bibfield  {title} {\bibinfo {title} {{Theory of the perpendicular
  magnetoresistance in magnetic multilayers}},\ }\href
  {https://doi.org/10.1103/PhysRevB.48.7099} {\bibfield  {journal} {\bibinfo
  {journal} {Phys. Rev. B}\ }\textbf {\bibinfo {volume} {48}},\ \bibinfo
  {pages} {7099} (\bibinfo {year} {1993})}\BibitemShut {NoStop}%
\bibitem [{Sup()}]{SupplementalMaterial}%
  \BibitemOpen
  \href@noop {} {}\bibinfo {note} {See Supplemental Material at [url] for
  supporting information---relations between our model and other SOC models,
  detail of the Boltzmann equation analysis in Sect.~II, microscopic derivation
  of the Boltzmann equation and transmission rate across the interface with
  arbitrary spin polarization in terms of the Keldysh Green's function, review
  of the Valet--Fert solution, derivation of the effective Boltzmann equation
  at the interface in Sect.~III, analytical expressions of electron
  distributions for Edelstein effect and its inverse, and dependence on the
  thickness of the three-dimensional metal, omitted in this main
  text.}\BibitemShut {Stop}%
\bibitem [{\citenamefont {Isshiki}\ \emph {et~al.}(2020)\citenamefont
  {Isshiki}, \citenamefont {Muduli}, \citenamefont {Kim}, \citenamefont
  {Kondou},\ and\ \citenamefont {Otani}}]{Isshiki2020}%
  \BibitemOpen
  \bibfield  {author} {\bibinfo {author} {\bibfnamefont {H.}~\bibnamefont
  {Isshiki}}, \bibinfo {author} {\bibfnamefont {P.}~\bibnamefont {Muduli}},
  \bibinfo {author} {\bibfnamefont {J.}~\bibnamefont {Kim}}, \bibinfo {author}
  {\bibfnamefont {K.}~\bibnamefont {Kondou}},\ and\ \bibinfo {author}
  {\bibfnamefont {Y.}~\bibnamefont {Otani}},\ }\bibfield  {title} {\bibinfo
  {title} {{Phenomenological model for the direct and inverse Edelstein
  effects}},\ }\href {https://doi.org/10.1103/PhysRevB.102.184411} {\bibfield
  {journal} {\bibinfo  {journal} {Phys. Rev. B}\ }\textbf {\bibinfo {volume}
  {102}},\ \bibinfo {pages} {184411} (\bibinfo {year} {2020})}\BibitemShut
  {NoStop}%
\bibitem [{\citenamefont {{\=O}nuki}\ \emph {et~al.}(2014)\citenamefont
  {{\=O}nuki}, \citenamefont {Nakamura}, \citenamefont {Uejo}, \citenamefont
  {Teruya}, \citenamefont {Hedo}, \citenamefont {Nakama}, \citenamefont
  {Honda},\ and\ \citenamefont {Harima}}]{Onuki2014}%
  \BibitemOpen
  \bibfield  {author} {\bibinfo {author} {\bibfnamefont {Y.}~\bibnamefont
  {{\=O}nuki}}, \bibinfo {author} {\bibfnamefont {A.}~\bibnamefont {Nakamura}},
  \bibinfo {author} {\bibfnamefont {T.}~\bibnamefont {Uejo}}, \bibinfo {author}
  {\bibfnamefont {A.}~\bibnamefont {Teruya}}, \bibinfo {author} {\bibfnamefont
  {M.}~\bibnamefont {Hedo}}, \bibinfo {author} {\bibfnamefont {T.}~\bibnamefont
  {Nakama}}, \bibinfo {author} {\bibfnamefont {F.}~\bibnamefont {Honda}},\ and\
  \bibinfo {author} {\bibfnamefont {H.}~\bibnamefont {Harima}},\ }\bibfield
  {title} {\bibinfo {title} {{Chiral-structure-driven split Fermi surface
  properties in TaSi$_2$, NbSi$_2$, and VSi$_2$}},\ }\href
  {https://doi.org/10.7566/JPSJ.83.061018} {\bibfield  {journal} {\bibinfo
  {journal} {J. Phys. Soc. Jpn.}\ }\textbf {\bibinfo {volume} {83}},\ \bibinfo
  {pages} {061018} (\bibinfo {year} {2014})}\BibitemShut {NoStop}%
\bibitem [{\citenamefont {{Yu A. Bychkov}}\ and\ \citenamefont
  {Rashba}(1984)}]{Bychkov1984}%
  \BibitemOpen
  \bibfield  {author} {\bibinfo {author} {\bibnamefont {{Yu A. Bychkov}}}\ and\
  \bibinfo {author} {\bibfnamefont {{\'E}.~I.}\ \bibnamefont {Rashba}},\
  }\bibfield  {title} {\bibinfo {title} {{Properties of a 2D electron gas with
  lifted spectral degeneracy}},\ }\href@noop {} {\bibfield  {journal} {\bibinfo
   {journal} {Pis'ma Zh. Eksp. Teor. Fiz.}\ }\textbf {\bibinfo {volume} {39}},\
  \bibinfo {pages} {66} (\bibinfo {year} {1984})},\ \bibinfo {note} {[JETP
  Lett. \textbf{39}, {78} ({1984})]}\BibitemShut {NoStop}%
\bibitem [{Note2()}]{Note2}%
  \BibitemOpen
  \bibinfo {note} {Our SOC model can also be converted to a 2D system with both
  Rashba and Dresselhaus SOCs under appropriate rotations~\cite
  [Sect.~S1]{SupplementalMaterial}.}\BibitemShut {Stop}%
\bibitem [{\citenamefont {Silsbee}(2001)}]{Silsbee2001}%
  \BibitemOpen
  \bibfield  {author} {\bibinfo {author} {\bibfnamefont {R.~H.}\ \bibnamefont
  {Silsbee}},\ }\bibfield  {title} {\bibinfo {title} {{Theory of the detection
  of current-induced spin polarization in a two-dimensional electron gas}},\
  }\href {https://doi.org/10.1103/PhysRevB.63.155305} {\bibfield  {journal}
  {\bibinfo  {journal} {Phys. Rev. B}\ }\textbf {\bibinfo {volume} {63}},\
  \bibinfo {pages} {155305} (\bibinfo {year} {2001})}\BibitemShut {NoStop}%
\bibitem [{Note3()}]{Note3}%
  \BibitemOpen
  \bibinfo {note} {Such an extremely slowly decaying mode stems from the
  conservation of charge.}\BibitemShut {Stop}%
\bibitem [{\citenamefont {Szolnoki}\ \emph {et~al.}(2017)\citenamefont
  {Szolnoki}, \citenamefont {D{\'{o}}ra}, \citenamefont {Kiss}, \citenamefont
  {Fabian},\ and\ \citenamefont {Simon}}]{Szolnoki2017}%
  \BibitemOpen
  \bibfield  {author} {\bibinfo {author} {\bibfnamefont {L.}~\bibnamefont
  {Szolnoki}}, \bibinfo {author} {\bibfnamefont {B.}~\bibnamefont
  {D{\'{o}}ra}}, \bibinfo {author} {\bibfnamefont {A.}~\bibnamefont {Kiss}},
  \bibinfo {author} {\bibfnamefont {J.}~\bibnamefont {Fabian}},\ and\ \bibinfo
  {author} {\bibfnamefont {F.}~\bibnamefont {Simon}},\ }\bibfield  {title}
  {\bibinfo {title} {{Intuitive approach to the unified theory of spin
  relaxation}},\ }\href {https://doi.org/10.1103/PhysRevB.96.245123} {\bibfield
   {journal} {\bibinfo  {journal} {Phys. Rev. B}\ }\textbf {\bibinfo {volume}
  {96}},\ \bibinfo {pages} {245123} (\bibinfo {year} {2017})}\BibitemShut
  {NoStop}%
\bibitem [{\citenamefont {Elliott}(1954)}]{Elliott1954}%
  \BibitemOpen
  \bibfield  {author} {\bibinfo {author} {\bibfnamefont {R.~J.}\ \bibnamefont
  {Elliott}},\ }\bibfield  {title} {\bibinfo {title} {{Theory of the effect of
  spin-orbit coupling on magnetic resonance in some semiconductors}},\ }\href
  {https://doi.org/10.1103/PhysRev.96.266} {\bibfield  {journal} {\bibinfo
  {journal} {Phys. Rev.}\ }\textbf {\bibinfo {volume} {96}},\ \bibinfo {pages}
  {266} (\bibinfo {year} {1954})}\BibitemShut {NoStop}%
\bibitem [{\citenamefont {Yafet}(1963)}]{Yafet1963}%
  \BibitemOpen
  \bibfield  {author} {\bibinfo {author} {\bibfnamefont {Y.}~\bibnamefont
  {Yafet}},\ }\bibfield  {title} {\bibinfo {title} {{g Factors and spin-lattice
  relaxation of conduction electrons}},\ }in\ \href@noop {} {\emph {\bibinfo
  {booktitle} {Solid State Physics}}},\ \bibinfo {series} {Advances in Research
  and Applications}, Vol.~\bibinfo {volume} {14},\ \bibinfo {editor} {edited
  by\ \bibinfo {editor} {\bibfnamefont {F.}~\bibnamefont {Seitz}}\ and\
  \bibinfo {editor} {\bibfnamefont {D.}~\bibnamefont {Turnbull}}}\ (\bibinfo
  {publisher} {Academic Press},\ \bibinfo {address} {New York and London},\
  \bibinfo {year} {1963})\ pp.\ \bibinfo {pages} {1--98}\BibitemShut {NoStop}%
\bibitem [{\citenamefont {D'yakonov}\ and\ \citenamefont
  {Perel'}(1971)}]{Dyakonov1972}%
  \BibitemOpen
  \bibfield  {author} {\bibinfo {author} {\bibfnamefont {M.~I.}\ \bibnamefont
  {D'yakonov}}\ and\ \bibinfo {author} {\bibfnamefont {V.~I.}\ \bibnamefont
  {Perel'}},\ }\bibfield  {title} {\bibinfo {title} {Spin relaxation of
  conduction electrons in noncentrosymmetric semiconductors},\ }\href@noop {}
  {\bibfield  {journal} {\bibinfo  {journal} {Fiz. Tverd. Tela}\ }\textbf
  {\bibinfo {volume} {13}},\ \bibinfo {pages} {3581} (\bibinfo {year}
  {1971})},\ \bibinfo {note} {[Sov. Phys. Solid State \textbf{13}, 3023
  (1972)]}\BibitemShut {NoStop}%
\bibitem [{\citenamefont {Overhauser}(1953)}]{Overhauser1953}%
  \BibitemOpen
  \bibfield  {author} {\bibinfo {author} {\bibfnamefont {A.~W.}\ \bibnamefont
  {Overhauser}},\ }\bibfield  {title} {\bibinfo {title} {{Paramagnetic
  relaxation in metals}},\ }\href {https://doi.org/10.1103/PhysRev.89.689}
  {\bibfield  {journal} {\bibinfo  {journal} {Phys. Rev.}\ }\textbf {\bibinfo
  {volume} {89}},\ \bibinfo {pages} {689} (\bibinfo {year} {1953})}\BibitemShut
  {NoStop}%
\bibitem [{\citenamefont {\ifmmode \check{Z}\else
  \v{Z}\fi{}uti\ifmmode~\acute{c}\else \'{c}\fi{}}\ \emph
  {et~al.}(2004)\citenamefont {\ifmmode \check{Z}\else
  \v{Z}\fi{}uti\ifmmode~\acute{c}\else \'{c}\fi{}}, \citenamefont {Fabian},\
  and\ \citenamefont {{Das Sarma}}}]{Zutic2004}%
  \BibitemOpen
  \bibfield  {author} {\bibinfo {author} {\bibfnamefont {I.}~\bibnamefont
  {\ifmmode \check{Z}\else \v{Z}\fi{}uti\ifmmode~\acute{c}\else \'{c}\fi{}}},
  \bibinfo {author} {\bibfnamefont {J.}~\bibnamefont {Fabian}},\ and\ \bibinfo
  {author} {\bibfnamefont {S.}~\bibnamefont {{Das Sarma}}},\ }\bibfield
  {title} {\bibinfo {title} {{Spintronics: Fundamentals and applications}},\
  }\href {https://doi.org/10.1103/RevModPhys.76.323} {\bibfield  {journal}
  {\bibinfo  {journal} {Rev. Mod. Phys.}\ }\textbf {\bibinfo {volume} {76}},\
  \bibinfo {pages} {323} (\bibinfo {year} {2004})}\BibitemShut {NoStop}%
\bibitem [{\citenamefont {Seemann}\ \emph {et~al.}(2015)\citenamefont
  {Seemann}, \citenamefont {K{\"{o}}dderitzsch}, \citenamefont {Wimmer},\ and\
  \citenamefont {Ebert}}]{Seemann2015}%
  \BibitemOpen
  \bibfield  {author} {\bibinfo {author} {\bibfnamefont {M.}~\bibnamefont
  {Seemann}}, \bibinfo {author} {\bibfnamefont {D.}~\bibnamefont
  {K{\"{o}}dderitzsch}}, \bibinfo {author} {\bibfnamefont {S.}~\bibnamefont
  {Wimmer}},\ and\ \bibinfo {author} {\bibfnamefont {H.}~\bibnamefont
  {Ebert}},\ }\bibfield  {title} {\bibinfo {title} {{Symmetry-imposed shape of
  linear response tensors}},\ }\href
  {https://doi.org/10.1103/PhysRevB.92.155138} {\bibfield  {journal} {\bibinfo
  {journal} {Phys. Rev. B}\ }\textbf {\bibinfo {volume} {92}},\ \bibinfo
  {pages} {155138} (\bibinfo {year} {2015})}\BibitemShut {NoStop}%
\bibitem [{\citenamefont {Rammer}\ and\ \citenamefont
  {Smith}(1986)}]{Rammer1986}%
  \BibitemOpen
  \bibfield  {author} {\bibinfo {author} {\bibfnamefont {J.}~\bibnamefont
  {Rammer}}\ and\ \bibinfo {author} {\bibfnamefont {H.}~\bibnamefont {Smith}},\
  }\bibfield  {title} {\bibinfo {title} {{Quantum field-theoretical methods in
  transport theory of metals}},\ }\href
  {https://doi.org/10.1103/RevModPhys.58.323} {\bibfield  {journal} {\bibinfo
  {journal} {Rev. Mod. Phys.}\ }\textbf {\bibinfo {volume} {58}},\ \bibinfo
  {pages} {323} (\bibinfo {year} {1986})}\BibitemShut {NoStop}%
\bibitem [{\citenamefont {Levanda}\ and\ \citenamefont
  {Fleurov}(2001)}]{Levanda2001}%
  \BibitemOpen
  \bibfield  {author} {\bibinfo {author} {\bibfnamefont {M.}~\bibnamefont
  {Levanda}}\ and\ \bibinfo {author} {\bibfnamefont {V.}~\bibnamefont
  {Fleurov}},\ }\bibfield  {title} {\bibinfo {title} {{A Wigner
  quasi-distribution function for charged particles in classical
  electromagnetic fields}},\ }\href {https://doi.org/10.1006/aphy.2001.6170}
  {\bibfield  {journal} {\bibinfo  {journal} {Ann. Phys. (N.Y.)}\ }\textbf
  {\bibinfo {volume} {292}},\ \bibinfo {pages} {199} (\bibinfo {year}
  {2001})}\BibitemShut {NoStop}%
\bibitem [{\citenamefont {Kita}(2001)}]{Kita2001}%
  \BibitemOpen
  \bibfield  {author} {\bibinfo {author} {\bibfnamefont {T.}~\bibnamefont
  {Kita}},\ }\bibfield  {title} {\bibinfo {title} {{Gauge invariance and Hall
  terms in the quasiclassical equations of superconductivity}},\ }\href
  {https://doi.org/10.1103/PhysRevB.64.054503} {\bibfield  {journal} {\bibinfo
  {journal} {Phys. Rev. B}\ }\textbf {\bibinfo {volume} {64}},\ \bibinfo
  {pages} {054503} (\bibinfo {year} {2001})}\BibitemShut {NoStop}%
\bibitem [{\citenamefont {Onoda}\ \emph {et~al.}(2006)\citenamefont {Onoda},
  \citenamefont {Sugimoto},\ and\ \citenamefont {Nagaosa}}]{Onoda2006}%
  \BibitemOpen
  \bibfield  {author} {\bibinfo {author} {\bibfnamefont {S.}~\bibnamefont
  {Onoda}}, \bibinfo {author} {\bibfnamefont {N.}~\bibnamefont {Sugimoto}},\
  and\ \bibinfo {author} {\bibfnamefont {N.}~\bibnamefont {Nagaosa}},\
  }\bibfield  {title} {\bibinfo {title} {{Theory of non-equilibirum states
  driven by constant electromagnetic fields}},\ }\href
  {https://doi.org/10.1143/ptp.116.61} {\bibfield  {journal} {\bibinfo
  {journal} {Prog. Theor. Phys.}\ }\textbf {\bibinfo {volume} {116}},\ \bibinfo
  {pages} {61} (\bibinfo {year} {2006})}\BibitemShut {NoStop}%
\bibitem [{\citenamefont {D'yakonov}\ and\ \citenamefont
  {Khaetskii}(1984)}]{Dyakonov1984}%
  \BibitemOpen
  \bibfield  {author} {\bibinfo {author} {\bibfnamefont {M.~I.}\ \bibnamefont
  {D'yakonov}}\ and\ \bibinfo {author} {\bibfnamefont {A.~V.}\ \bibnamefont
  {Khaetskii}},\ }\bibfield  {title} {\bibinfo {title} {{Relaxation of
  nonequilibrium carrier-density matrix in semiconductors with degenerate
  bands}},\ }\href@noop {} {\bibfield  {journal} {\bibinfo  {journal} {Zh.
  Eksp. Teor. Fiz}\ }\textbf {\bibinfo {volume} {86}},\ \bibinfo {pages} {1843}
  (\bibinfo {year} {1984})},\ \bibinfo {note} {[Sov. Phys. JETP \textbf{59},
  {1072} ({1984})]}\BibitemShut {NoStop}%
\bibitem [{\citenamefont {Khaetskii}(2006)}]{Khaetskii2006}%
  \BibitemOpen
  \bibfield  {author} {\bibinfo {author} {\bibfnamefont {A.}~\bibnamefont
  {Khaetskii}},\ }\bibfield  {title} {\bibinfo {title} {{Nonexistence of
  intrinsic spin currents}},\ }\href
  {https://doi.org/10.1103/PhysRevLett.96.056602} {\bibfield  {journal}
  {\bibinfo  {journal} {Phys. Rev. Lett.}\ }\textbf {\bibinfo {volume} {96}},\
  \bibinfo {pages} {2} (\bibinfo {year} {2006})}\BibitemShut {NoStop}%
\bibitem [{\citenamefont {Shytov}\ \emph {et~al.}(2006)\citenamefont {Shytov},
  \citenamefont {Mishchenko}, \citenamefont {Engel},\ and\ \citenamefont
  {Halperin}}]{Shytov2006}%
  \BibitemOpen
  \bibfield  {author} {\bibinfo {author} {\bibfnamefont {A.~V.}\ \bibnamefont
  {Shytov}}, \bibinfo {author} {\bibfnamefont {E.~G.}\ \bibnamefont
  {Mishchenko}}, \bibinfo {author} {\bibfnamefont {H.-A.}\ \bibnamefont
  {Engel}},\ and\ \bibinfo {author} {\bibfnamefont {B.~I.}\ \bibnamefont
  {Halperin}},\ }\bibfield  {title} {\bibinfo {title} {{Small-angle impurity
  scattering and the spin Hall conductivity in two-dimensional semiconductor
  systems}},\ }\href {https://doi.org/10.1103/PhysRevB.73.075316} {\bibfield
  {journal} {\bibinfo  {journal} {Phys. Rev. B}\ }\textbf {\bibinfo {volume}
  {73}},\ \bibinfo {pages} {075316} (\bibinfo {year} {2006})}\BibitemShut
  {NoStop}%
\bibitem [{\citenamefont {Kailasvuori}(2009)}]{Kailasvuori2009}%
  \BibitemOpen
  \bibfield  {author} {\bibinfo {author} {\bibfnamefont {J.}~\bibnamefont
  {Kailasvuori}},\ }\bibfield  {title} {\bibinfo {title} {{Boltzmann approach
  to the spin Hall effect revisited and electric field modified collision
  integrals}},\ }\href {https://doi.org/10.1088/1742-5468/2009/08/P08004}
  {\bibfield  {journal} {\bibinfo  {journal} {J. Stat. Mec.}\ }\textbf
  {\bibinfo {volume} {2009}},\ \bibinfo {pages} {P08004} (\bibinfo {year}
  {2009})}\BibitemShut {NoStop}%
\bibitem [{\citenamefont {Jungwirth}\ \emph {et~al.}(2002)\citenamefont
  {Jungwirth}, \citenamefont {Niu},\ and\ \citenamefont
  {MacDonald}}]{Jungwirth2002}%
  \BibitemOpen
  \bibfield  {author} {\bibinfo {author} {\bibfnamefont {T.}~\bibnamefont
  {Jungwirth}}, \bibinfo {author} {\bibfnamefont {Q.}~\bibnamefont {Niu}},\
  and\ \bibinfo {author} {\bibfnamefont {A.~H.}\ \bibnamefont {MacDonald}},\
  }\bibfield  {title} {\bibinfo {title} {{Anomalous Hall effect in
  ferromagnetic semiconductors}},\ }\href
  {https://doi.org/10.1103/PhysRevLett.88.207208} {\bibfield  {journal}
  {\bibinfo  {journal} {Phys. Rev. Lett.}\ }\textbf {\bibinfo {volume} {88}},\
  \bibinfo {pages} {207208} (\bibinfo {year} {2002})}\BibitemShut {NoStop}%
\bibitem [{\citenamefont {Onoda}\ and\ \citenamefont
  {Nagaosa}(2002)}]{Onoda2002}%
  \BibitemOpen
  \bibfield  {author} {\bibinfo {author} {\bibfnamefont {M.}~\bibnamefont
  {Onoda}}\ and\ \bibinfo {author} {\bibfnamefont {N.}~\bibnamefont
  {Nagaosa}},\ }\bibfield  {title} {\bibinfo {title} {{Topological nature of
  anomalous Hall effect in ferromagnets}},\ }\href
  {https://doi.org/10.1143/JPSJ.71.19} {\bibfield  {journal} {\bibinfo
  {journal} {J. Phys. Soc. Jpn.}\ }\textbf {\bibinfo {volume} {71}},\ \bibinfo
  {pages} {19} (\bibinfo {year} {2002})}\BibitemShut {NoStop}%
\bibitem [{\citenamefont {Kohn}\ and\ \citenamefont
  {Luttinger}(1957)}]{Kohn1957}%
  \BibitemOpen
  \bibfield  {author} {\bibinfo {author} {\bibfnamefont {W.}~\bibnamefont
  {Kohn}}\ and\ \bibinfo {author} {\bibfnamefont {J.~M.}\ \bibnamefont
  {Luttinger}},\ }\bibfield  {title} {\bibinfo {title} {{Quantum theory of
  electrical transport phenomena}},\ }\href
  {https://doi.org/10.1103/PhysRev.108.590} {\bibfield  {journal} {\bibinfo
  {journal} {Phys. Rev.}\ }\textbf {\bibinfo {volume} {108}},\ \bibinfo {pages}
  {590} (\bibinfo {year} {1957})}\BibitemShut {NoStop}%
\bibitem [{\citenamefont {Inoue}\ \emph {et~al.}(2004)\citenamefont {Inoue},
  \citenamefont {Bauer},\ and\ \citenamefont {Molenkamp}}]{Inoue2004}%
  \BibitemOpen
  \bibfield  {author} {\bibinfo {author} {\bibfnamefont {J.-i.}\ \bibnamefont
  {Inoue}}, \bibinfo {author} {\bibfnamefont {G.~E.~W.}\ \bibnamefont
  {Bauer}},\ and\ \bibinfo {author} {\bibfnamefont {L.~W.}\ \bibnamefont
  {Molenkamp}},\ }\bibfield  {title} {\bibinfo {title} {{Suppression of the
  persistent spin Hall current by defect scattering}},\ }\href
  {https://doi.org/10.1103/PhysRevB.70.041303} {\bibfield  {journal} {\bibinfo
  {journal} {Phys. Rev. B}\ }\textbf {\bibinfo {volume} {70}},\ \bibinfo
  {pages} {041303} (\bibinfo {year} {2004})}\BibitemShut {NoStop}%
\end{thebibliography}

\end{document}